\documentclass[ba]{imsart}

\RequirePackage[authoryear]{natbib}   
\usepackage{bibunits}                 

\defaultbibliographystyle{ba}         
\defaultbibliography{references}      

\RequirePackage{amsthm,amsmath,amsfonts,amssymb}
\RequirePackage[authoryear]{natbib}
\usepackage[colorlinks,citecolor=blue,urlcolor=blue,breaklinks=true]{hyperref}

\usepackage{breakurl}
\RequirePackage{graphicx}

\usepackage{url}            
\usepackage{booktabs}       
\usepackage{float}
\usepackage{subcaption}
\usepackage{xcolor}         
\usepackage{tikz}
\usetikzlibrary{shapes,snakes}
\usepackage{cancel}
\usepackage{makecell} 
\usepackage{scalerel,stackengine}
\usepackage[section]{placeins}
\stackMath
\newcommand\reallywidehat[1]{%
\savestack{\tmpbox}{\stretchto{%
  \scaleto{%
    \scalerel*[\widthof{\ensuremath{#1}}]{\kern.1pt\mathchar"0362\kern.1pt}%
    {\rule{0ex}{\textheight}}
  }{\textheight}%
}{2.4ex}}%
\stackon[-6.9pt]{#1}{\tmpbox}%
}

\usepackage{siunitx}
\usepackage[acronym]{glossaries}
\newacronym{lalme}{LALME}{Linguistic Atlas of Late Mediaeval English}
\newacronym{lp}{LP}{Linguistic Profile}

\startlocaldefs
\theoremstyle{plain}

\newtheorem{theorem}{Theorem}[section]
\newtheorem{lemma}[theorem]{Lemma}
\newtheorem{corollary}[theorem]{Corollary}
\theoremstyle{definition}

\theoremstyle{remark}

\def\PMSE{{\mbox{PMSE}}}

\def\P{{\mathcal{P}}}
\def\D{{\mathcal{D}}}
\def\L{{\mathcal{L}}}
\def\E{{\mathbb{E}}}

\def\bara{m} 
\def\barA{M} 
\def\ELBO{\mbox{ELBO}}
\def\aELBO{\text{A-ELBO}}
\def\I{{\mathcal{I}}}
\def\X{{\mathcal{X}}}
\def\F{{\mathcal{F}}}
\def\B{{\mathcal{B}}}
\def\C{{\mathcal{C}}}
\def\K{{\mathcal{K}}}
\def\U{{\mathcal{U}}}
\def\Q{{\mathcal{Q}}}
\def\tpsi{{\psi}}
\def\lf{$\langle$}
\def\rf{$\rangle$}
\endlocaldefs

\begin{document}
\begin{bibunit}
\setcounter{secnumdepth}{3}

\begin{frontmatter}
\title{Amortising over hyperparameters in Generalised Bayesian Inference}
\runtitle{Amortising over hyperparameters in GBI}

\begin{aug}
\author[A]{\fnms{Laura}~\snm{Battaglia}\ead[label=e1]{laura.battaglia@stats.ox.ac.uk}},
\author[B]{\fnms{Chris U.} \snm{Carmona}\ead[label=e2]{ccarmona@berkeley.edu}},
\author[C]{\fnms{Ross A.} \snm{Haines}\ead[label=e3]{ross.a.haines@gmail.com}},
\author[A]{\fnms{Max} \snm{Anderson Loake}\ead[label=e4]{max.andersonloake@keble.ox.ac.uk}},
\author[D]{\fnms{Michael} \snm{Benskin}\ead[label=e5]{michael.benskin@ilos.uio.no}}
\and
\author[A]{\fnms{Geoff K.} \snm{Nicholls}\ead[label=e6]{nicholls@stats.ox.ac.uk}}
\address[A]{Department of Statistics, University of Oxford, \printead{e1,e4}}
\address[A]{\printead{e6}}
\address[B]{Amazon, Berlin, \printead{e2}}
\address[C]{Palantir Technologies, \printead{e3}}
\address[D]{Department of Literature, Area Studies and European Languages, University of Oslo, \printead{e5}}
\runauthor{Battaglia et al.}
\end{aug}

\begin{abstract}
 In Bayesian inference prior hyperparameters are chosen subjectively or estimated using empirical Bayes methods.  Generalised Bayesian Inference (GBI) also has a learning rate hyperparameter. This is compounded in Semi-Modular Inference (SMI), a GBI framework for multiple datasets (multi-modular problems). As part of any GBI workflow it is necessary to check sensitivity to the choice of hyperparameters, but running MCMC or fitting a variational approximation at each of the hyperparameter values of interest is impractical. Simulation-based Inference has been used by previous authors to amortise over data and hyperparameters, fitting a posterior approximation targeting the forward-KL divergence. However, for GBI and SMI posteriors, it is not possible to amortise over data, as there {\it is no generative model}. 
 Working with a variational family parameterised by a conditional normalising flow, we give a direct variational approximation for GBI and SMI posteriors, targeting the reverse-KL divergence, and amortised over prior and loss hyperparameters at fixed data. This can be sampled efficiently at different hyperparameter values without refitting, and supports efficient robustness checks and hyperparameter selection.
 We show that there exist amortised conditional normalising-flow architectures which are universal approximators. We illustrate our methods with an epidemiological example well known in SMI work and then give the motivating application, a spatial location-prediction task for linguistic-profile data. SMI gives improved prediction with hyperparameters chosen using our amortised framework. The code is available online.
\end{abstract}


\begin{keyword}
\kwd{Generalised Bayes}
\kwd{Semi-Modular Inference}
\kwd{Variational Inference}
\kwd{Normalising Flow}
\kwd{Hyperparameters}
\kwd{Linguistic Profiles}
\end{keyword}

\end{frontmatter}

\section{Introduction}\label{sec:intro}

Generalised Bayesian Inference (GBI, \cite{bissiri_general_2016,grunwald_inconsistency_2017}) has recently emerged as a flexible framework for updating belief in settings where model misspecification is an issue. Semi-Modular Inference (SMI, \cite{carmona_semi-modular_2020, frazier23posterior}) extends this to handle models that combine multiple modules, some of which may be misspecified. This is achieved by introducing a learning-rate parameter $\eta$ which balances the contribution of individual data modules to shared latent variables. Alongside these learning-rate parameters, SMI retains prior hyperparameters that control belief about latent structure before observing data.

Both types of hyperparameters, model hyperparameters and learning rates, can significantly shape the resulting posterior and downstream inferential conclusions in GBI and SMI. It is therefore important to assess the sensitivity of inference to their values, yet this process can be computationally expensive. Evaluating posterior expectations or predictive quantities across a grid of hyperparameter settings typically involves repeating full inference procedures, whether based on MCMC or variational methods, for each value. In practice, this makes principled loss-based hyperparameter tuning prohibitive.

To address this challenge, we propose an amortised variational framework that enables fast exploration of hyperparameter effects. We use variational inference with a normalising flow family, following \citet{carmona_scalable_2022}, who condition a variational approximation on the learning rate (so their approximation is amortised over $\eta$). We build on this idea and propose a more expressive and general amortised variational architecture that conditions on the full set of hyperparameters, including both prior parameters and learning rates. Once trained, this model defines a single variational approximation to the entire family of SMI posteriors, enabling fast exploration of how the posterior varies across the hyperparameter space.

Once such a model is available, it becomes straightforward to define and compare different notions of hyperparameter optimality. The marginal likelihood, for instance, is a classical choice for hyperparameter selection, and can be approximated directly from the trained variational model. However, this criterion is known to perform poorly in the presence of model misspecification \citep{grunwald_inconsistency_2017, gelman2020holes, fortuin2022priors}. Motivated by this, we also consider different loss-aware objectives, such as the Expected Log Pointwise Predictive Density (ELPD), which prioritises predictive accuracy, and the Posterior Mean Squared Error (PMSE), which targets parameter estimation. This loss is available in settings like supervised classification, where the parameters we wish to estimate, the class labels, are available for some observations and can be held out.

Although our focus is on SMI, the proposed framework applies more broadly to settings in which variational inference is preferred over other sampling-based approaches. This includes situations where MCMC is computationally prohibitive due to the high dimensionality or complexity of the model, and where a reverse-KL objective maximising the evidence lower bound (ELBO) is convenient.  
Forward-KL approaches such as Simulation-based Inference (SBI) have proven effective for Bayesian inference, with the useful property of not requiring a tractable likelihood \citep{radev2020bayesflow, sharrock2022sequential, siahkoohi2023reliable, radev2023jana, wildberger2023flow, schmitt2024consistency, elsemuller2024sensitivityaware}. However, they generally rely on sampling parameter-data pairs by simulating the observation model.
In GBI and SMI the normalised likelihood is replaced by an unnormalised exponential loss, and the generative model at the heart of SBI is not available. There is a growing line of work on SBI for GBI \citep{pacchiardi2024generalized,gao2023generalized}. However, these typically rely on using MCMC to target an approximate posterior. In our setting, the ELBO objective is tractable, and we amortise over hyperameters, not data, so the absence of a generative model is not an obstacle.

\section{Background and motivation} \label{sec:background-motivation}

\subsection{Semi-modular inference}\label{sec:smi}

``Since all models are wrong the scientist cannot obtain
a `correct' one by excessive elaboration'' \citep{box76}. 
GBI offers a way to update beliefs when the model is misspecified. The likelihood is replaced by a ``generalised likelihood function'', $\exp({-\eta\,\ell(\theta;Y)})$ which need not be normalised over the data $Y$. Here $\ell(\theta;Y)$ is a loss function expressing the utility of parameters $\theta\in\mathbb{R}^p$ given data $Y$, and $\eta \in \mathbb{R}^+$ is an inverse temperature or ``learning rate'' hyperparameter.
If the prior for $\theta$ is $p(\theta|\varphi)$ with hyperparameters $\varphi\in\mathbb{R}^{d_\varphi}$ and $\tpsi=\{\eta,\varphi\},\ \tpsi\in \Psi=\mathbb{R}^{d_\varphi}\times \mathbb{R}^+$ then the belief update is given by the GBI posterior 
\begin{align}\label{eq:GB-basic-posterior}
p(\theta \mid Y,\tpsi)\propto \exp({-\eta\,\ell(\theta;Y)})\,p(\theta|\varphi).
\end{align}
Practitioners can choose loss functions that are robust to particular forms of model misspecification \citep{jewson2022general,kelly2025simulation}. In this paper our loss is the negative log-likelihood loss. Because $\eta$ is not fixed to one as it would be in Bayesian inference, this makes \eqref{eq:GB-basic-posterior} a \emph{power posterior} \citep{grunwald_inconsistency_2017}.

SMI is a variant of GBI designed for combining multiple datasets using modular generative models that share parameters, and accounting for the fact that some modules might be misspecified. Consider a model with two modules $\{Z, \delta\}$ and $\{Y, \theta, \delta\}$, including data $Y = (Y_1, \dots, Y_n)$ and $Z = (Z_1, \dots, Z_m)$, a shared parameter $\delta$ and prior hyperparameters $\varphi$. The generative model is
\begin{align}
    \delta &\sim p(\delta; \varphi), \\
    Z &\sim p(Z \mid \delta), \\
    \theta &\sim p(\theta \mid \delta; \varphi), \\
    Y &\sim p(Y \mid \theta, \delta). \label{eq:SMI_graph_as_gen_model}
\end{align}
The full Bayesian posterior can be written as
\begin{align}
    p(\delta, \theta \mid Z, Y; \varphi) = p(\delta \mid Z, Y; \varphi)\, p(\theta \mid \delta, Y; \varphi). \label{eq:bayes_posterior_factors}
\end{align}
This decomposition resembles Bayesian Multiple Imputation (BMI, \cite{meng94}): the posterior distribution of $\delta$ is imputed using its marginal posterior given both datasets, and this random variable is passed to the conditional for $\theta$ for analysis. 

SMI as originally formulated by \cite{carmona_semi-modular_2020} supposes that the $Z$-module is well-specified but the $Y$-module has potential misspecification. The influence of $Y$ on $\delta$ is tempered using an auxiliary variable $\tilde{\theta}$ and a learning rate $\eta \in [0, 1]$. The resulting SMI posterior modifies the imputation stage in Equation \ref{eq:bayes_posterior_factors} to be
\begin{align}
    p_{\text{smi}}(\delta, \theta, \tilde{\theta} \mid Y, Z; \tpsi) 
    &:= p_{\text{pow}}(\delta, \tilde{\theta} \mid Y, Z; \tpsi) \, p(\theta \mid \delta, Y; \varphi),
    \label{eq:smi-posterior}
\end{align}
and the power posterior is
\begin{align}
    p_{\text{pow}}(\delta, \tilde{\theta} \mid Y, Z; \tpsi) 
    &:= \frac{p(Z \mid \delta)\, p(Y \mid \delta, \tilde{\theta})^{\eta} \, p(\delta, \tilde{\theta}; \varphi)}{p(Z, Y; \eta)}.
    \label{eq:SMI_power_posterior}
\end{align}
Here, $\tilde{\theta}$ acts as an auxiliary copy of $\theta$ that connects $Y$ to $\delta$ during imputation. The learning rate $\eta$ controls how much information flows from the $Y$-likelihood to $\delta$. Figure \ref{fig:SMI_diagram} illustrates the process. SMI interpolates between the Bayes posterior \eqref{eq:bayes_posterior_factors} (at $\eta=1$) and the Cut posterior, (\cite{lunn2009combining, plummer_cuts_2015, jacob_better_2017, pompe2021asymptotics, yu-nott-cutvi23}, at $\eta=0$) where information from $Y$ is cut and $\delta$ is informed by $Z$ alone. The power posterior in \eqref{eq:SMI_power_posterior} is a variant of GBI in \eqref{eq:GB-basic-posterior} in which $\eta\,\ell(\theta;Y)$ is replaced by $\log(p(Z|\delta))-\eta\log(p(Y|\tilde\theta,\delta))$, reflecting our belief that the observation model for $Z$ is accurate but the likelihood for $Y$ may need to be downweighted.  
\begin{figure}[htb]
    \centering
    \resizebox{0.9\textwidth}{!}{%
    \begin{tikzpicture}[scale=1, every node/.style={transform shape}]

    \node at (-4, -1.5) {(a) Imputation stage of $\delta$};
    \node at (4, -1.5) {(b) Analysis stage of $\theta$};

    \node[rectangle, draw, inner sep=0pt, minimum size=1cm] (Z1) at (-6, 0) {$Z$};
    \node[rectangle, draw, inner sep=0pt, minimum size=1cm] (Y1) at (-2, 0) {$Y$};
    \node[circle, draw, inner sep=0pt, minimum size=1cm] (delta1) at (-6, 2) {$\delta$};
    \node[circle, draw, inner sep=0pt, minimum size=1cm] (thetatilde1) at (-2, 2) {$\tilde{\theta}$};
    \node[rectangle, draw, inner sep=2pt, minimum size=1cm] (psi1) at (-4, 3.5) {$\psi$};

    \draw[->, >=stealth] (psi1) -- (delta1);
    \draw[->, >=stealth] (psi1) -- (thetatilde1);
    \draw[->, >=stealth] (delta1) -- (thetatilde1);
    \draw[->, >=stealth] (delta1) -- (Y1.west);
    \draw[->, >=stealth] (delta1.south) -- (Z1.north);
    \draw[->, >=stealth] (thetatilde1.south) -- (Y1.north);

    \draw[->, red, thick] (-3.5,0.85) -- (-4.5,0.85); 
    \node[red] at (-3.8,1.05) {$\eta$};

    \node[rectangle, draw, inner sep=0pt, minimum size=1cm] (Y2) at (6, 0) {$Y$};
    \node[regular polygon, regular polygon sides=5, draw, inner sep=0pt, minimum size=1.2cm] (delta2) at (2, 2) {$\delta$}; 
    \node[circle, draw, inner sep=0pt, minimum size=1cm] (theta2) at (6, 2) {$\theta$};
    \node[rectangle, draw, inner sep=2pt, minimum size=1cm] (psi2) at (4, 3.5) {$\psi$};

    \draw[->, >=stealth] (psi2) -- (delta2);
    \draw[->, >=stealth] (psi2) -- (theta2);  
    \draw[->, >=stealth] (delta2) -- (theta2);
    \draw[->, >=stealth] (delta2) -- (Y2.west);
    \draw[->, >=stealth] (theta2.south) -- (Y2.north);

    \end{tikzpicture}
    }%
    \caption{Semi-modular Inference: in (a) the posterior for $\delta$ is imputed making use of auxiliary $\tilde{\theta}$, where the influence of the $Y$-module is controlled by $\eta$ and then in (b) the posterior for $\theta$ is computed conditional on the distribution of $\delta$. Circles denote parameters that are being inferred; squares denote quantities treated as fixed during inference; the pentagon indicates that the {\it distribution} of $\delta$ is fixed in (b), inherited from (a) and not further updated.}
    \label{fig:SMI_diagram}
\end{figure}

This setup already has useful generality. For example, it arises naturally when ``experimental'' data $Z$ informing $\delta$ are paired with ``observational'' data $Y$ which also depend on $\delta$ (as in Section~\ref{sec:hpv} below and in the Covid-prevalence model in \cite{Nicholson22}) and it has the right structure for the more complex analysis in Section~\ref{sec:LP-data}. 
Other ways of interpolating between the Cut posterior and Bayes give variants of SMI \citep{nicholls_valid_2022}. \citet{chakraborty_modularized_2022, frazier23posterior} develop on an approach based on linear opinion pooling, for which \citet{frazier23posterior} provide formal guarantees of more accurate inferences than either the cut or exact posterior, particularly under local model misspecification. Further developments and applications can be found in \citet{liu_general_2022, frazier24cutting, carmona_scalable_2022}, with the latter providing the foundation for the approach taken in this paper.

In SMI, the choice of learning rate $\eta$ and the hyperparameters of the prior heavily impact inference.  Work on learning rate estimation for GBI has been conducted separately from work on model hyperparameter selection (see Section~\ref{sec:related-work} for discussion). However, there are advantages to doing this jointly: as \cite{frazier23posterior} highlight in the context of SMI, it is important to use a high-level loss $\L:\Psi\to\mathbb{R}$ expressing inferential goals (such as prediction of new data, or parameter estimation) to guide the choice of learning rate; in this paper we select all the hyperparameters $\tpsi=(\eta,\varphi)$ jointly to minimise $\L(\tpsi)$, so we optimise over a larger space.

\subsection{Variational inference amortising hyperparameters}\label{sec:VI-ELBO-AH}

Investigating hyperparameter influence and performing sensitivity analysis often requires model refits at a grid of hyperparameter values.  We reduce this computational burden by training a single model for the SMI posterior that can be evaluated at any hyperparameter value $\tilde \psi$, so we \textit{amortise over the hyperparameters}. For ease of exposition, in this Section we present our method as targeting this single posterior $p(\theta|Y; \psi)$ (which could be Bayes or GBI) rather than SMI as presentation is simpler and the same issues arise. See Section \ref{sec:SMI-VMP-further-details} for details for SMI.

Throughout this paper we use variational inference (VI; \cite{jordan1999introduction}, \cite{wainwright2008graphical}, \cite{blei2017variational}) with a normalising flow parameterisation \citep{dinh_density_2016,rezende2015variational} for the variational family, as set out in Section~\ref{sec:VMP}. Classical VI minimises the reverse-KL divergence $D_{KL}(q||p; \tpsi)$ over a family of densities $q(\theta;\lambda),\ \lambda\in\Lambda$ at fixed hyperparameters $\tpsi$. This is equivalent to finding the ${\lambda\mathstrut\!}^*_{\tpsi}$ that maximises the Evidence Lower Bound (ELBO), 
\begin{equation}
    \label{eq:ELBO}
    \ELBO(\lambda;\tpsi) = \underset{\theta \sim q(\theta;\lambda)}{\mathbb{E}} \left[ \text{log} \frac{p(\theta, Y| \tpsi)}{q(\theta;\lambda)}\right].
\end{equation}
We call the fitted $q(\theta;{\lambda\mathstrut\!}^*_{\tpsi})$ approximating $p(\theta|Y;\tpsi)$ the \emph{Variational Posterior} (VP).   

If we change $\tpsi$ in the posterior in \eqref{eq:ELBO}, we have to re-fit the VP, making exploration over $\tpsi$ cumbersome. This is not necessary. We can instead estimate a single set of parameters $\lambda^\ast$ for a conditional variational family \( q(\theta; \lambda, \tpsi) \), trained to approximate \( p(\theta \mid Y, \tpsi) \) across \( \tpsi \) by maximising the expected conditional (or amortised) ELBO
\begin{equation}
    \label{eq:A-ELBO}
    \aELBO(\lambda) = \underset{\tpsi \sim \rho(\tpsi)}{\mathbb{E}}\left[\underset{\theta \sim q(\theta;\lambda, \tpsi)}{\mathbb{E}} \left[ \text{log} \frac{p(\theta, Y| \tpsi)}{q(\theta;\lambda, \tpsi)}\right]\right].
\end{equation}
The outer expectation is taken with respect to an exogenous training distribution $\rho(\tpsi)$. Following \cite{carmona_scalable_2022}, we call $q(\theta;\lambda,\tpsi),\ \tpsi\in\Psi$ the \emph{Variational Meta Posterior} (VMP).

This $\aELBO$ framework can be understood as targeting a joint distribution $p(\theta,\tpsi \mid Y)=p(\theta\mid Y; \tpsi)\rho(\tpsi)$ using a joint variational family restricted to the form $q(\theta,\tpsi;\lambda)=q(\theta;\lambda, \tpsi)\rho(\psi)$. Factors of $\rho(\tpsi)$ cancel and we are left with \eqref{eq:A-ELBO}. 
This is not the same as fitting a general joint variational density $q(\theta, \tpsi; \lambda_h)$ approximating $p(\theta,\tpsi \mid Y)$ and then conditioning on $\tpsi$ in $q$, as that approach has no constraint on the functional form of $q(\theta,\tpsi;\lambda)$. The ELBO for fitting the hierarchical model would be
\begin{equation}
    \label{eq:H-ELBO}
    \text{H-ELBO}(\lambda_h) = \underset{\tpsi \sim q(\tpsi;\lambda_h)}{\mathbb{E}}\left[\underset{\theta \sim q(\theta|\tpsi;\lambda_h)}{\mathbb{E}} \left[ \text{log} \frac{p(\theta, Y| \tpsi)\rho(\tpsi)}{q(\theta,\psi;\lambda_h)}\right]\right].
\end{equation}
One disadvantage of using the H-ELBO in \eqref{eq:H-ELBO} is that it unnecessarily approximates both the marginal, $q(\tpsi;\lambda_h)\simeq \rho(\tpsi)$, which we already know, and the conditional $q(\theta;\lambda, \tpsi)$; in \eqref{eq:A-ELBO} we instead focus computational effort on accurately fitting the target conditional and use the exogenous $\rho(\tpsi)$ solely to weight the training of the amortised map. The choice of $\rho$ is discussed in Section~\ref{sec:rho-choice}. 

However, an advantage of learning the joint via the H-ELBO and then conditioning is that some variational families are known to be universal approximators \citep{huang18,draxler2022whitening,ishikawa23,draxler2024universality}. These results don't apply directly when $q(\theta,\tpsi;\lambda)$ is constrained to the form $q(\theta;\lambda, \tpsi)\rho(\tpsi)$ as in \eqref{eq:A-ELBO}. In Section~\ref{sec:UA-NF-bayes-amortised} we show that at least some variational families remain universal approximators within the constrained parameterisation used for the $\aELBO$, and relate this to the flow parameterisation we use in our software.

We considered using a forward-KL objective or equivalently minimising the log-likelihood loss $-\mathbb{E}_{p(\theta, Y|\tpsi)\rho(\tpsi)}[\log q(\theta; \lambda_f,Y,\tpsi)]$ as this loss has desirable mass-covering properties. This loss can be evaluated using samples from the joint for $\theta, Y$ and $\tpsi$. In SBI with a standard Bayesian posterior these samples are obtained by simulating the generative model for the data. The conditional $q(\theta \mid Y, \tpsi ; \lambda)$ is trained once but can be evaluated at any unseen $Y$, so it is said to \textit{amortise over the data}.  \citet{elsemuller2024sensitivityaware} do this for Bayesian inference in a very general way, amortising over data, hyperparameters and other model settings. However, in GBI and in SMI there is no straightforward generative model for the data so no convenient way to get samples $(\theta, Y, \tpsi)$ with the property that $\theta|Y,\tpsi\sim p(\theta|Y,\tpsi)$. 

In order to make this concrete, consider the joint for the GBI posterior in \eqref{eq:GB-basic-posterior}, $p(\theta, Y \mid \tpsi)\propto \exp({-\eta\,\ell(\theta;Y)})\,p(\theta|\varphi)$ and treat the exponential loss as if it were a generative model for $Y$.
The marginal for $\theta$ we get by integrating $Y$ is $p(\theta|\tpsi)\propto p(\theta|\varphi)\,c_{\eta}(\theta)$, with $c_{\eta}(\theta)=\int \exp({-\eta\,\ell(\theta;y)})\, dy$ an intractable function of $\theta$. The conditional for $Y$ is $p(Y|\theta,\tpsi)=\exp({-\eta\,\ell(\theta;Y)})/{c_{\eta}(\theta)}$.
These are the (intractable) sampling distributions that we would need to use if we wanted to do SBI for GBI. Even if the loss is a negative log-likelihood, there is no straightforward way to sample $p(Y|\theta)^\eta$. For this reason, no direct variational approximation to the posterior in GBI is currently available.
Existing methods giving SBI for GBI learn an approximation to $\exp(-\eta\ell(\theta;Y))$ amortised over $Y$ (or a related score function) and follow this with a non-amortised sampling step (e.g., MCMC) to obtain the final posterior \citep{cherief2020mmd, gao2023generalized, dellaporta2022robust, pacchiardi2024generalized, matsubara2022robust}. By using the $\aELBO$, and amortising over $\psi$, we avoid repeated MCMC runs, and get a direct variational approximation to the posterior in GBI. Moreover, the mode-seeking behavior of reverse KL may actually be helpful, especially in high dimensional problems of the kind we consider in Section~\ref{sec:LP-SMI} \citep{mittal2025forwardVreverseKL}. 

\subsection{Choice of amortising density for hyperparameters}  \label{sec:rho-choice}

The amortising density $\rho(\tpsi)$ does not function as a regular prior for $\tpsi$ in the $\aELBO$ \eqref{eq:A-ELBO}. First of all, our variational approximation $q(\theta,\tpsi;\lambda)=q(\theta;\lambda,\tpsi)\rho(\tpsi)$ targets $p(\theta,\tpsi|Y)=p(\theta|Y,\tpsi)\rho(\tpsi)$. The first factor is normalised, so the marginal posterior distribution for $\tpsi$ is still just $\rho(\tpsi)$. If we used $\rho$ as a prior then the posterior distribution of $\tpsi$ would depend on $Y$.
Secondly, suppose the variational family $q,\ \lambda\in\Lambda$ 
actually included $p(\theta|Y,\tpsi)$. The $\aELBO$ is maximised when $q$ is equal to $p(\theta|Y,\tpsi)$, irrespective of the choice of $\rho$ (subject only to a condition on its support), so in the limit of an expressive variational family, we expect results to be insensitive to $\rho$. It is more common to amortise over data. In that setting the outer expectation in \eqref{eq:A-ELBO} is over the prior predictive $y\sim p(\cdot)$ and the user seeks a variational approximation at each $Y\in\mathcal Y$. Working with the reverse KL (ie, maximising the $\aELBO$) allows the same freedom in choosing the outer expectation over $Y$ \citep{golinski19}.

In practice we use of a small number of stacked coupling-layer conditioners (see Section \ref{sec:cond-auto-flow}) so the variational family we implement doesn't prefectly express $p(\theta|Y,\tpsi)$. As a consequence the fitted $q(\theta;{\lambda\mathstrut\!}^*, \tpsi)$ tends to be more accurate at values of $\tpsi$ where $\rho(\tpsi)$ is larger, as the penalty for error imposed by the $\aELBO$ is larger for errors at such $\tpsi$. However, precisely because $\rho$ is not a prior, it can be informed by the data, so in practice we can make a first fit, use it to estimate the high-level loss $\L(\tpsi)$ for the inference and revise $\rho$ to focus on regions of $\tpsi\in\Psi$ that seem important (where $\L$ is small).

It is computationally efficient to amortise, but there is a price. The difference in accuracy between approximating $p(\theta|Y,\tpsi)$ at all $\tpsi\in\Psi$ with a single shared parameterisation for $q_{\theta|\tpsi}$ (via the $\aELBO$) compared to fitting a separate and independently parameterised approximation at each $\tpsi$ (via the ELBO in \eqref{eq:ELBO}) leads to what is often called the \emph{amortisation gap}~\citep{wu2020meta, margossian2023amortized}. In our experiments, we evaluate this gap and find that the amortised model retains good approximation accuracy across \(\Psi\).
See for example Figure~\ref{fig:amortisation-gap} of the Supplementary Material \citep{battaglia2025supplement}. There is some evidence (in Table~\ref{tab:WDs}) that the VMP agrees slightly less well with the MCMC benchmark than does a VP fitted at each hyperparameter.

\section{Related work} \label{sec:related-work}

\paragraph{Learning hyperparameters.}
SMI, like many other GBI approaches, requires tuning both prior hyperparameters and loss-specific parameters, such as learning rates. For prior hyperparameters, a common strategy is to evaluate a chosen loss (e.g., predictive or decision-based) over a grid of values, though this typically requires costly model refits. In contrast, in the variational inference literature, it is common to optimise the marginal likelihood (or its ELBO approximation) jointly with the variational parameters during training, yielding point estimates of the hyperparameters. This kind of Empirical Bayes approach \citep{robbins56} may be found in the original papers on Variational Auto-Encoders \citep{kingma2014auto} and normalising flows \citep{rezende2015variational}.  However, the marginal likelihood is known to be inappropriate when the model is misspecified \citep{grunwald_inconsistency_2017,gelman2020holes,fortuin2022priors}, and this usually applies in GBI and SMI. Tuning methods for GBI focus on estimating the learning rate parameter (see \cite{wu23} for a review). Methods differ according to the loss they use to express inference goals. SafeBayes \citep{grunwald_inconsistency_2017} achieves convergence over data to the pseudo-true at optimal rate. However, it uses a sequential-predictive loss which is very computationally demanding. \cite{holmes2017assigning} and \cite{lyddon2019general} adjust the learning rate to match the expected information gain between Bayes and generalised Bayes updates. \cite{syring2019calibrating} and \cite{winter2023sequential} tune the learning rate so that posterior credible sets approximately achieve nominal frequentist coverage. \citet{carmona_semi-modular_2020} and \cite{lee2025bayesianinferencelearningrate} target predictive criteria, the Watanabe-Akaike information criterion (WAIC) \citep{watanabe2010asymptotic,vehtari_practical_2017} and the posterior predictive density for held out data. We are guided by inference-goals in our choice of loss. However, it is certainly convenient to have a loss we can evaluate rapidly and optimise using Stochastic Gradient Descent (SGD). See Section~\ref{sec:hyperparam_selection} for futher discussion of these points.

\paragraph{Reverse and forward KL. } 

We approximate posterior distributions $p(\theta \mid Y,\tpsi)$ by optimising the reverse-KL objective of variational inference: $D_{KL}(q \mid \mid p)$. Posteriors can also be approximated by minimising the SBI forward-KL objective $D_{KL}(p \mid \mid q)$ or equivalently minimising the log-likelihood loss $-\mathbb{E}_{p(\theta)p(Y \mid \theta)}[\log q(\theta \mid Y)]$. 

Notable examples of SBI with amortised neural architectures include \citet{pmlr-v89-ambrogioni19a}, and with normalising flows specifically  
\citet{ardizzone2018analyzing,radev2020bayesflow,radev2023jana}. \cite{ardizzone2019guided,kruse21} and \cite{siahkoohi2023reliable} fit this description and use the same conditional normalising-flow setup as our own. More recent Neural Conditional Density Estimator (NCDE) approaches such as flow matching \citep{lipman2023flow, wildberger2023flow}, diffusion models \citep{ho2020denoising,song2021scorebased} and free flow models \citep{draxler2024free} also rely on mass-covering forward-KL type objectives and can be used to amortise over data. \citet{frazier2024statistical} provides a recent review on NCDE methods for SBI and studies their statistical accuracy.

As discussed in Section~\ref{sec:VI-ELBO-AH}, the forward KL is not a practical training objective for GBI and SMI and the reverse KL is to some extent imposed by the setting. Nevertheless, the problem of dealing with misspecified simulators in SBI is an active field of research (see \cite{kelly2025simulation} for a comprehensive review), and a number of studies employ generalised Bayes-type losses to achieve robust SBI. However, these approaches do not directly target the posterior distribution itself, but focus instead on approximating a robust loss function. Examples include Maximum Mean Discrepancy (MMD) \citep{cherief2020mmd, dellaporta2022robust} and more general energy and kernel scores \citep{gao2023generalized,pacchiardi2024generalized}. With the exception of \citet{chakraborty2023modularized}, which relies on a global Gaussian mixture approximation that does not scale well to high dimensions,
 none of these treat multi-modular losses. More importantly for our purpose, they rely on non-amortised variational \citep{matsubara2022robust}, MCMC \citep{gao2023generalized,pacchiardi2024generalized} or Bayesian non-parametric learning \citep{dellaporta2022robust} schemes to sample from a posterior based on the estimated loss. Of these \cite{gao2023generalized} is particularly interesting to us as they mention amortising over the learning rate, as in \cite{carmona_scalable_2022}. However, they ultimately run MCMC to gather final samples.

\paragraph{Sensitivity analysis. } 
Sensitivity to likelihood and prior is well studied in Bayesian inference (see \citet{depaoli2020importance, ruggeri2005robust} for an overview). In non-amortised Bayesian inference, \citet{giordano_covariances_2018} use linear approximation techniques for local sensitivity analysis in variational inference. Similar ideas are applied by \citet{giordano2019swiss} to study sensitivity to data reweighting, and extended by \citet{giordano_evaluating_2022} to assess the impact of prior perturbations in specific Bayesian nonparametric models with stick-breaking constructions.
\citet{mikkola2024preferential} uses normalising flows to parameterise prior elicitation from expert preferences expressed as noisy comparisons. The learning objective for the flow is in effect a tilted marginal likelihood.
Broader in scope, amortised and more similar to our work is \cite{elsemuller2024sensitivityaware}. They consider sensitivity to data, model and hyperparameters (the context) in an SBI setting where the likelihood is not available, minimising a forward-KL objective targeting the posterior as previously discussed. As noted above, this framework doesn't handle GBI or SMI contexts. Specifically on likelihood and prior choices, they quantitatively explore robustness in terms of difference in KL divergence between distributions under different context choices. We instead explore sensitivity to hyperparameter in the sense of performance of the posterior approximation against specific losses of choice, e.g. MSE or ELPD. Finally, we provide amortised universal approximator results for our conditional-flow framework.

\paragraph{Amortised Variational Inference.} The idea of amortising is also well
known in the context of variational inference minimising a reverse-KL objective, and again, typically amortising over the data; see, e.g., \citet{zammit2024neural, ganguly2023amortized} for a recent review.
Amortised VI (A-VI, \citet{gershman2014amortized, margossian2023amortized} ) traditionally considers factorised variational families that estimate a different latent factor per data point (commonly used in Variational Autoencoders, VAE, \cite{kingma2014auto}), and learns instead a single conditional distribution that can be evaluated at any data point. This amounts to minimising a reverse-KL objective in expectation over the data: $\mathbb{E}_Y[D_{KL}(q(\theta \mid Y) \mid p(\theta \mid Y))]$, where $q(\theta \mid Y)$ is parameterised with a function $f(y)$ that returns the variational parameters for every $y$. The concept of learning an inverse map $f(y)$ from observations to parameters of the variational posterior dates back to \citet{dayan2000helmholtz} where it was called an inference network or recognition model. \citet{rezende2015variational} showcase its use in the context of variational inference with normalising flows and  
\citet{kingma2016improved, durkan_neural_2019} use more flexible inverse autoregressive flows.
In this case, the inference network outputs the base distribution and a shared context vector $f(y)$ that conditions each flow transformation. 
We are interested in minimising a reverse-KL objective in expectation over hyperparameters, to learn a variational approximation that is amortised over hyperparameters, and use a different architecture for amortisation.

\paragraph{Amortising and hyperparameters} 
Some recent works have explored amortisation beyond data inputs. Notably, the Amortised Conditioning Engine (ACE) \citep{chang2024amortized} learns a single transformer-based model to perform both inference and prediction across tasks such as SBI, image completion, and Bayesian optimisation. Given a context set of observed inputs, such as simulated data in SBI, partially observed images, or function evaluations, it predicts distributions over target quantities, which may be unobserved data points or latent parameters. In the SBI setting, ACE is trained on simulated $(\theta,Y)$ pairs from a known generative model using a forward-KL objective, learning a conditional density estimator $q(\theta \mid Y)$, as in the standard amortised SBI framework under a well-specified simulator. Prior information can optionally be passed at test time as histograms over latent variables, rather than being encoded parametrically and tuned during inference. This differs from our setting, where inference is governed by (generalised) Bayes rule and priors are specified through hyperparameters that can be varied or optimised explicitly. Unlike our method, ACE is not designed to study posterior sensitivity or robustness, but instead serves as a general-purpose amortised inference and prediction engine.

In the context of Gaussian Processes, where hyperparameters are typically selected by maximising the log marginal likelihood—a closed-form but costly and dataset-specific objective—there is also work that amortises hyperparameter inference over datasets. \citet{liu2020task} learn a deterministic mapping from data to kernel hyperparameters, trained to maximise the marginal likelihood on synthetic tasks. \citet{bitzer2023amortizedinferencegaussianprocess} extend this by additionally conditioning on symbolic representations of structured kernels, enabling inference amortised over both datasets and kernel compositions.

However, approaches that amortise variational inference over hyperparameters using a reverse-KL objective, closer to our own approach, are less common. \citet{wu2020meta} propose to learn a variational approximation doubly amortised over data and data-generating distributions, using synthetic datasets drawn from known generative models to simulate variation across tasks. Their approach targets task variation rather than explicit hyperparameter inputs and does not focus on sensitivity analysis. \cite{carmona_scalable_2022} gave an amortised variational scheme with normalising flows to optimise for the learning rate in SMI. We build on their work and develop a more general architecture that amortises over all hyperparameters at once. We use the learnt conditional flow for sensitivity analysis and explore a different range of losses that practitioners might consider for tuning purposes.  

\section{The Variational Meta-Posterior for approximate Bayesian inference} \label{sec:VMP}

\subsection{Conditional autoregressive flow}\label{sec:cond-auto-flow}

We define an amortised variational distribution \( q(\theta; \lambda, \psi) \), optimised via the $\aELBO$ (Equation~\ref{eq:A-ELBO}), adapting a neural autoregressive normalising flow built from user-supplied transformer and conditioner functions \citep{huang18, jaini2019sum}. Normalising flows \citep{tabak2010density, tabak2013family, rezende2015variational, papamakarios21} have become a standard tool for constructing flexible densities, with autoregressive variants \citep{papamakarios2017masked, kingma2016improved, huang18, dinh_density_2016} being amongst the most widely used flow architectures. We begin by defining a conditional flow suitable for approximating the GBI posterior in \eqref{eq:GB-basic-posterior} and then give the extension to a flow approximating the SMI posterior in \eqref{eq:smi-posterior}. Our flow parameterisation conditions on hyperparameters but has a similar high level structure as \cite{kruse21} where the authors condition on data.

A conditional autoregressive flow defines, for each $\tpsi\in\Psi$, an invertible, continuously differentiable map \( T(\cdot; \lambda, \tpsi): \mathbb{R}^p \to \mathbb{R}^p \). We take a simple base distribution \( q(\epsilon) \) (e.g., uniform or standard normal), set \( \theta = T(\epsilon; \lambda, \tpsi) \) and define \( q(\theta; \lambda, \tpsi) \) via change of variables.
Each component is autoregressively transformed as:
\begin{align}
    \theta_1&=\tau(\epsilon_1,h_1),&h_1&=c_1(\tpsi;\lambda_1),\nonumber\\
    \theta_i&=\tau(\epsilon_i,h_i),&h_i&=c_i(\epsilon_{<i},\tpsi;\lambda_i),\ i=2,\dots,p.\label{eq:amortised-conditioner}
\end{align}
where \( \tau \) is a strictly monotone transformer (e.g., spline as in \citet{durkan_neural_2019}) and \( c_i \) is a conditioner (e.g., MLP) outputting parameters for \( \tau \), with inputs $\epsilon_{<i}=(\epsilon_1,\dots,\epsilon_{i-1})$ and \(\tpsi\). The inclusion of \(\tpsi\) in the conditioner is the only change from the standard setup \citep{papamakarios21}.
The resulting density is
\begin{equation}\label{eq:vmp-basic-variational-conditional-family}
q(\theta; \lambda, \tpsi) = q(\epsilon) \left| \det J_T(\epsilon; \lambda, \tpsi) \right|^{-1}, \quad \text{where } \epsilon = T^{-1}(\theta; \lambda, \tpsi).
\end{equation}
Here, \(\lambda\) and \(\tpsi\) are treated as fixed parameters of the flow \(T\) and its induced density.

In practice in our software we parameterise $T$ using coupling layers \citep{dinh_density_2016, durkan_neural_2019}. These provide an efficient balance between expressiveness and tractable Jacobian determinants. A coupling layer permutes the components of $\epsilon^{(k-1)}$, partitions them into two subvectors $(\epsilon_a,\epsilon_b)$, and defines the transformation
\begin{equation}
\theta_a = \epsilon_a, \qquad 
\theta_b = \tau\!\left(\epsilon_b;\,c(\epsilon_a,\tpsi;\lambda)\right),
\label{eq:coupling-layer}
\end{equation}
where $c$ is a conditioner network producing the parameters of the monotone transformer $\tau$. Conditional coupling layers (CCLs), when stacked, have theoretical guarantees \citep{draxler2022whitening,draxler2024universality} and have proven to be empirically effective on both high-dimensional data \citep{kingma2018glow} and in a range of complex Bayesian applications \citep{bellagente2022understanding, radev2021outbreakflow, von2022mental}. 

In the context of SMI we target \eqref{eq:smi-posterior}. We consider a variational approximation parameterised by flows of the form
\begin{align}
    q(\theta, \tilde{\theta}, \delta; \lambda, \tpsi) &= q_1(\delta;\lambda_1, \tpsi)q_2({\theta};\delta;\lambda_2, \tpsi)
    q_3(\tilde\theta;\delta,\lambda_3, \tpsi)\nonumber\\
    &= p(\epsilon_1) \mid J_{T_1} \mid ^{-1}p(\epsilon_2 \mid \epsilon_1) \mid J_{T_2} \mid ^{-1}p(\epsilon_3 \mid \epsilon_1) \mid J_{T_2} \mid ^{-1}\label{eq:smi-vmp-variational-density}
\end{align}
where $T_1, T_2$ and $T_3$ are normalising flows with 
flow parameters $\lambda = \{\lambda_1, \lambda_2, \lambda_3\}$ and $\epsilon_1 = T_1^{-1}(\delta; \lambda_1, \tpsi)$, $\epsilon_2 = T_2^{-1}(\theta; \lambda_2, \delta, \tpsi)$ and $\epsilon_3 = T_3^{-1}(\tilde\theta; \lambda_3, \delta, \tpsi)$. This allows us to enforce the conditional independence structure in \eqref{eq:smi-posterior}. As explained in \cite{carmona_scalable_2022} and \cite{yu-nott-cutvi23}, flow-fitting for SMI and Cut models must be done in two stages in order to stop the variational approximation from introducing uncontrolled information flow between modules. See Section~\ref{sec:SMI-VMP-further-details} for the main idea and the cited papers for detail. 

Our set-up generalises \cite{carmona_scalable_2022}, who sought a practical way to explore SMI-posterior approximations at different $\eta$ values. They used coupling layers with an additive conditioner of the form 
$c^{(1)}(\epsilon_{a};\lambda^{(1)})+c^{(2)}(\tpsi;\lambda^{(2)})$ in place of the joint conditioner in \eqref{eq:coupling-layer}.
This additive conditioner has many more variational parameters (due to the different structure of the MLP layers in the flow conditioner) and drops co-dependence between $\tpsi$ and $\epsilon_{a}$. As we discuss in Section \ref{sec:UA-NF-bayes-amortised}, the conditional autoregressive flow in \eqref{eq:vmp-basic-variational-conditional-family} is a universal approximator, while the counterpart with an additive conditioner is not. In the examples we looked at, the two architectures gave similar results.

\subsection{Amortised universal approximator for conditional distributions}\label{sec:UA-NF-bayes-amortised}

We now give an example of a class of normalising flows which universally approximate conditional densities. Our outline proof omits some regularity conditions which may be found in the cited literature. 

A Universal Approximator (UA) for a class of densities $\D$ is a class $\Q$ of $q$ parameterisations with the property that, for any $p\in \D$, there exists a sequence of densities $q^{(k)}\in\Q,\ k\ge 1$ of increasing complexity converging weakly to $p$. We need $\D$ to be some useful class of densities including the joint GBI-posteriors $p(\theta,\tpsi|Y)=p(\theta|Y,\tpsi)\rho(\tpsi)$ of interest to us. Existing theory for autogressive normalising flows \citep{huang18} and CCLs \citep{papamakarios21,draxler2022whitening,draxler2024universality,ishikawa23} would meet this requirement. Since weak convergence of the joint implies weak convergence of conditionals, these families are UAs for the conditionals $p(\theta|Y,\tpsi)$ of ultimate interest. 
However, we do not parameterise a joint  $q(\theta, \tpsi;\lambda)$ to target a joint distribution, as in \eqref{eq:H-ELBO}, and then condition to get $q(\theta;\tpsi,\lambda)$. Instead, we directly parameterise a conditional variational family $\Q_{\theta|\tpsi}$ targeting a class of conditional distributions $\D_{\theta|\tpsi}$, as in \eqref{eq:A-ELBO}. 

We now give sufficient conditions for $\Q_{\theta|\tpsi}$ to be a UA for $\D_{\theta|\tpsi}$. Suppose it holds that \\[-0.3in]
\begin{enumerate}
\item[](A) $Q$ is a UA for $\D$, \\[-0.25in]
\item[](B) for any $f\in\D_{\theta|\tpsi}$ there is $p\in\D$ such that $f=p/\int p\, d\theta$ and \\[-0.25in]
\item[](C) for $q\in\Q$ and $\psi\in\Psi$ the conditional $q_{\theta|\tpsi}=q/\int q d\theta$ satisfies $q_{\theta|\tpsi}\in \Q_{\theta|\tpsi}$. \\[-0.3in]
\end{enumerate}
In this case the simple argument works.
\begin{lemma}\label{lem:ua-amortised}
   If conditions (A-C) hold then $Q_{\theta|\tpsi}$ is a UA for distributions in $\D_{\theta|\tpsi}$.\\[-0.3in]
\end{lemma}
\begin{proof}
For any $f\in \D_{\theta|\tpsi}$ there is $p_{\theta,\tpsi}\in\D$ with conditional $p_{\theta|\tpsi}=f$ by (B) and by (A) there is $q^{(k)}_{\theta,\tpsi}$ converging weakly to $p_{\theta,\tpsi}$, so the conditionals of $q^{(k)}_{\theta,\tpsi}$ converge to $p_{\theta|\tpsi}=f$; since by (C) the conditionals $q^{(k)}_{\theta|\tpsi}$ are in $\Q_{\theta|\tpsi}$ it follows that for any $f\in \D_{\theta|\tpsi}$ there is a sequence $q^{(k)}_{\theta|\tpsi}\in \Q_{\theta|\tpsi},\ k\ge 1$ converging weakly to $f$, so $Q_{\theta|\tpsi}$ is a UA for $\D_{\theta|\tpsi}$.
\end{proof}
\vspace*{-0.15in}
It follows that the conditional autoregressive normalising flow in Section~\ref{sec:cond-auto-flow} is a UA for a useful class of densities. 
\begin{corollary}\label{cor:UA-ANF-Huang}
  Let $\D$ be a class of densities which are continuous and strictly positive on some bounded rectangular region of $\mathbb{R}^p\times \Psi$ and let $\D_{\theta|\tpsi}$ be the class of densities on $\theta$ we can get by conditioning a density in $\D$ on any value of $\tpsi$ in its domain.
  Subject to regularity conditions \citep{huang18} on the conditioner and transformer functions and the base density $q(\epsilon)$, the class $\Q_{\theta|\tpsi}$ of autoregressive normalising flow defined in \eqref{eq:amortised-conditioner} and \eqref{eq:vmp-basic-variational-conditional-family} is a UA for $\D_{\theta|\tpsi}$. \\[-0.3in]
\end{corollary}
\begin{proof}
 The flows in $\Q$ realising $(\tpsi,\theta)$ are given by \eqref{eq:amortised-conditioner}, extended from $\Q_{\theta|\tpsi}$ to realise both $\tpsi$ and $\theta$: the variables input to the flow are $\tpsi_1,\ldots,\tpsi_{d_{\tpsi}},\epsilon_1,\ldots,\epsilon_p$ in that order. \cite{huang18} shows that, subject to regularity conditions on the conditioner and transformer functions and the extended base density $q(\tpsi,\epsilon)$, this class of extended flows is a UA for $\D$ so condition (A) holds and (B) holds by the definition of $\D_{\theta|\tpsi}$. Now $q\in\Q$ parameterised in this way factorises into a density $q_{\tpsi}$ realised in the first $d_{\tpsi}$ steps and a density $q_{\theta|\tpsi}$ realised in the next $p$ steps. However, the class of maps defining $\Q_{\theta|\tpsi}$ in \eqref{eq:amortised-conditioner} is just the mapping in the last $p$ steps so (C) is satisfied and so the corollary holds by Lemma~\ref{lem:ua-amortised}.
\end{proof}
\vspace*{-0.15in}

It would be good to apply this to other classes of conditioned flows and base Corollary~\ref{cor:UA-ANF-Huang} on UA results with stronger mode of convergence \citep{draxler2022whitening}. However, Condition (C) can be an obstacle. For example, \cite{draxler2024universality} has shown that CCLs (see Section \ref{sec:cond-auto-flow}) are UAs, and this gives us (A) and (B).   
However, CCLs get expressive power by permuting outputs across inputs between layers, so when we extend a class of conditional CCLs $\Q_{\theta|\tpsi}$ to a joint class $\Q$ we have to allow general permutations across layers to get a UA. Now (C) doesn't hold, as the conditionals of $\Q$ are no longer in $\Q_{\theta|\tpsi}$.  The conditions in Lemma~\ref{lem:ua-amortised} are only sufficient, not necessary.
However, \cite{papamakarios21} points out that CCLs can be stacked to reproduce a general autoregressive flow, and so the conditional variational family we implement in our software is an approximation to a family we have proven to be a UA for conditional densities. Although the approximation we actually implement won't perfectly approximate the target, it is at least reassuring to know that we have a variational family which can be expanded to make the approximation error as small as we wish.

\section{Hyperparameter selection} \label{sec:hyperparam_selection}

The VMP $q(\theta;{\lambda\mathstrut\!}^*,\tpsi)$ allows us to produce posterior samples for any given set of hyperparameters $\tpsi$ in an efficient manner, taking $\epsilon\sim q(\cdot)$ and setting $\theta=T(\epsilon;{\lambda\mathstrut\!}^*,\tpsi)$. We leverage this to select $\tpsi^*$ optimising a loss to inference $L(\tpsi)$. The choice of loss depends on the aims of the inference. Here we give two loss functions relevant in misspecified settings and SMI. In the example in Section~\ref{app:synthetic-data} of the Supplementary Material \citep{battaglia2025supplement} we 
have standard Bayesian inference with a well-specified model so the marginal likelihood is a valid utility and so we can follow \cite{rezende2015variational} and choose $\tpsi^*$ to maximise the ELBO in \eqref{eq:ELBO}.

When the model is not well-specified and the goal of the inference is prediction we follow \cite{carmona_semi-modular_2020} and use the Expected Log Pointwise Predictive Density (ELPD) for new data \citep{vehtari_practical_2017}. 
The loss for $\tpsi$ is 
\[
\L(\tpsi)=-\int p^*(y')\log(p(y' \mid y;\tpsi))dy'
\]
where $p^*$ is the true generative model for the data and $p(y' \mid y;\tpsi)$ is the posterior predictive distribution for one component of new data (hence ``pointwise''). \cite{mclatchie2024predictive} observe that, when $\eta$ is not close to zero and the data size is moderately large, the ELPD is insensitive to $\eta$. In contrast, here and in \cite{carmona_semi-modular_2020} and \cite{carmona_scalable_2022} we see significant dependence and a clear maximum for some hyperparameters. 
This may be due to the data-to-parameter dimension scaling. In our examples, the dimension of the parameter space increases as the number of observations increases: this is outside the setting considered in \cite{mclatchie2024predictive}. \cite{cooper24} show that the ELPD can perform poorly when there is serial dependence across components of the data. They suggest a modification which would be straightforward to implement here if needed. 

When we have two datasets, as in Figure~\ref{fig:SMI_diagram}, we must select one or both as the target for prediction. If the target is $Y$ then the LOOCV estimator is
\begin{equation}
  \hat\L_Y(\tpsi)=-\sum_{i=1}^n \log(\reallywidehat{p(Y_i|Y_{-i},Z;\tpsi)})  
  \label{eq:LOOCV-elpd}
\end{equation}
where $Y_{-i}$ is the data with the $i$'th sample removed and
\[
\reallywidehat{p(Y_i|Y_{-i},Z;\tpsi)}=\E_{\delta\sim q_1(\cdot;\widehat{{\,\lambda\mathstrut\!}^*}_{\!\!1},\tpsi),\, \theta\sim q_2(\cdot;\delta,\widehat{{\,\lambda\mathstrut\!}^*}_{\!\!2},\tpsi)}\bigl(p(Y_i|\theta,\delta)\bigr).
\]
We optimise over parameters of the model and inference taking $(\widehat{\varphi^*},\widehat{\,\eta^*})=\widehat{\tpsi^*}$ with
\[
\widehat{\tpsi^*}=\arg\!\min_{\tpsi} \hat\L_Y(\tpsi).
\]
The approximating VMP from \eqref{eq:smi-vmp-variational-density} is 
$q_1(\delta;\widehat{{\,\lambda\mathstrut\!}^*}_{\!\!1},\widehat{\tpsi^*})\,q_2(\theta;\delta,\widehat{{\,\lambda\mathstrut\!}^*}_{\!\!1},\widehat{\tpsi^*})\simeq p(\delta,\theta|Y,Z;\tpsi^*)$. The loss $\hat\L_Z(\tpsi)$ when the target is $Z$ is defined in a similar way. We used LOOCV, rather than following \cite{carmona_semi-modular_2020} and using the WAIC \citep{watanabe2010asymptotic}, as we expect LOOCV to be more reliable than WAIC on the small dataset we analyse in Section~\ref{sec:hpv}, and VMP simulation is fast. 

If the goal of the inference is parameter estimation then we can target the square-loss to the true parameter value. One generic setting where we can access this loss arises when the parameters we wish to estimate are physical quantities $X_i,\ i\in \mathcal{P}$ (notation anticipating the application in Section~\ref{sec:LP-data}) which are well-defined whether or not the model is misspecified. Consider a model of the form
\begin{align}
&\alpha \sim p(\cdot;\tpsi), \nonumber\\
&X_i \sim p(\cdot;\tpsi), \label{eq:missing_data_model} \\
&Y_i \sim p(\cdot \mid X_i,\alpha; \tpsi),\nonumber
\end{align}
jointly independent for $i \in \mathcal{P}$. A subset $A\subset \P$ of the $X$'s are perfectly observed (the ``anchors'') so that $X_a=x_a$ is given for $a\in A$ and the values of $X_{\bara},\ \bara\in \barA,\ \barA=\P\setminus A$ are missing. The (regular Bayesian) posterior is
\begin{align}\label{eq:post-labeled-unlabeled}
p(\alpha, X_{\barA}\! \mid\! y_\mathcal{P}, x_A;\tpsi)&\propto p(\alpha;\tpsi)\,p(X_{\barA};\tpsi)\!\!\prod_{\bara\in\barA} p(y_{\bara}\! \mid\! \alpha,X_{\bara};\tpsi)\prod_{a\in A} p(y_{a}\! \mid\! \alpha,x_{a};\tpsi).
\end{align}
Denote by $x_{\bara}$ the unknown true value of $X_{\bara},\ \bara\in \barA$. We would like to target the Posterior Mean Squared Error
\begin{equation*}
    \PMSE(\tpsi) =\frac{1}{ \mid \barA \mid }\sum_{{\bara}\in\barA}\mathbb{E}_{\alpha,X_{\bara}}( \mid \mid x_{\bara}- {X}_{\bara} \mid \mid ^2 \mid y_\mathcal{P}, x_A;\tpsi)
\end{equation*}
and set $\tpsi^* = \arg\!\min_{\tpsi}\PMSE(\tpsi)$
but $x_{\barA}$ is unknown.
However, if the pairs $(X_i,Y_i)_{i\in\P}$ are exchangeable, we can target a cross-validated objective, holding out a subset $B\subset A$ of anchors and computing 
\begin{equation}
\label{eq:OS-PMSE-LMO}
    \widehat\PMSE_{LMO}(\tpsi) =\frac{1}{ \mid {B} \mid }\sum_{a\in B}\mathbb{E}_{\alpha,X_{ a}}( \mid \mid x_a- {X}_{a} \mid \mid ^2 \mid y_\mathcal{P}, x_{A\setminus B};\tpsi),
\end{equation}
and setting ${\widehat{\tpsi}^*}=\arg\!\min_{\tpsi} \widehat\PMSE_{LMO}(\tpsi)$.
In \eqref{eq:OS-PMSE-LMO} we treat the labels with indices in $B$ just as if they were more missing labels alongside $\barA$, so the VMP we use to estimate the expectation in \eqref{eq:OS-PMSE-LMO} is $q(\alpha,X_{B},X_{\barA};\widehat{{\lambda\mathstrut\!}^*},\tpsi)$. 

In order to apply this to SMI, we treat the module for labeled data as well-specified and identify it with the $Z$-module in Figure~\ref{fig:SMI_diagram}. We treat the model for unlabeled data as misspecified and identify it with the $Y$-module, so $Y_A$ is $Z$ in Section~\ref{sec:smi} and $Y_{\barA}$ is $Y$ there. The posterior in \eqref{eq:post-labeled-unlabeled} factorises as
\[
p(\alpha,X_{\barA}|y_\P,x_A;\varphi)=p(X_{\barA}|\alpha,y_{\barA};\varphi)\,p(\alpha|y_A,y_{\barA},x_A;\varphi)
\]
This matches the factorised Bayes posterior in \eqref{eq:bayes_posterior_factors} with the identification $\alpha\to\delta$ and $X_{\barA}\to \theta$ giving the same setup as in \eqref{eq:SMI_graph_as_gen_model}, so the VMP setup for SMI carries over. The duplicated auxiliary variable $\tilde\theta$ is a second copy of $X_{\barA}$ needed to define the likelihood for $Y_{\barA}$ in the power posterior. The power posterior version of \eqref{eq:post-labeled-unlabeled} 
is
\[
p_{pow}(\alpha, \tilde X_{\barA}|y_\mathcal{P}, x_A;\tpsi)\propto p(\alpha;\varphi)\,p(\tilde X_{\barA};\varphi)\prod_{\bara\in\barA} p(y_{\bara}|\alpha, \tilde X_{\bara};\varphi)^\eta \prod_{a\in A} p(y_{a}|\alpha,x_{a};\varphi)
\]
and the analysis-posterior is $p(X_{\barA}|\alpha,y_{\barA};\varphi)$.
The variational approximation to the marginal SMI posterior for $\alpha, X_{\barA}$ (integrating $\tilde X_{\barA}$) is $q_1(\alpha;\lambda_1,\tpsi)q_2(X_{\barA};\alpha,\lambda_2,\tpsi)$. For the loss estimated using a held out sample $B\subset A$ of the labeled data, the held out samples are added to the unlabeled data. The SMI posterior is
\begin{align*}
p_{smi}(\alpha, \tilde X_{\barA},\tilde X_B, X_{\barA}, X_B|Y_{A},Y_{\barA},x_{A\setminus B};\tpsi)&=p_{pow}(\alpha, \tilde X_{\barA},\tilde X_B|Y_{A},Y_{\barA}, x_{A\setminus B};\tpsi)\\
&\quad\times \quad p(X_{\barA},X_B|Y_{\barA},Y_B;\tpsi).
\end{align*}
In this held-out setting we recover the SMI notation with $X_{A\setminus B}\to Z$, $(Y_{\barA},Y_B)\to Y$, $\alpha\to\delta$ and $(X_{\barA},X_B)\to \theta$.
The SMI posterior is approximated with a VMP $q_1(\alpha;\lambda_1,\tpsi)q_2(X_{\barA},X_B;\alpha,\lambda_2,\tpsi)$ for $\alpha,X_{\barA}$ and $X_B$ and we estimate the PMSE with
\begin{equation}\label{eq:OS-PMSE-LMO-SMI}
\widehat\PMSE_{LMO}(\tpsi) =\frac{1}{|{B}|}\sum_{a\in B}\mathbb{E}_{\alpha\sim q_1,X_{a}\sim q_2}(||x_a- {X}_{a}||^2|y_\mathcal{P}, x_{A\setminus B};\tpsi).
\end{equation}
Finally $\tpsi^*=\arg\!\min_{\tpsi} \widehat\PMSE_{LMO}(\tpsi)$. Optimisation is carried out using automatic differentiation with respect to the conditioner inputs to the flow.

\section{Experiments} \label{sec:experiments}

We illustrate amortised hyperparameter analysis using two datasets, one for each of the selection criteria \eqref{eq:LOOCV-elpd} and \eqref{eq:OS-PMSE-LMO-SMI} given in Section~\ref{sec:hyperparam_selection}: in Section~\ref{sec:hpv} we analyse epidemiological data in an SMI setting and target the ELPD; in Section~\ref{sec:LP-data} we analyse a dataset of Linguistic Profiles (LP) using SMI and targeting the PMSE. This last example was the motivating problem for the work presented in this paper. 
In addition, Section~\ref{app:synthetic-data} of the Supplementary Material \citep{battaglia2025supplement} presents a synthetic example of a well-specified Bayesian setting, where we nonetheless use the VMP in order to illustrate the ELBO as a third criterion.


\subsection{Epidemiological data}\label{sec:hpv}

\subsubsection{Model and data}

In this section we give an example targeting the LOOCV estimator for the ELPD to estimate $\tpsi^*$ for an epidemiological model. This model is analysed in \cite{plummer_cuts_2015} and is based on data from human papillomavirus (HPV) prevalence surveys and cancer registries for age group cohorts (55-64 years) over $n=13$ countries \citep{maucort2008international}. Outcome $Y_i$
denotes the count of cancer instances observed over $T_i$ woman-years of monitoring in the $i$-th country, while 
$Z_i$ indicates the number of high-risk HPV instances identified in a sample comprising $N_i$
women from the same population. The model has a Poisson module for cancer incidence and a Binomial module for HPV prevalence,
\begin{align}
    Z_i &\sim Binomial(N_i, \delta_i),\nonumber\\
    Y_i &\sim Poisson(\mu_i),\label{eq:hpv-generative-model}\\
    \mu_i &= T_i \text{ exp}(\theta_1 + \theta_2 \delta_i),\nonumber
\end{align}
for $i=1,\dots,n$ with priors
\begin{align}
    \delta_i &\sim Beta(c_1, c_2)\nonumber\\
    \theta_1 &\sim N(m, s^2)\nonumber\\
    \theta_2 &\sim Gamma(g_1, g_2)\nonumber.
\end{align}
The model for the survey data $Z$ is well specified but cancer incidence $Y$ is misspecified. 

\subsubsection{SMI with the VMP and amortised prior hyperparameters} \label{sec:hpv-smi}

In this section we fit the VMP of Section~\ref{sec:VI-ELBO-AH} to the SMI-posterior in \eqref{eq:smi-posterior} for the generative model in \eqref{eq:hpv-generative-model}. The VMP $q(\theta,\delta;\psi)$ approximates the posterior for $\theta=(\theta_1,\theta_2)$ and $\delta=(\delta_1,\dots,\delta_n)$. It is amortised over $\psi=(\eta,c_1,c_2)$. Further hyperparameters $m,s,g_1$ and $g_2$ are fixed to $0, 100, 1, 0.1$ respectively. 
We could learn a VMP for all the hyperparameters and carry out a sensitivity analysis. However, the data informs them weakly, as they control a single parameter.
Working with fixed $c_1,c_2$, \cite{carmona_semi-modular_2020} found that the WAIC selects $\eta = 1$ (Bayes) when targeting prediction of $Y$, while $\eta = 0$ (Cut-posterior) was preferred for $Z$. We get similar results for $\eta$ in joint estimation with $c_1$ and $c_2$. 

We fit the VMP by estimating ${\lambda\mathstrut\!}^*$ using the $\aELBO$ measures determined by the divergences in \eqref{eq:smi-loss-13} and \eqref{eq:smi-loss-2}. We amortise over $\eta,c_1$ and $c_2$ using $\rho$-distributions $c_1,c_2\sim \mbox{Uniform}(0,15)$ $\eta\sim \mbox{Uniform}(0,1)$. We arrived at this after some experiments which we omit, aiming for support on all a priori plausible hyperparameter values. We then estimate $\tpsi^*$ by minimising $\hat\L(\tpsi)$ in \eqref{eq:LOOCV-elpd}, using the $Y$ or $Z$ data depending on which ELPD we target. The optimisation traces are shown in Figure~\ref{fig:HPV-SMI-trace} in Section~\ref{app:experimental-results-HPV} of the Supplementary Material \citep{battaglia2025supplement}. The SGD can get stuck at sub-optimal values so we take the hyperparameters achieving the least loss over multiple SGD initialisations. This gives $\tpsi^*_Y = \{0.87, 13.04, 15\}$ when targeting $Y$, and $\tpsi^*_Z = \{0.02, 0.97, 14\}$ when targeting $Z$, in line with earlier work (Cut for $Z$ and Bayes for $Y$). Hyperparameters $\eta$ and $c_1$ are well-informed by the data. For $c_2$ we simply learn that large values are favoured. 

\begin{figure}[t!]
    \centering
    \includegraphics[width=0.5\textwidth]{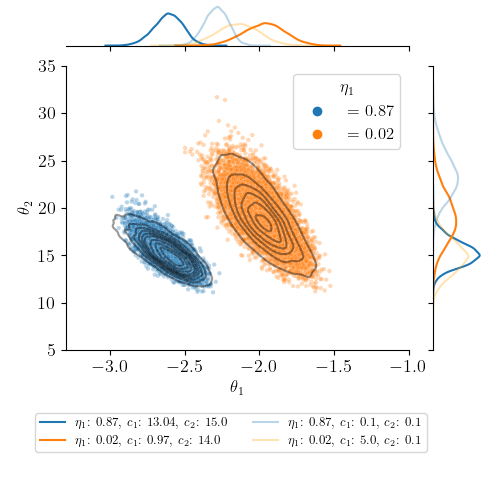}
    \caption{SMI analysis of HPV data. VMP samples for $\theta_1$ and $\theta_2$ at $\eta = 0.87$ (blue) and $\eta =0.02$ (orange) at optimal (full opacity) and alternative (reduced opacity) prior hyperparameters; overlayed with KDE contours for corresponding MCMC distributions. 
    }
    \label{fig:HPV-SMI-theta}
\end{figure}

The posterior is sensitive to the choice of hyperparameters. 
This is illustrated in Figure \ref{fig:HPV-SMI-theta}, which shows samples for $\theta$ at $\eta = 0.87$ (blue) and $\eta = 0.02$ (orange) at the optimal $c_1$ and $c_2$ values. The univariate densities along the two axes show variation in the posterior for $\theta$ if we vary $c_1$ and $c_2$ keeping $\eta$ fixed. The overlaid KDE contours for MCMC samples targeting the SMI posterior at $\tpsi_Y^*$ and $\tpsi_Z^*$ agree with the VMP samples. On the $Y-$module, the optimal $\tpsi_Y^*$ values lie at the boundary of the support of the $\rho-$distribution where the VMP has few training samples. This can be resolved by adjusting the $\rho-$distribution. In a real VMP application we wouldn't have the MCMC benchmark so we have left this unadjusted. 

Figures \ref{fig:HPV-SMI-delta-eta1} and \ref{fig:HPV-SMI-delta-eta0} in Section~\ref{app:experimental-results-HPV} show distributions for selected $\delta_i$'s at $\eta=0.87$ and $\eta=0.02$ respectively. Although these posterior distributions are sensitive to $c_1$ and $c_2$, the VMP is able to follow the dependence and VMP approximations match MCMC distributions very well.

\subsection{Linguistic Profiles data}\label{sec:LP-data}

We now illustrate our approach on a spatial model with almost 10,000 parameters (see Table~\ref{tab:LP-parameter-dimensions}). The model is fairly complex, and the inferential task is novel, but the idea is simple. In spatial statistics the classical setup \cite{diggle_model-based_2007, diggle_spatial_2013} has observations of spatial fields at known locations and the task is to fit a model for the fields and estimate field values at new selected locations. In contrast we have noisy observations of field values but only one third of the measurement locations are known. We fit a model for the fields and simultaneously estimate the missing location values as in \citep{giorgi_inverse_2015}.

\subsubsection{Background to the data} \label{sec:LP-background-data}

In this section we illustrate hyperparameter estimation using the VMP to target the PMSEs. Our data are 367 ``Linguistic Profiles'' (LPs) recording the presence and absence of 741 late-medieval dialect spellings and grammatical word forms across 367 documents from a 200km x 180km region in central Great Britain (rescaled to $\X=[0,1] \times [0,0.9]$). A map of the locations of the LPs can be found in Section \ref{app:experimental-results-LPs} of the Supplementary Material \citep{battaglia2025supplement}. These data are a subset of the Linguistic Atlas of Late Medieval English (LALME) \citep{mcintosh1986linguistic}, originally including 1044 LPs from text samples written 1350CE-1450CE in England, part of Wales and southern Scotland. Each LP records the presence ($y_{p,i,f}=1$) or absence ($y_{p,i,f}=0$) of a form $f\in \F_i$ for an item $i\in \I$ in document $p\in\P$, across all 367 LPs.
For example, the singular noun ``brother'' is an item, with distinct forms including
``BROTHER'', ``BROTHer'', ``BROyer'', ``BROyEr'', ``BROTHYR'' and ``BRUTHIR'' (\cite{haines_simultan_2016} and note ``E'' and ``e'' etc represent different letters in the manuscripts, not a change of case) and any given document will only contain one or two of these forms (if the item is used at all).

 Among the 367 LPs, 120 have precise location data ($x_p\in\X,\ p\in A$). The remaining 247 LPs have missing locations $x_p\in\X,\ p\in\barA$ (with $\P=A\cup\barA$). The Atlas authors call profiles in $A$ and $\barA$ the $\textit{anchor}$ and $\textit{floating}$ profiles respectively. They provide estimates for the locations of the floating profiles using a non-statistical method described in the Section~\ref{sec:lalme_data_main} of the Supplementary Material \citep{battaglia2025supplement}. In our subset of the Atlas data we have 71 items so $\I=\{1,2,\ldots,71\}$. If $\F=\cup_{i\in\I}\F_i$ is the set of all forms appearing then we see in all $|\F|=741$ forms. The present analysis re-estimates the locations of the floating LPs and compares them to the values given by the Atlas authors. 
When we estimate a location for a floating profile we expect it to be placed in a region where the dialect forms it contains appear frequently.

\subsubsection{Observation model} \label{sec:LP-background-model}
 
We work in the GP-GLM modelling framework of \citet{diggle_model-based_1998,savitsky_variable_2011}. Details of the model are set out in Section~\ref{sec:LP-background-expanded} of the Supplementary Material \citep{battaglia2025supplement}.  
In document $p\in \P$, item $i\in\I$ is used a Poisson number of times with item-dependent mean $\mu_i$. Given item $i$ is used it takes form $f\in\F_i$ with a location-dependent form-usage probability $\phi_{i,f}(x_p)$ satisfying $\sum_{f\in\F_i}\phi_{i,f}(x_p)=1$. The binary data  $y_p=(y_{p,i,f})_{f\in\F_i}^{i\in\I}$ for profile $p$ at location $x_p$ are generated by taking independent Poisson random variables $z_{p,i,f}$ with means $\mu_i\phi_{i,f}(x_p)$, zero-inflating them with independent Bernoulli variables $z'_{p,i,f}\sim \mbox{Bernoulli}(1-\zeta_i)$ and thresholding them, so that $y_{p,i,f}=\mathbb{I}_{z'_{p,i,f}z_{p,i,f}>0}$. Here $\zeta_i$ is the item-dependent zero-inflation probability. 

Denote by $\Phi(x)=(\phi_{i,f}(x))_{f\in\F_i}^{i\in\I}$ the $d_\phi$-dimensional latent spatial form-frequency field $\Phi: \X \rightarrow [0,1]^{d_\phi}$, with $d_\phi=741$ component fields, one for each form. \cite{haines_simultan_2016} gives a prior for $\Phi$ which proved to be misspecified. We give the following improved model for $\Phi$. In order to reduce the dimensionality of the problem and induce correlation across items and forms at a given location, we parameterise the fields as a (soft-maxed) linear combination of a small number of shared basis fields $\gamma_b:\X\to R,\ b=1,\dots,B$, taking $B$ equal to 10. To induce correlation across locations, the basis fields are modeled as Gaussian Processes (GPs) with a common kernel $k(\cdot,\cdot)$ modeling spatial decay of correlation.
\begin{align*}
     \gamma_b(x_p) &\sim GP(0,k(\cdot,\cdot)) \quad b \in \B=\{1,\ldots,B\} \\
    \gamma_{i,f}(x_p) &= a_{i,f} + \sum_{b=1}^B\gamma_b(x_p)w_{i,f,b}
    \\
    \phi_{i,f}(x_p) &= \frac{\text{exp} (\gamma_{i,f}{(x_p)})}{\sum_{f' \in \F_i}\text{exp}(\gamma_{i,f'}{(x_p))}}
\end{align*}
The kernel $k(\cdot,\cdot)$ is an exponential kernel with amplitude $\sigma_k$ and length-scale $\ell_k$. Denote by $\alpha=(\mu,\zeta,a,W)$ the parameters $\mu,\zeta$ for the zero-inflated Poisson observation model, the weights $W = \{w_{i,f,b}\}$ and their offsets $a = \{a_{i,f}\}$. Let $\Gamma(X_\P)=(\gamma_b(x_p))_{b\in\B}^{p\in\P}$ denote the set of GP basis-field values at profile locations (including anchor locations $X_A=(x_p)_{p\in A}$ and the locations $X_\barA=(x_p)_{p\in\barA}$ of floating profiles). The prior for $X_\barA$ is uniform in $\X^{ \mid \barA \mid }$. The weights $W$ and offsets $a$ are given zero-centred Laplace and zero-centred Gaussian independent priors with scales $\sigma_w$ and $\sigma_a$ respectively. The priors for $\mu_i,\zeta_i\ i\in\I$ given in \eqref{eq:mu-prior} and \eqref{eq:zeta-prior} in Section~\ref{sec:LP-background-expanded} of the Supplementary Material \citep{battaglia2025supplement} have no unknown hyperparameters. The prior hyperparameters are $\varphi=(\sigma_a,\sigma_w,\sigma_k,\ell_k)$. 

The exact joint distribution of the data and parameters, given hyperparameters, is
\begin{align}
p(Y_{\mathcal{P}},\alpha, \Gamma(X_{\mathcal{P}}), X_{\barA}) \mid \varphi)
=& p(\alpha \mid \sigma_a,\sigma_w)\, p(Y_{A} \mid \alpha, \Gamma(X_{A}))\,p(Y_{\barA} \mid \alpha, \Gamma(X_{\barA}))\nonumber\\
&\times p(X_{\barA})\; p(\Gamma(X_A),\Gamma(X_{\barA}) \mid X_{\barA},\sigma_k,\ell_k)
\label{eq:LP-exact-joint}
\end{align}

We go on to make a widely-used approximation, parameterising the GPs at a fixed lattice of $121$ inducing points $\U\subset\X$ rather than at the profile locations and approximating the joint distribution of the GPs at profile locations given the DP values at inducing points. This is convenient for variational work and further reduces the parameter dimension. The final approximate posterior $p(\alpha, \Gamma(X_{\U}),X_{\barA} \mid Y_{\mathcal{P}},\varphi)$ is given in Equation \eqref{eq:LP-final-approx-post} in Section~\ref{sec:LP-background-expanded} of the Supplementary Material \citep{battaglia2025supplement}.

The aim of the analysis is to estimate the locations, $X_\barA$, for the floating profiles. In this analysis all other parameters including the form-frequency fields $\Phi$ are latent nuisance parameters estimated jointly with $X_\barA$. We use the VMP to adjust hyperparameters to give the posterior which most accurately locates a small set of held-out anchor profiles, so we target the PMSE loss in \eqref{eq:OS-PMSE-LMO}.

Bayesian inference for missing locations is explored in \cite{giorgi_inverse_2015}.
They develop an MCMC procedure to target missing locations in a rectangular region in $\mathbb{R}^2$, assuming normally distributed observations with a single spatial field parameterising the mean.
They focus on cases where the number of floating observations is at most two.
The sparsity of our LP data-vectors $y_p$ and the high dimension of $\Phi$ make inference sensitive to misspecification (more model than data) so SMI is needed. 

\subsubsection{Tuning hyperparameters for the LALME data} \label{sec:LP-SMI}

In Section~\ref{sec:LP-MCMC} of the Supplementary Material \citep{battaglia2025supplement} we compare the VMP $q(\cdot;\lambda,\psi)$ (amortised over $\psi$, with $\lambda$ maximising the $\aELBO$ in \eqref{eq:A-ELBO}) with the VP $q(\cdot;\lambda_\psi)$ (fit separately at each $\psi$, with $\lambda_\psi$ maximising the ELBO in \eqref{eq:ELBO}) and MCMC targeting the posterior $p(\alpha,\Gamma(X_\U),X_{\barA}|Y_\P,\varphi)$. 

We benchmark on a subset of the full LALME data  set so that MCMC is feasible. As in our earlier example in Section~\ref{sec:hpv} (and also in the additional example in Section~\ref{app:synthetic-data}) of the Supplementary Material \citep{battaglia2025supplement}, we find that the VMP is able to learn a posterior approximation to the MCMC baseline of similar quality to that provided by the VP. The target distribution is far more complex than the earlier examples but the amortised flow is sufficiently general to express the target. 

In this section we use the VMP to estimate hyperparameters for all the LALME data in the study region $\X$.
The posterior in \eqref{eq:LP-final-approx-post} is a density over 9997 dimensions, and MCMC is infeasible. See Table~\ref{tab:LP-parameter-dimensions} in Section~\ref{app:experimental-results-LPs} of the Supplementary Material \citep{battaglia2025supplement} for a complete breakdown of parameter dimensions. Furthermore, as \cite{haines_simultan_2016} shows, the model in Section~\ref{sec:LP-background-model} is misspecified.  These issues motivated us to apply SMI,
using the VMP to estimate $\eta$ and the locations $X_{\barA}$ of floating profiles. Here we show that prior hyperparameters can be treated in the same framework. 

Denote by $\tpsi = \{\varphi, \eta \}$ the set of all hyperparameters. The SMI set-up has the same modular structure as \eqref{eq:SMI_graph_as_gen_model}, indentifying
\begin{align*}
    &Z \equiv Y_A, &Y \equiv Y_{\barA}\\
    &\theta \equiv X_{\barA} & \delta \equiv \alpha, \Gamma(X_\U).
\end{align*}
These choices identify the anchor-module $\{Y_A, \alpha, \Gamma(X_\U)\}$ as well specified and the float-module $\{Y_{\barA}, X_{\barA}, \alpha, \Gamma(X_\U)\}$ as misspecified, reflecting the fact that anchor-texts and non-anchor texts (i.e. floating profiles) are somewhat different types of texts. The SMI-posterior replacing the Bayes posterior $p(\alpha, \Gamma(X_{\U}),X_{\barA} \mid Y_{\mathcal{P}},\psi)$ is 
\begin{align}
\label{eq:smi_missing_data}
    p_{smi}(\alpha, \Gamma(X_\U), X_{\barA}, \tilde{X}_{\barA} \mid Y_\P; \tpsi) &\equiv p(X_{\barA} \mid \alpha, \Gamma(X_\U), Y_{\barA}; \psi)\\
    &\quad\times\quad p_{pow}(\alpha, \Gamma(X_\U),\tilde{X}_{\barA} \mid Y_\P; \tpsi),  \nonumber
\end{align}
where $\tilde X_{\barA}$ are the auxiliary parameters introduced in SMI to access information from the observation model for $Y$ in the ``imputation'' stage. The variational posterior is 
\begin{align}
q(X_{\barA}, \tilde{X}_{\barA}, \alpha, \Gamma(X_\U); \lambda,\tpsi) &= q_1(X_{\barA} \mid \alpha,\Gamma(X_\U);\lambda_1, \psi)q_2(\tilde{X}_{\barA} \mid \alpha,\Gamma(X_\U);\lambda_2, \tpsi)\nonumber\\
    &\quad\times\quad q_3(\alpha,\Gamma(X_\U);\lambda_3, \tpsi)\label{eq:LP-VMP-density}
\end{align}
and we use the VMP as per Section \ref{sec:VMP} as variational family. 

For $\rho$ distributions for $\tpsi$, we take Gamma(10, 1), Gamma(5, 1), Uniform(0.1, 0.4), Uniform(0.2, 0.5) and Beta(0.5, 0.5) for $\sigma_a, \sigma_w, \sigma_k,\ell_k$ and $\eta$ respectively. These choices are guided by the same physical considerations we would use to do prior elicitation. For example, we are unlikely to see significant dialect changes in less than 20km but 100km should be enough, translating to a range $[0.1,0.5]$ for $\ell_k$ after scaling. Some experiments showed that focusing on $[0.2,0.5]$ concentrated training on the range of $\ell_k$ values actually favoured by the data. We choose a $\rho$ for $\eta$ that concentrates on values of $\eta$ close to $0$ and $1$, corresponding to strongly or negligibly misspecified. In fact we will see that a moderate $\eta$ is favoured. The VMP performed well despite using a search focus covering but directed away from the optimal value. We estimate ${\lambda\mathstrut\!}^*$ minimising the SMI objectives given in \eqref{eq:smi-loss-13} and \eqref{eq:smi-loss-2}. This gives the fitted VMP $q(X_{\barA}, \tilde{X}_{\barA}, \alpha, \Gamma(X_\U); {\lambda\mathstrut\!}^*,\tpsi)$.

The goal of the whole analysis is to estimate the float-locations $X_\barA$, so we use the PMSE outlined in Section \ref{sec:hyperparam_selection} to select $\tpsi$.  
The observed anchor locations $X_{A}$ are an obvious proxy for the $X_\barA$: if we are good at predicting the locations of held-out anchors then we will be good at predicting the locations of floating profiles. We take a subset $\K\subset A$ of $K=40$ randomly chosen anchor profiles and make a single train-test split, leaving $ \mid A \mid - K=80$ in the training set. The optimal hyperparameters are given by
\begin{equation}
\label{eq:LP-MSE}
    \hat{\psi}^* = \underset{\tpsi}{\text{argmin }} \underset{{X}_p \sim \hat{q}^{\text{train}}_{\tpsi}}{\mathbb{E}}\left(\sum_{p \in \K} \mid \mid x_p- {X}_p \mid \mid ^2\right)
\end{equation} 
where $x_p,\ p\in \K$ are the true locations of the held out anchors, $X_p$ are posterior locations for the same quantities, and $\hat{q}^{\text{train}}_{\tpsi} = q(X_{\barA \cup \K}, \tilde{X}_{\barA \cup \K}, \alpha, \Gamma(X_\U); {\lambda\mathstrut\!}^*,\tpsi)$
is a VMP trained with held-out anchors in $\K$ treated as missing (so $X_A$ becomes $X_{A\setminus\K}$). This is evaluated at $\psi$ and approximates $p_{smi}(\alpha, \Gamma(X_\U),X_{\barA \cup \K}, \Tilde{X}_{\barA \cup \K} \mid Y_\P; \tpsi)$ in \eqref{eq:smi_missing_data}. We minimise the PMSE in Equation \eqref{eq:LP-MSE} using Stochastic Gradient Descent, taking derivatives over $\psi$ via automatic differentiation. 

\begin{figure}[htb]
    \centering
    \includegraphics[width=0.8\textwidth]{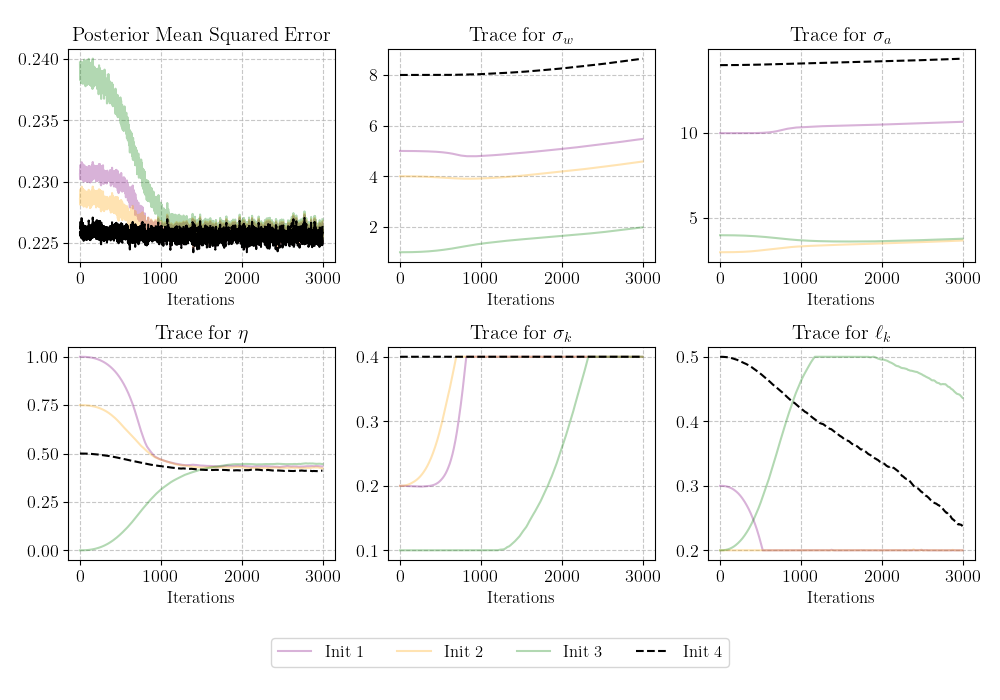}
    \caption{LALME data. Training loss and trace plots for hyperparameter optimisation of each component of $\psi$ at different initialisations. The run achieving the lowest average loss over the last 20 iterations is highlighted in black. Hyperparameters were selected as the average values over the 20 last iterations of this run.}
    \label{fig:LP-SMI-hps-optim}
\end{figure}
Optimisation traces for four initialisations are shown in Figure~\ref{fig:LP-SMI-hps-optim}. The rapidly converging learning rate $\eta$ drives the decrease in PMSE with $\tpsi^*$; we estimate $\hat\eta=0.42$. 

The drop in PMSE after fixing the misspecification corresponds to a root-PMSE gain of $7$km in prediction accuracy.
The large optimal $\sigma_k$ and small $\ell_k$ values in Figure \ref{fig:LP-SMI-hps-optim} favor GP fields that vary by relatively large amounts over short spatial scales. The PMSE is not sensitive to $\sigma_w$ and $\sigma_a$ (different values give the same loss), as long as they are large enough to admit the weights $w$ and biases $a$ for the mixture of Gaussian processes that the data support.  Based on this sensitivity analysis, we identify the hyperparameter vector $\psi^* = \{\sigma_a = 10, \sigma_w = 5, \sigma_k = 0.4, \ell_k = 0.2, \eta = 0.42\}$ as suitable. 

\begin{figure}[p]
\centering
\begin{subfigure}{\textwidth}
\centering
\includegraphics[width=\linewidth]{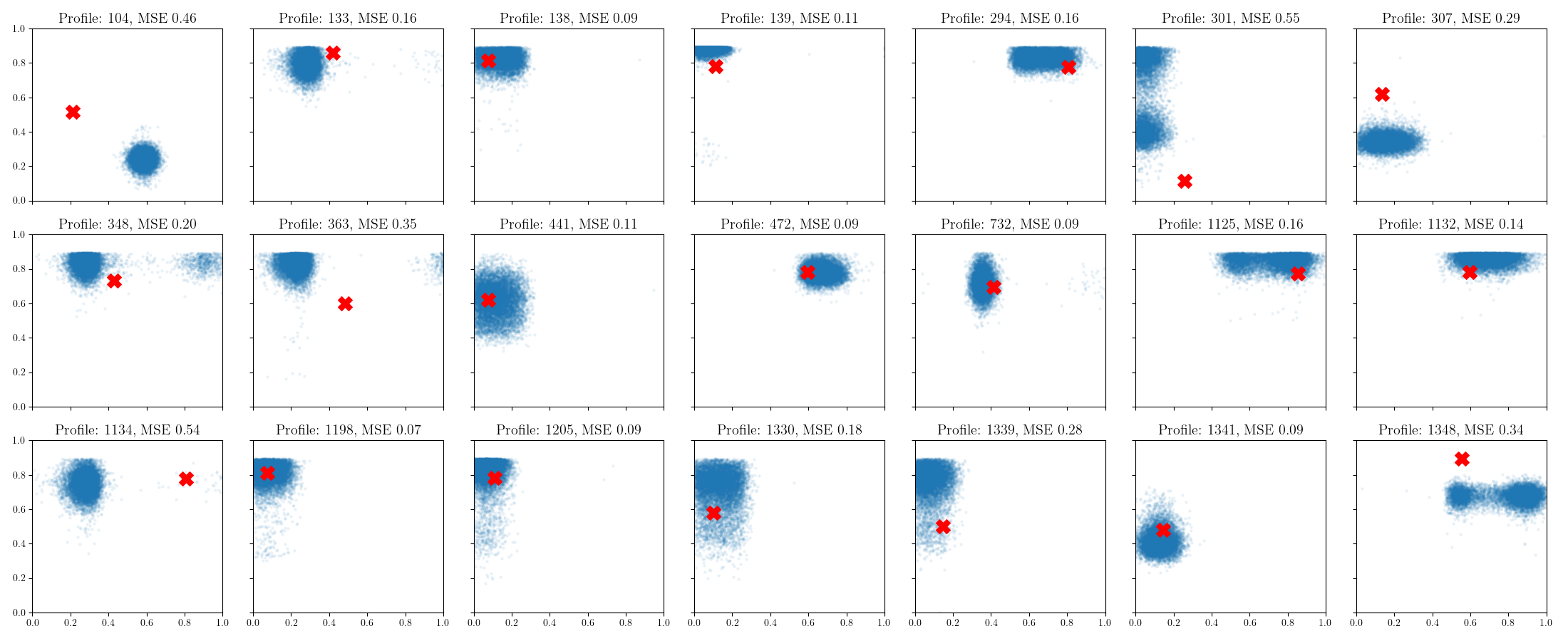}
\caption{$\eta = 1$}
\end{subfigure}
\begin{subfigure}{\textwidth}
\centering
\includegraphics[width=\linewidth]{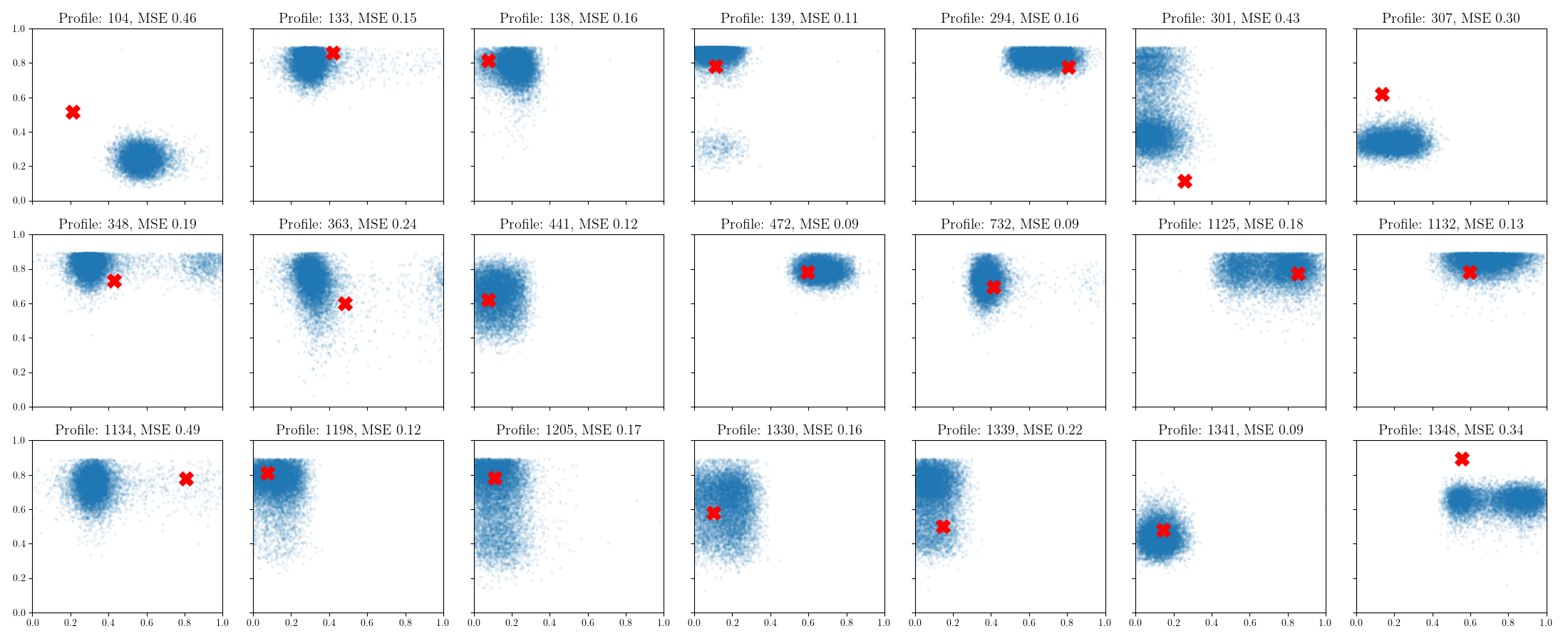}
\caption{$\eta = 0.42$}
\end{subfigure}
\begin{subfigure}{\textwidth}
\centering
\includegraphics[width=\linewidth]{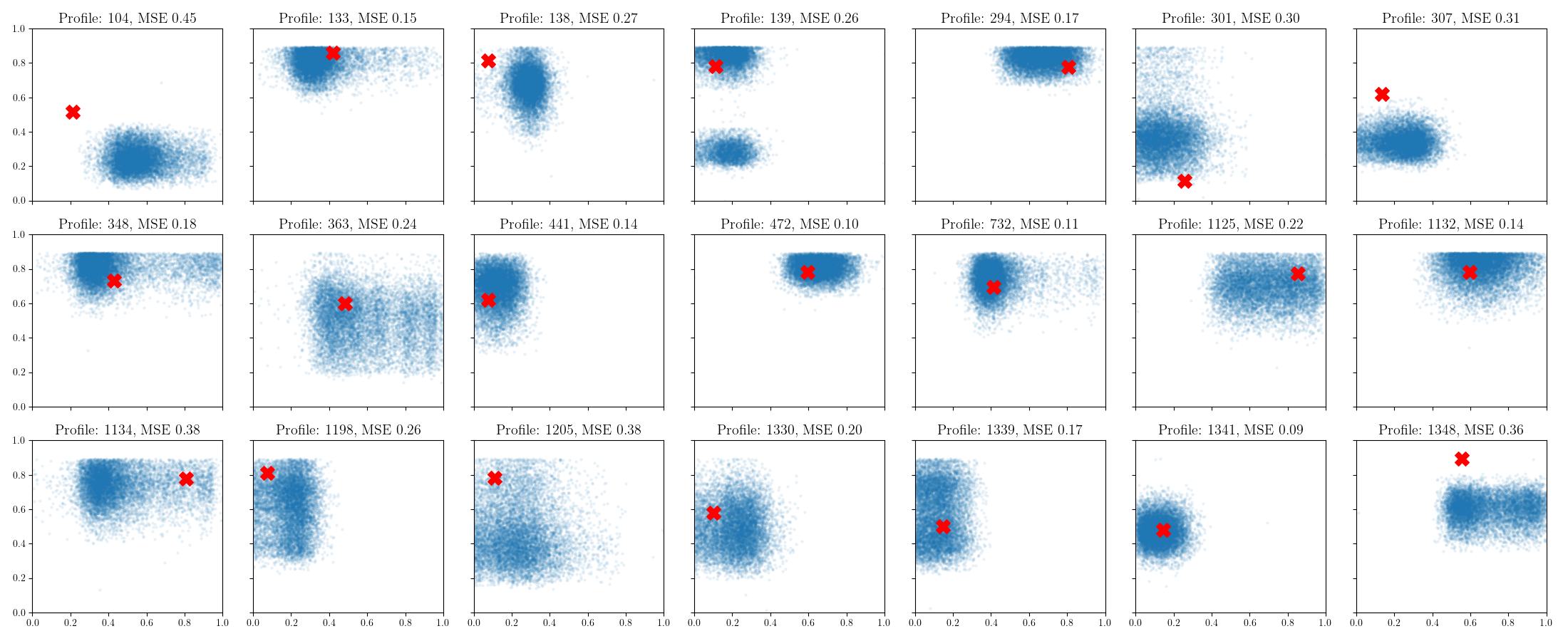}
\caption{$\eta = 0$}
\end{subfigure}
\caption{LALME data. Posterior distributions of locations for 21 held-out anchor profiles using the VMP at optimal prior hyperparameters $\sigma_a = 5$, $\sigma_w = 10$, $\sigma_k = 0.4$, $\ell_k = 0.2$.}
\label{fig:LP-SMI-21-val-compare}
\end{figure}

Figure \ref{fig:LP-SMI-21-val-compare} shows posterior distributions for $21$ of the $40$ held-out anchor profiles at $\eta=0,0.42,1$, keeping the other hyperparameters $\psi$ fixed at $\psi^*$. Modulating the information flow from the misspecified $Y_{\barA}$-module via SMI helps for many profiles (distributions shift closer to true values). However, cutting all that information ($\eta=0$) gives overdispersed results (e.g. profiles 138, 1205); this bias-variance trade-off is behind the gain in PMSE at $\eta$ somewhere between cut and Bayes.
In Figure~\ref{fig:LP-SMI-floating-compare} of Section~\ref{sec:LP-final-float-analysis} of the Supplementary Material \citep{battaglia2025supplement}
 we use the trained VMP at $\tpsi^*$ to estimate the locations $X_{\barA}$ of floating profiles. This was the main goal of the analysis. We show VMP-posterior distributions for 10 randomly selected floating profiles and compare them with the Atlas locations (which are themselves estimates), making use of the trained VMP evaluated at $\psi^*$. These are simply our original fit with held out anchors; it would be natural to refit a VP at the estimated optimal $\psi^*$ using all the anchors though the gains are small. 

Figure \ref{fig:amortisation-gap} shows how the VP training losses (negative SMI-ELBO) of the fitted VP and VMP vary with $\eta$; let $L_{VP}(\eta,\psi)$ and $L_{VMP}(\eta,\psi)$ denote these losses; we might expect $L_{VP}(\eta;\psi)<L_{VMP}(\eta;\psi)$ (an amortisation gap) as the ``local'' VP fits at fixed $\eta,\psi$ while the VMP is one fit at all $\eta,\psi$. We find the gap is too small to measure. See Section~\ref{sec:LP-final-float-analysis} of the Supplementary Material \citep{battaglia2025supplement} for further detail.

\section{Conclusions}

We have set out a variational framework for exploring posterior hyperparameter dependence in GBI and SMI by fitting a normalising flow amortised over hyperparameter values. The fitted VMP can be used to check for sensitivity to the choice of hyperparameters or to estimate hyperparameters by minimising a suitable loss. In our examples the learning rate and some hyperparameters controlling many parameters are well informed by the selection criteria we used and can be estimated; for other hyperparameters, we learn that posterior distributions are robust to any reasonable value assignment. The posterior mean square error for parameter estimates is particularly well suited when it is available (as it was in the example in Section~\ref{sec:LP-data}). 

When exploring sensitivity over $\psi$, the VMP can be used whenever a standard VP is appropriate. Suppose we want to check the posterior at $S$ $\tpsi$-values. The wall-clock time $c$ for fitting a VMP is essentially the same as fitting a VP at a single $\psi$, since the additional cost $d\ll c$ of sampling $\psi$ and computing the empirical estimate of $\aELBO$ in \eqref{eq:A-ELBO} is slight. MCMC must also be rerun for each $\psi$ at a cost $g$ say.
It follows that the total cost for VMP analysis is $c+dS$, for VP it is $(c+d)S$ and for MCMC it is $gS$. VMP dominates efficiency comparisons at large $S$. In Section~\ref{sec:LP-data} we consider a substantial real problem where a full-data MCMC run was not feasible even once, let alone across many $\tpsi$. Implementation details and runtime summaries are provided in the Supplementary Material (Section~\ref{app:hyperparameters}; \cite{battaglia2025supplement}).

In terms of architecture, even simple normalising flows are relatively expressive, and this reduces problems with under-dispersed variational approximation based on the reverse KL (a point made in \cite{rezende2015variational}). We compare VMP against MCMC baselines in Figure~\ref{fig:HPV-SMI-theta}, in the synthetic study in Section \ref{app:synthetic-data} of the Supplementary Material \citet{battaglia2025supplement}, and in the more challenging Linguistic Profile setting on a reduced example where MCMC is feasible (Section~\ref{sec:LP-MCMC}). Coverage is good across these comparisons; the main exception is that the variational posteriors for 
$\mu$ in Figure~\ref{fig:LP-small-mu-compare-VMP} are under-dispersed. If these were important for the inference then further experimentation with flow architecture would be necessary.

Recent work by \cite{lipman2023flow} on flow matching methods for fitting continuous normalising flows show very good expressive power with numerically stable fitting procedures. The corresponding flow matching methods for posterior approximation given in \cite{wildberger2023flow} inherit these properties. It is at least conceptually straightforward to extend the amortisation to cover prior hyperparameters and this should deliver greater approximation accuracy. However, the approach is necessarily amortised over data as the flow training needs target samples. It is unclear how these methods might be used to approximate posteriors in GBI and SMI, as there is no generative model for the data in that case, and a solution to this problem would be very useful. It is possible to target the forward KL without simulating the joint distribution: \cite{jerfel21_FKL_IS} and \cite{glockler2022variational} 
estimate the forward KL using importance sampling and it may be possible to adapt this method to our setting.

The code to reproduce all the experiments and figures of this paper can be found online at \url{https://github.com/llaurabatt/amortised-variational-flows.git}.

\begin{acks}[Acknowledgments]
The authors thank Shahine Bouabid for valuable insight and advice.
We also thank the Google TPU Research Cloud (TRC) program for supporting the project.
\end{acks}

\begin{funding}
LB was supported by a Clarendon Scholarship from the University of Oxford. 
The compute was provided by Google through the TPU Research Cloud (TRC) program.
\end{funding}

\begin{supplement}

\noindent
\textbf{Estimation of the SMI posterior.}
Additional background on the estimation of SMI posteriors. It includes a description of the nested MCMC scheme used for comparison and details on the two-stage variational optimisation procedure ensuring modularity.

\medskip
\noindent
\textbf{Additional experimental results on synthetic data and epidemiological data.}
Experimental details for synthetic data and further experimental details for epidemiological data. Presents full training traces, hyperparameter optimisation plots, and posterior comparisons between MCMC, the Variational Posterior (VP), and the Variational Meta-Posterior (VMP) across different dataset sizes and inference settings.

\medskip
\noindent
\textbf{The Linguistic Atlas of Late Mediaeval
English (LALME) Data.}
Detailed specification of the linguistic profiles model, including generative assumptions, Gaussian-process spatial priors, sparse inducing-point approximation, and additional posterior diagnostics. Provides quantitative and visual comparisons between MCMC, VP, and VMP for both small and full datasets.

\medskip
\noindent
\textbf{Architecture, hyperparameters, and runtime.}
Description of the neural-flow architecture and training setup for the VMP and VP models. Includes network and optimiser settings, training hyperparameters, and runtime comparisons across all experiments.

\end{supplement}

\putbib     
\end{bibunit}

\newpage
\setcounter{page}{1}

\begin{supplement}

\begin{bibunit}
\appendix

\section{Estimation of the SMI posterior}
\label{sec:SMI-VMP-further-details}

In this section we add some further background on the estimation of SMI posteriors.

MCMC targeting the SMI posterior in \eqref{eq:smi-posterior} is not straightforward as $p(\theta|\delta, Y; \varphi)$ has a marginal factor $p(Y|\delta;\varphi)$ in the denominator (which doesn't cancels out as it does in the Bayes posterior in \eqref{eq:bayes_posterior_factors}). However, a form of nested MCMC can be used, targeting $p_{pow}(\delta, \tilde{\theta}|Y,Z; \tpsi)$ and sampling 
$p(\theta|\delta, Y; \varphi)$ for each sampled $\delta$. \cite{carmona_semi-modular_2020} explain why this seemingly inefficient scheme is often quite practical, though it does suffer from ``double asymptotics'' and has the usual strengths and weaknesses of MCMC. While variants of MCMC have been proposed to address these issues for SMI-posteriors and cut-posteriors \citep{plummer_cuts_2015,jacob20,liu_stochastic_2022}, better guarantees in terms of sample quality have often come at the expense of speed and scalability. We use a form of nested MCMC \citep{plummer_cuts_2015} as ground truth in small problems, for comparison purposes.

When we come to fit the variational approximation in \eqref{eq:smi-vmp-variational-density} we seek a set of variational parameters $\lambda^\ast$ maximising a utility that
rewards for the variational approximation being close to the SMI-posterior at any value of the hyperparameters $\tpsi$. Following \cite{carmona_scalable_2022}, this must be done in
two stages in order to stop the variational approximation from introducing uncontrolled
information flow between modules. 
We first set $\lambda_1^\ast, \lambda_3^\ast$ to minimise the expectation over the KL divergence to the power-posterior
\begin{align}\label{eq:smi-loss-13}
    \lambda_1^*, \lambda_3^*  &= \underset{(\lambda_1, \lambda_3)}{\text{argmin}}\underset{\tpsi \sim \rho}{\mathbb{E}}\left[ D_{KL}(q_{1}(\delta;\lambda_1, \tpsi)q_{3}(\Tilde{\theta}|\delta;\lambda_3, \tpsi)||p_{pow}(\delta, \Tilde{\theta}|Y,Z;\tpsi)))\right]
\end{align}
and then choose $\lambda^\ast_2$ to minimise the KL divergence to the part of the posterior constituting the analysis stage
\begin{align}\label{eq:smi-loss-2}
    {\lambda\mathstrut\!}^*_2 & = \underset{\lambda_2}{\text{argmin }}  \underset{\tpsi \sim \rho}{\mathbb{E}}\left[D_{KL} (q_{1}(\delta;\lambda^\ast_1, \tpsi)q_2(\theta|\delta;\lambda_2, \tpsi) ||p(\theta|Y,\delta; \varphi))\right]
\end{align}
These can be expanded to give the usual ELBO format maximisation target, using the reparameterisation trick and automatic differentiation. 

As \cite{carmona_scalable_2022} show, this setup gives a variational approximation in which information from $Y$ into $\delta$ is completely removed at $\eta=0$, which exactly matches standard variational Bayes using $q_1(\delta;\lambda_1,{\varphi})q_2({\theta};\delta;\lambda_2, {\varphi})$ to target $p(\delta,\theta|Y;\varphi)$ at $\eta=1$ and which approximates SMI at all $\eta$ in the sense that $q(\theta, \tilde{\theta}, \delta; \lambda, \tpsi)\to p_{smi}(\phi, \theta, \Tilde{\theta}|Y,Z; \tpsi)$ in the limit that the variational family encompasses $p_{smi}$. 
Each of the two optimisation stages fits a VMP by selecting $\lambda_1, \lambda_2$ and $\lambda_3$ to maximise the A-ELBO in \eqref{eq:A-ELBO} and the discussion of its properties as a universal approximator in Section~\ref{sec:UA-NF-bayes-amortised} carries over. The two-stage optimisation process set out above can be implemented in a single step by making use of the stop-gradient operator.
We refer the reader to \cite{carmona_scalable_2022} for further detail. 

\section{Additional experimental results on synthetic data and epidemiological data}\label{app:experimental-results}

\subsection{Synthetic data}\label{app:synthetic-data}

In this Section we consider synthetic data and standard Bayesian inference with no model misspecification so the marginal likelihood $p(Y|\psi)$ is a valid target for hyperparameter selection. We fit the VMP $q(\theta;{\lambda\mathstrut\!}^*,\psi)$, estimating ${\lambda\mathstrut\!}^*$ by taking $\widehat{{\lambda\mathstrut\!}^*}$ to maximise a sample based estimate of the A-ELBO in \eqref{eq:A-ELBO} and then estimate hyperparameters by choosing them to maximise the ELBO (a proxy for the marginal likelihood). We check estimated hyperparameters match true values. This is sanity check on a simple example but some of the things we learn carry over into the misspecified settings of interest to us in this paper.

The data generating process for the random effects model is
\begin{align}
    \mu_j &\sim N(m, s^2)\nonumber\\
    \sigma_j &\sim \text{IG}(g_1, g_2),\ j=1,\dots,J
    \label{eq:hierarchy-normal-simple}\\
    Y_{ij} &\sim N(\mu_j, \sigma_j),\hspace{0.075in} i=1,\dots,I.
    \nonumber
\end{align}
with hyperparameters $\psi=(m,s^2,g_1,g_2)$, true values $\psi_{True}=(0,1,1.5,0.5)$ and \text{IG} the Inverse Gamma distribution with shape $g_1$ and scale $g_2$. The posterior we target is $p(\theta|Y,\psi)$ with parameters $\mu=(\mu_j)_{j=1}^J$, $\sigma=(\sigma_j)_{j=1}^J$ and $\theta=(\mu,\sigma)$.
When $I$ and $J$ are large, hyperparameter estimation is ``easy'': the hyperameters each control the distribution of $J$ exchangeable parameters which are informed directly from the data.

The VMP is trained by maximising the $\aELBO$ in \eqref{eq:A-ELBO} under the true generative model in \eqref{eq:hierarchy-normal-simple}. The hyperparameter distribution $\rho$ used for amortisation were Normal(0, 3) (for $\mu$), Gamma(shape=2, scale=1) (for $\sigma$) and Gamma(0.5, 2) (for both $g_1$ and $g_2$). Once we have estimated ${\lambda\mathstrut\!}^*$ we select hyperparameters by maximising the ELBO-estimator as a proxy for maximising the marginal likelihood $p(Y \mid \psi)$. The ELBO is a good approximation to the marginal likelihood $p(y \mid \tpsi)$ when the fitted VMP $q(\theta;\widehat{{\lambda\mathstrut\!}^*},\tpsi)$ is close to the target $p(\theta \mid Y,\tpsi)$. \cite{rezende2015variational} and later authors make a joint maximisation of the ELBO over $\tpsi$ and $\lambda$ to estimate ${\lambda\mathstrut\!}^*$ and $\tpsi^*$. By contrast we amortise over $\tpsi$ when we fit the VMP, we don't immediately maximise. 

If we want a point estimate for $\tpsi$ we use the fitted VMP to estimate the ELBO at any $\tpsi$-value,
\begin{equation}\label{eq:elbo-estimate}
\widehat{\ELBO}(\widehat{{\lambda\mathstrut\!}^*},\tpsi)=\frac{1}{N}\sum_{i=1}^N \log\left( \frac{p(T(\epsilon^{(i)};\widehat{{\lambda\mathstrut\!}^*},\tpsi),Y \mid \tpsi)}{q(\epsilon^{(i)})\left\vert \det(J_{T}(\epsilon^{(i)};\widehat{{\lambda\mathstrut\!}^*},\tpsi))\right\vert^{-1}}\right)
\end{equation}
where $\epsilon^{(i)}\sim q(\cdot),\ i=1,\ldots,N$ are independent samples from the base distribution of the amortised flow.
If the ELBO is our utility, then we set
\[
{\widehat{\tpsi}}^*=\arg\!\max_{\tpsi} \widehat{\ELBO}(\widehat{{\lambda\mathstrut\!}^*},\tpsi).
\]

We consider a small dataset where $I=8$ and $J=10$ and a larger dataset where $I=50$ and $J=50$. Training loss and hyperparameter trace plots from the second-stage $\psi$-optimisation are shown in Figure~\ref{fig:synth-small-hparam-optim} (small data set) and Figure~\ref{fig:synth-large-hparam-optim} (large data set) of Section \ref{app:additional-results-synth}. In Table \ref{tab:synth-results} we give the hyperparameter values which maximise $\widehat\ELBO({\lambda\mathstrut\!}^*,\psi)$. For the large dataset these are very close to the true values. In Figures \ref{fig:synth-large-mu} and \ref{fig:synth-large-sigma} of  Section \ref{app:additional-results-synth} we plot the posteriors we estimate for $\mu$ and $\sigma$ on the large data set using three different methods: MCMC at the true values $\psi_{True}$ (the oracle); the VP fitted at $\psi_{True}$; the VMP posterior at hyperparameters $\psi^\ast$ estimated using the VMP. These are all essentially indistinguishable. On the small dataset, estimated hyperparameter values in Table~\ref{tab:synth-results} differ a bit more from true values and the difference in the reconstructed posteriors is visible in Figures~\ref{fig:synth-small-posteriors}, though slight. There is little under-dispersion (despite the reverse-KL) due to the relatively simple target and the expressive power of a normalising flow. 

\begin{table}[H]
\centering
\begin{tabular}{lccc}
\toprule
& True & \multicolumn{2}{c}{Estimated Optimum} \\
\cmidrule(lr){3-4}
& Value & Small & Large \\
 \midrule
m & 0.0 & 0.4 & 0.01 \\
s & 1.0 & 0.85 & 0.98 \\
$g_1$ & 1.5 & 1.05 & 1.42 \\
$g_2$ & 0.5 & 0.43 & 0.5 \\
\bottomrule
\end{tabular}
\caption{True vs VMP-estimated optimal values for the synthetic dataset where $I=8$ and $J=10$ (small dataset) and that where $I=50$ and $J=50$ (large dataset).}
\label{tab:synth-results}
\end{table}

\begin{figure}[htb]
\vspace*{-0.1in}
\centering
\begin{subfigure}{\textwidth}
  \centering
  \includegraphics[width=\linewidth]{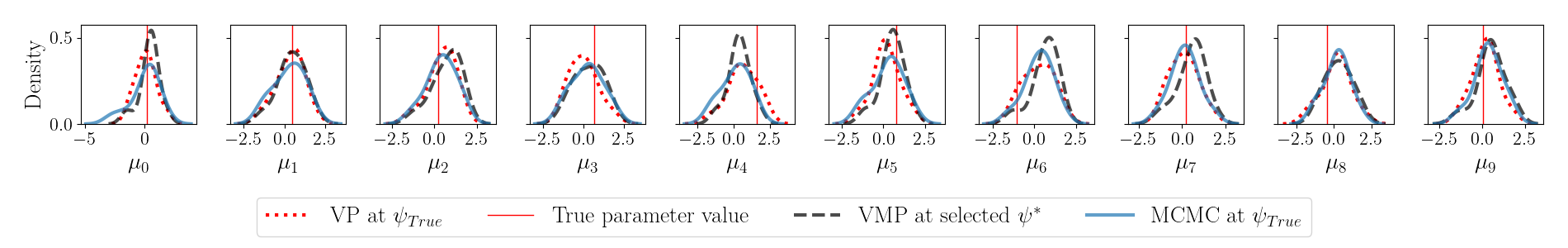}
  \caption{Posteriors for $\mu_i$}
  \label{fig:synth-small-mu}
\end{subfigure}%

\begin{subfigure}{\textwidth}
  \centering
  \includegraphics[width=\linewidth]{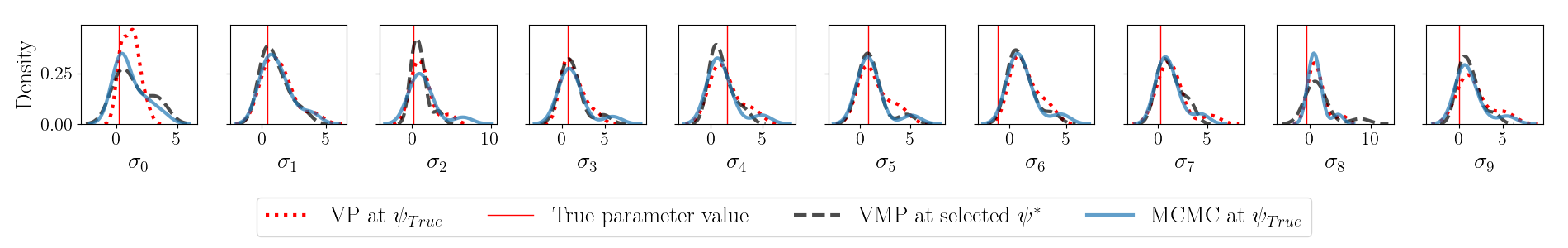}
  \caption{Posteriors for $\sigma_j$}
  \label{fig:synth-small-sigma}
\end{subfigure}
\caption{Synthetic small dataset. Ground truth posteriors compared to VMP posteriors at selected hyperparameters. See Figures \ref{fig:synth-large-mu} and \ref{fig:synth-large-sigma} for the large data set.}
\label{fig:synth-small-posteriors}
\end{figure}

\subsubsection{Additional information}
\label{app:additional-results-synth}

Figure \ref{fig:synth-small-simdata} gives a graphical illustration for the data and data generating process of the small synthetic dataset. Figure \ref{fig:synth-small-hparam-optim} shows the ELBO and the traces for the four hyperparameters. We can see that they converge to reasonable values, which we have seen to induce sensible posterior approximations in Figure \ref{fig:synth-small-posteriors}.

\begin{figure}[htb]
    \centering
    \includegraphics[width=0.7\linewidth]{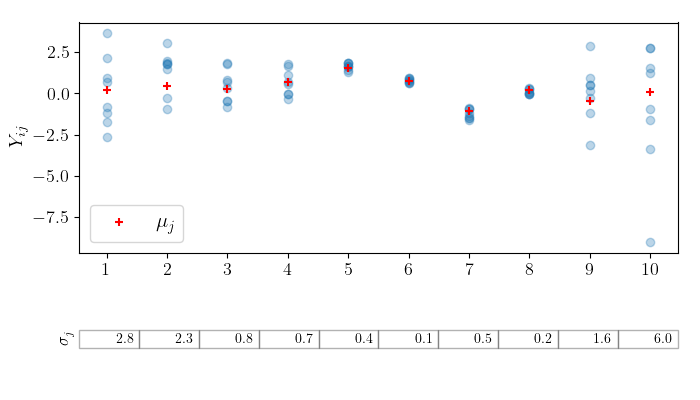}
    \caption{Synthetic data. Data generating process for dataset with $I=8$ and $J=10$.}
    \label{fig:synth-small-simdata}
\end{figure}

\begin{figure}[htb]
    \centering
    \includegraphics[width=0.7\textwidth]{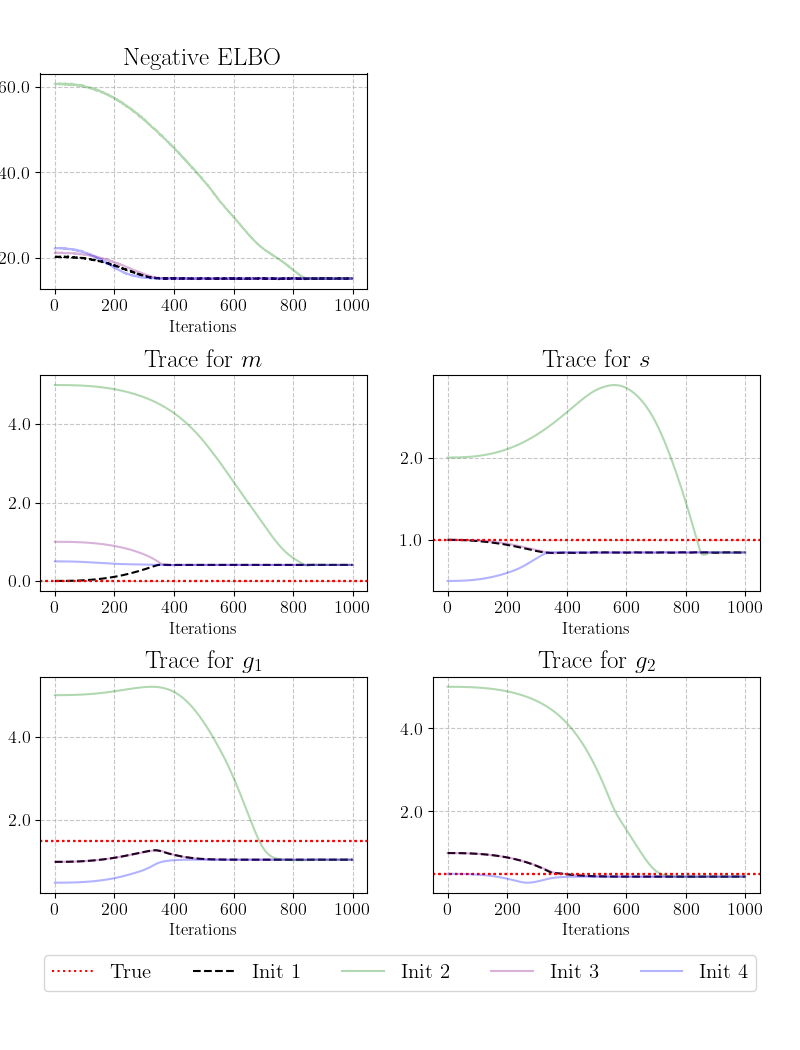}
    \caption{Synthetic data. Training loss and trace plots for hyperparameter optimisation on smaller dataset at different inital values. The run achieving the lowest average loss over the last 20 iterations is highlighted in black. Hyperparameters were selected as the average values over the 20 last iterations corresponding to the highlighted loss.}
    \label{fig:synth-small-hparam-optim}
\end{figure}

Figure \ref{fig:synth-large-simdata} gives a graphical illustration for the data and data generating process of the large synthetic dataset. Figure \ref{fig:synth-large-hparam-optim} shows the ELBO and the traces for the four hyperparameters. It is evident that they converge closely to the true values. The estimated posterior approximations are nearly indistinguishable from MCMC-derived posteriors (see Figures \ref{fig:synth-large-mu} and \ref{fig:synth-large-sigma}).
\begin{figure}[htb]
    \centering
    \includegraphics[width=\textwidth]{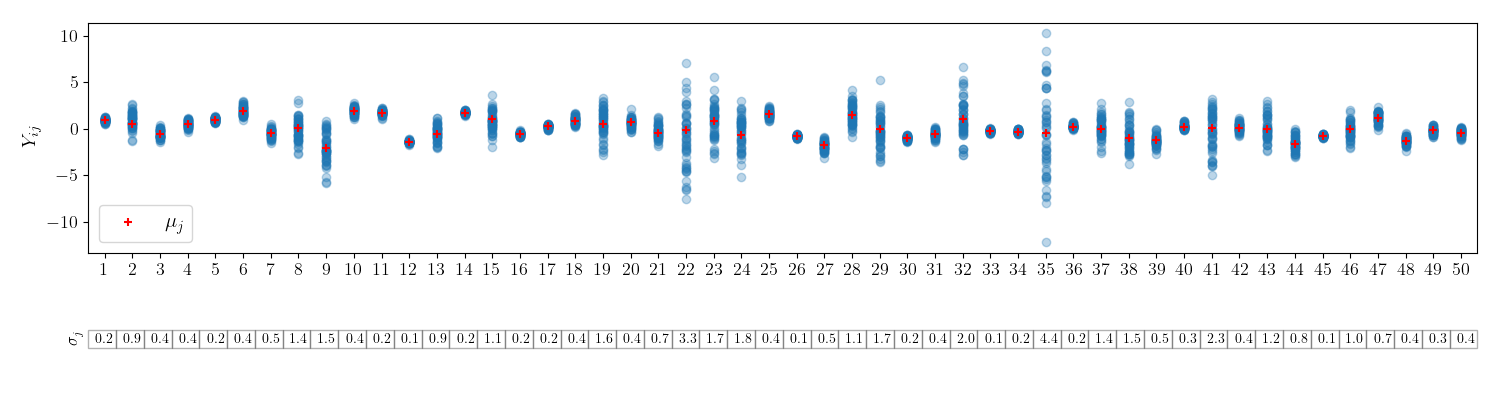}
    \caption{Synthetic data. Data generating process for larger dataset with $I=50$ and $J=50$.}
    \label{fig:synth-large-simdata}
\end{figure}

\begin{figure}[htb]
    \centering
    \includegraphics[width=0.65\textwidth]{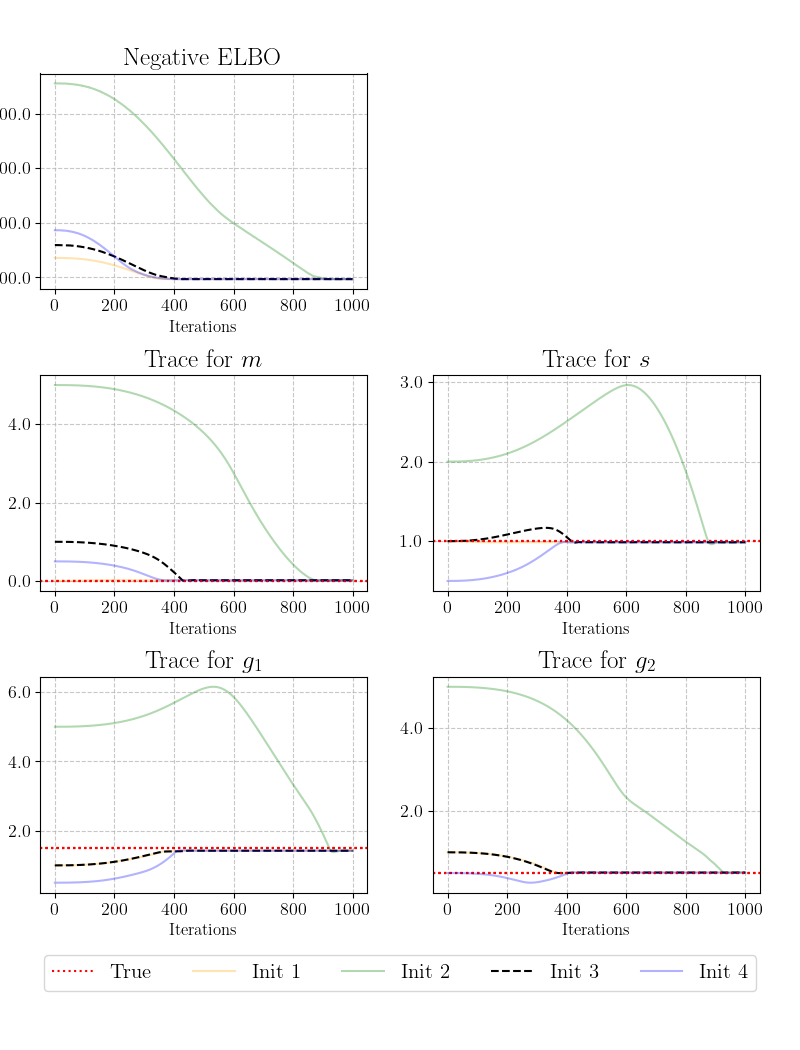}
    \caption{Synthetic data. Training loss and trace plots for hyperparameter optimisation on larger dataset at different inital values. The run achieving the lowest average loss over the last 20 iterations is highlighted in black. Hyperparameters were selected as the average values over the 20 last iterations corresponding to the highlighted loss.}
    \label{fig:synth-large-hparam-optim}
\end{figure}

\begin{figure}[htb]
\centering
\includegraphics[width=\linewidth]{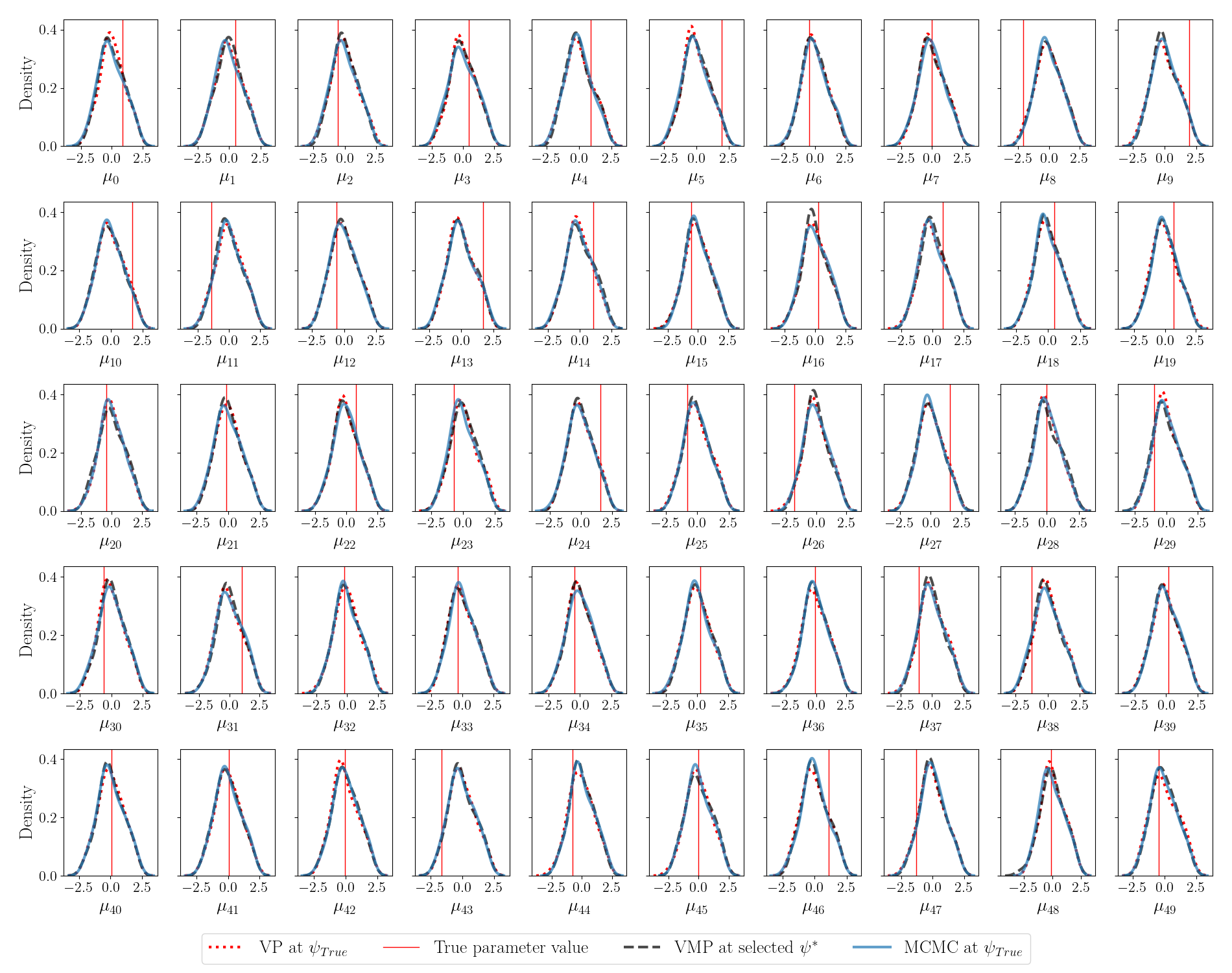}
\caption{Synthetic larger dataset. Ground truth  $\mu_j$ posteriors compared to estimated $\mu_j$ posteriors at selected hyperparameters.}
\label{fig:synth-large-mu}
\end{figure}

\begin{figure}[htb]
\centering
\includegraphics[width=\linewidth]{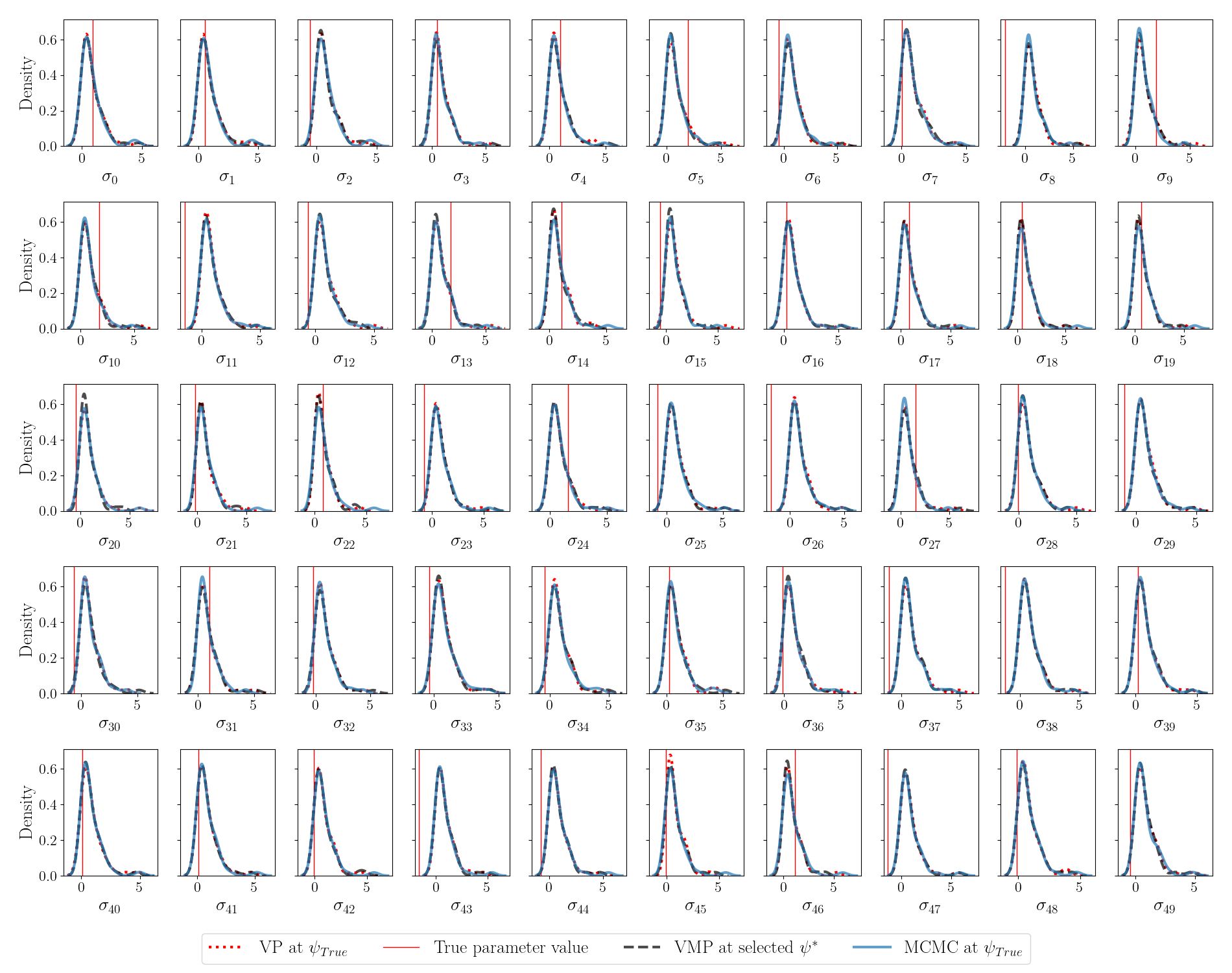}
\caption{Synthetic larger dataset. Ground truth  $\sigma_j$ posteriors compared to estimated $\sigma_j$ posteriors at selected hyperparameters.}
\label{fig:synth-large-sigma}
\end{figure}

\subsection{Epidemiological data}\label{app:experimental-results-HPV}

\subsubsection{Supplementary figure for Section~\ref{sec:hpv-smi}: SMI with the VMP and amortised prior hyperparameters}

In this Section we give Figure~\ref{fig:HPV-SMI-trace} illustrating convergence of the optimisation of the ELPD over $\tpsi$ described in Section~\ref{sec:hpv-smi}. We also give a comparison against MCMC of the fitted VMP targeting the SMI-posterior in \eqref{eq:smi-posterior} for the case of the generative model in \eqref{eq:hpv-generative-model}. We show the marginal posteriors for selected $\delta$-parameters in Figure~\ref{fig:HPV-SMI-delta}.

\begin{figure}[htb]
\centering
\begin{subfigure}{.5\textwidth}
  \centering
  \includegraphics[width=\linewidth,
  height=\linewidth]{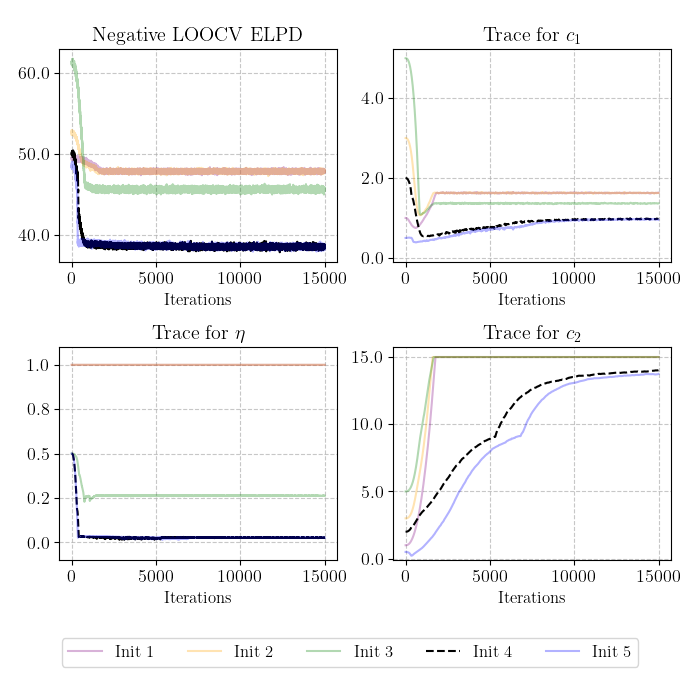}
  \caption{Z module}
  \label{fig:HPV-SMI-trace-Z}
\end{subfigure}%
\begin{subfigure}{.5\textwidth}
  \centering
  \includegraphics[width=\linewidth,height=\linewidth]{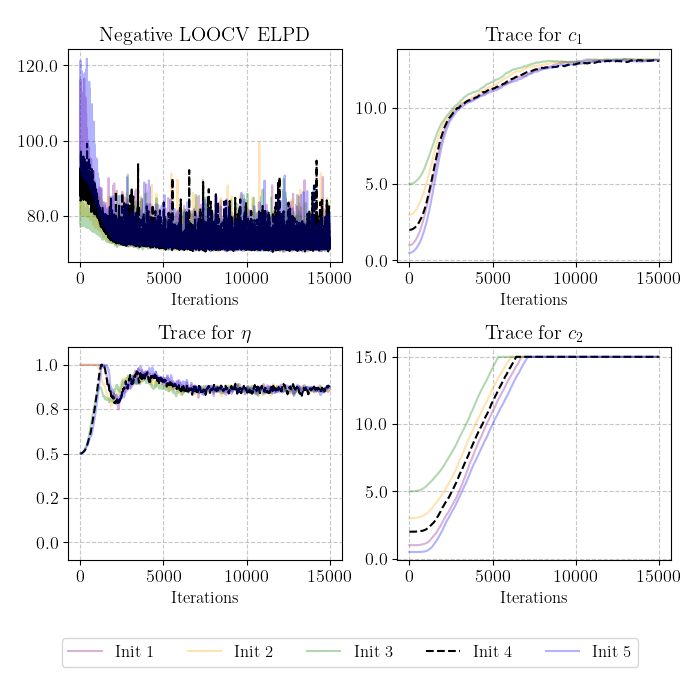}
  \caption{Y module}
  \label{fig:HPV-SMI-trace-Y}
\end{subfigure}
\caption{Epidemiological data in SMI setting of Section~\ref{sec:hpv-smi} in the main paper. LOOCV loss and trace plots for hyperparameter optimisation at different inital values. The run achieving the lowest average loss over the last 20 iterations is the highlighted dashed line in black. Hyperparameters were selected as the average values over the 20 last iterations corresponding to the highlighted loss.}
\label{fig:HPV-SMI-trace}
\end{figure}

\begin{figure}[htb]
\vspace*{-0.2in}
\centering
\begin{subfigure}{.5\textwidth}
  \centering
  \includegraphics[width=\textwidth]{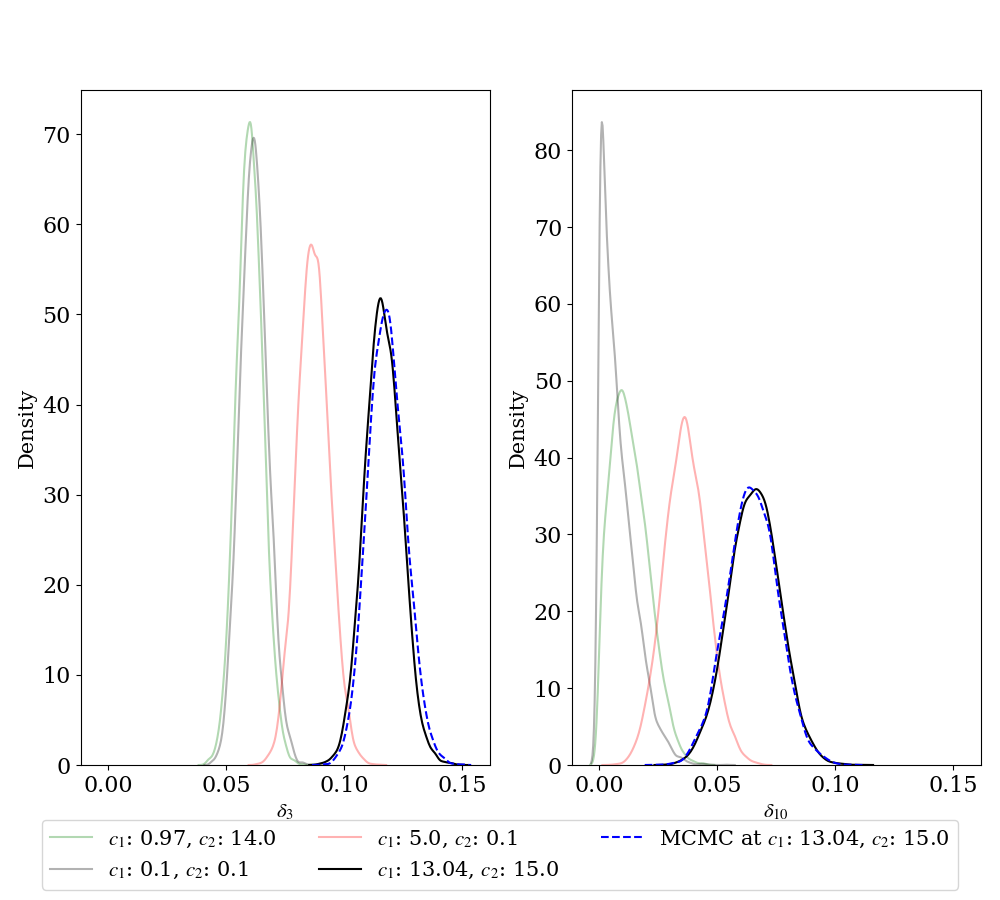}
  \caption{$\eta = 0.87$}
  \label{fig:HPV-SMI-delta-eta1}
\end{subfigure}%
\begin{subfigure}{.5\textwidth}
  \centering
  \includegraphics[width=\linewidth]{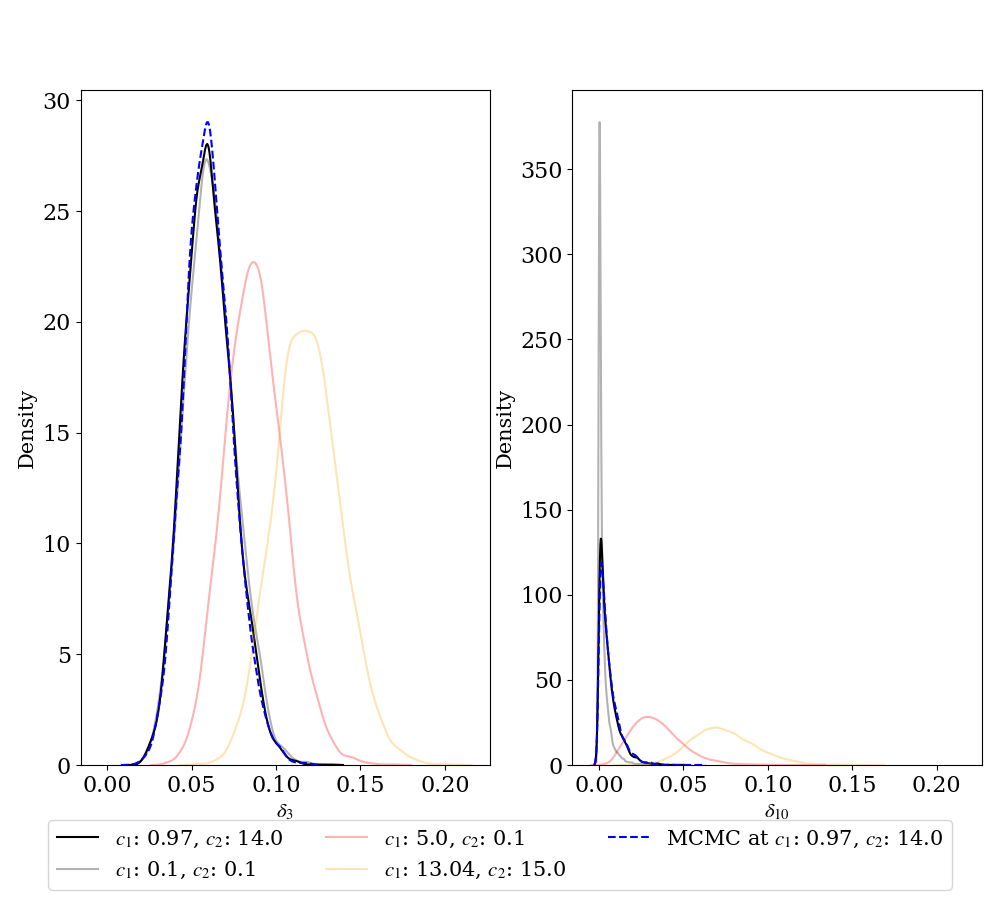}
  \caption{$\eta = 0.02$}
  \label{fig:HPV-SMI-delta-eta0}
\end{subfigure}
\caption{Epidemiological data in SMI setting. Solid lines: densities for VMP samples for $\delta_8$ and $\delta_{10}$ at different prior hyperparameters; optimal prior hyperparameters in black. Blue dashed lines: MCMC samples at optimal prior hyperparameters.}
\label{fig:HPV-SMI-delta}
\end{figure}

\section{The Linguistic Atlas of Late Mediaeval English (LALME) data}
\label{sec:lalme_data_main}

We now describe the data in detail. See Section~\ref{sec:data_background_LP} in the Supplementary Material for further background from the perspective of historical linguistics and \cite{haines_simultaneous_2016} for exploration of these data from a statistical perspective.

The \acrfull*{lalme}, 
originally published by
\citet{mcintosh_linguistic_1986} and later revised and supplemented \citep{benskin_electronic_2013}, 
presents a corpus of lexico-grammatical data, consisting of 1044 completed lexico-grammatical survey questionnaires. Each such inventory represents the internally consistent usage of a text which, whether original or a copy, is on the evidence of its handwriting the work of a single scribe; each such inventory constitutes a \emph{linguistic profile} (LP), and typically it identifies one and only one scribe \citep{McIntosh1974,McIntosh1975,benskin1981translations}. It should be emphasised that a dialect whether written or spoken may be identified as much by its combination of features shared with other dialects as by features that are found in that dialect alone, and \textit{a fortiori }this is true of an idiolect, the usage of an individual. The LALME corpus is in the first instance a record of idiolects, probably no one of which is identical with any other, and is hence the basis of an ``identikit'' not only for local dialects, but for individual scribes.

The Leeds Survey of Modern English Dialects \citep{orton_survey_1962} could choose its three hundred survey points as more or less evenly spaced across the land, but a survey of the mediaeval language has to make the best of what it can find. The Atlas’s ‘informants’ were its sample manuscripts, represented by 1044 linguistic profiles.
 Ideally the source material for a linguistic atlas would all be of the same date, so that its maps would be free of any chronological disruption. In practice, however, the language historian must use what material has survived, and for LALME to attain anything like the desired density and evenness of coverage, sources from within a date range of ca. 1350 to ca. 1450 had to be accepted \citep{mcintosh_new_1963}. Only towards the end of this period was a national written standard coming into being: LALME records the non-standard usage of individual scribes during the last hundred or so years when, in the main, written English was local or regional dialect as a matter of course. Largely the written language reflects variation in spoken language (scribes spelled as they spoke) but conventions of writing that are independent of speech may prove no less to be markers of local origin \citep{benskin1982letter}. The fixed-format survey questionnaire underlying the LPs is designed to elicit data of both kinds. 

 Large though it is, LALME's corpus represents less than a fifth of the potential sources investigated; prominent among the disqualifications is mixed usage, whether a copyist's rendering of an alien dialect, or an accommodation to regional or incipient national standard \citep{benskin1981translations,Samuels1963,Samuels1981,benskin1981translations,benskin2004chancery}. Nearly all of LALME's 1044 LPs belong within England, though its coverage by county is perforce still uneven; six are from within Wales, and eighteen from within Lowland Scotland (for which a separate and comprehensive atlas was later published, as \cite{Williamson2008}.

Each category in the survey questionnaire is an \emph{item} (loosely, a separate word or grammatical element, indicated by \{\}), which elicits the  \emph{forms} (loosely, different spellings, indicated below by \lf\rf) functionally equivalent to that item. We elaborate these definitions in Section~\ref{sec:data_background_LP}.
For example, the item \{AGAINST\} has 165 variant forms, including \lf{AGEINST}\rf, \lf{AYEINST}\rf, \lf{AGAINz}\rf ~and \lf{TOYENST}\rf
(lower case letters are used for manuscript symbols, abbreviations and mediaeval letters now obsolete).
A single item can have many forms, some as many as several hundred.
The Atlas authors start with a list of items. They have $P$ sample texts and take each text and go through each item in the list recording the forms of the item which appear in the text. Each text is then summarised by a record called a \acrfull*{lp} recording for each item the forms it contains.

Denote by $\I=\{1,2,...,I\}$ the set of all item labels (distinct words and grammatical concepts) and for each item $i\in \I$ let $\F_i=\{1,2,...,F_i\}$ give the set of forms it takes. For example if $i\in \I$ is the item label for \{AGAINST\} then each of the forms listed above will have a corresponding label $f\in \F_i$. As explained below, we focus on $I=71$ selected forms which we subject to re-registration in a linguistically informed coarsening process detailed below and in Section~\ref{sec:data_background_coarsening}.
There are in all $d_{\phi}=\sum_{i=1}^I F_i$ forms in these data. Let $\P=\{1,2,...,P\}$ give the set of profile labels. An observation $y_{p,i,f}=1$ indicates that form $f$ of item $i$ is present in the sample text summarised by profile $p$, and $y_{p,i,f}=0$ is observed otherwise. Let $y_{p,i}=(y_{p,i,f})_{f\in \F_i}$ and $y_p=(y_{p,i})_{i\in\I}$, so that data $y_p\in \{0,1\}^{d_\phi}$ records the presence and absence of forms for the \acrshort*{lp} for the text with label $p\in \P$.

The 1044 \acrshortpl*{lp} are divided into two categories: anchor profiles and floating profiles.
Anchor profiles are localised on the basis of non-linguistic evidence. 
In rare cases, a scribe can be identified and his biography is known, but most anchors are legal or administrative documents whose place of origin is explicitly stated, and which (though it is not a foregone conclusion) can be trusted at least provisionally to represent the local forms of language \citep{benskin_chancery_2004}. 
\citet{mcintosh_linguistic_1986} deduce the location of floating profiles using a sequential method they call the \emph{fit-technique}. 
For each form that appears in
a floating profile, anchor profiles and previously located floating profiles are used to delimit the  region in which this form was used. The unknown profile location should lie in the intersection of all such regions across all the forms that the floating profile displays.
A profile that cannot be localised in this way is likely to represent a dialect mixture; in many cases such mixtures result from uneven copying from a text written in a dialect that is not the scribe's own. \citep{benskin_translations_1981}
The localised float is then treated as an anchor and the process continued until all the floats are positioned. 
If someone wishes to locate a new text they treat it as the $M+1$'st float and apply the same method.
In one or two cases, anchors found after the Atlas was published have confirmed placings previously determined by ‘fit’.

Statistical modelling of these data is challenging, as a given item can have hundreds of forms, and many of these may be rare. In order to treat this a coarsened version of the data was made. This coarsening is described in Section~\ref{sec:data_background_coarsening}. Forms with very similar spellings were merged according to linguistic criteria developed by the authors of the Atlas. 
The coarsened data contains, for the purpose of location estimation, most of the information in the original data, but is significantly less sparse. In this paper we analyse the coarsened data.

We focus on the rectangular area displayed in Figure~\ref{fig:rectangular_region_anchorfloat}, spanning 200 \unit{\km} east-west and 180 \unit{\km} north-south.
The region contains $N=120$ anchor profiles with labels $A=\{1,...,N\}$. A further $M=247$ floating profiles with labels $\bar{A}=\{N+1,...,N_M\}$ were assigned to this region using the fit-technique (so that $P=367$ above).
We are assuming the Atlas float-locations are sufficiently accurate for this purpose. Our analysis below supports this, so the selection rule and analysis results are at least consistent.
The data $X_A, Y_A, Y_{\bar{A}}$ and missing data $X_{\bar{A}}$ are now as defined in the introduction. There are in all $d_{\phi}=741$ forms in the coarsened data displayed in the region of interest. As an example of dialectal variation, in Section~\ref{fig:SUCH_anchor_locations} we display the locations corresponding to usage of coarsened forms of the item \{SUCH\}.

\begin{figure}[ht]
  \centering
  \includegraphics[width=0.7\linewidth]{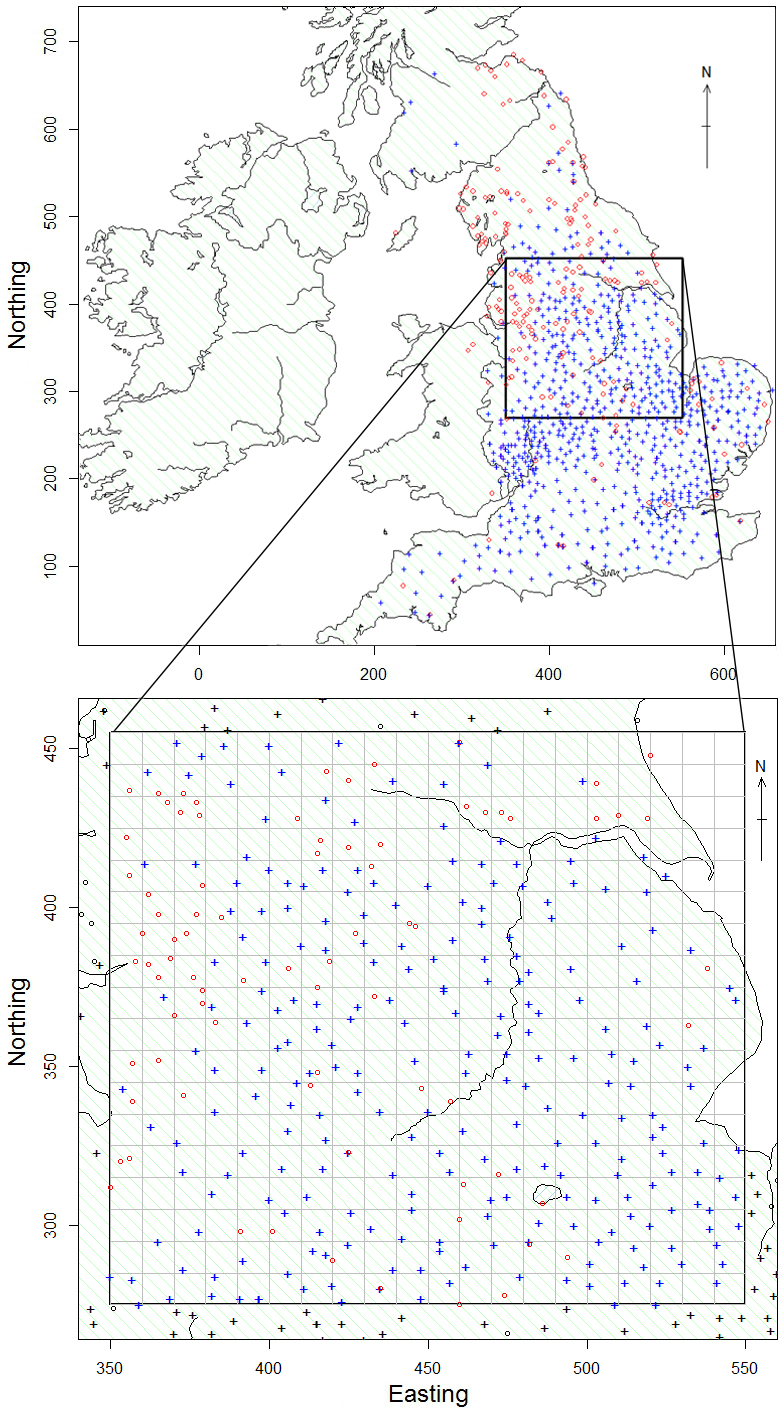}
  \caption{The locations of the anchor profiles (red) and the locations of the floating profiles estimated using the fit-technique (blue).
  LALME data: a map outlining the exact location of the 120 anchor linguistic profiles and the Atlas estimates for the location of the 247 floating linguistic profiles. The focus is on a a 200km $\times$ 180km region in central Great Britain. The Atlas contains many profiles located outside the highlighted region.
}
  \label{fig:rectangular_region_anchorfloat}
\end{figure}

Our analysis of the Atlas data itself is a proof-of concept. It is clearly at least possible to give a statistical analysis automating the placement of new profiles and potentially replacing the fit-technique. We estimate and quantify uncertainty in the location of floating profiles given the anchor profiles. This is the same problem the Atlas authors faced. We get good agreement between the credible regions we estimate for floating profiles and the Atlas locations estimated using the fit-technique.
Our approach is automated, and provides well-calibrated uncertainty measures for location.
The level of location precision achievable with these data has been in question and this is not easily quantified using the non-statistical fit-technique. Our analysis shows that location accuracy of order twenty to forty kilometres is possible, underpinning the value of the Atlas. One application of the Atlas is to locate new floating profiles. We validate our method by treating anchors as if they were floating profiles and predicting their locations. This demonstrates the feasibility of locating new texts.

\begin{figure}
  \centering
  \includegraphics[width=0.8\linewidth]{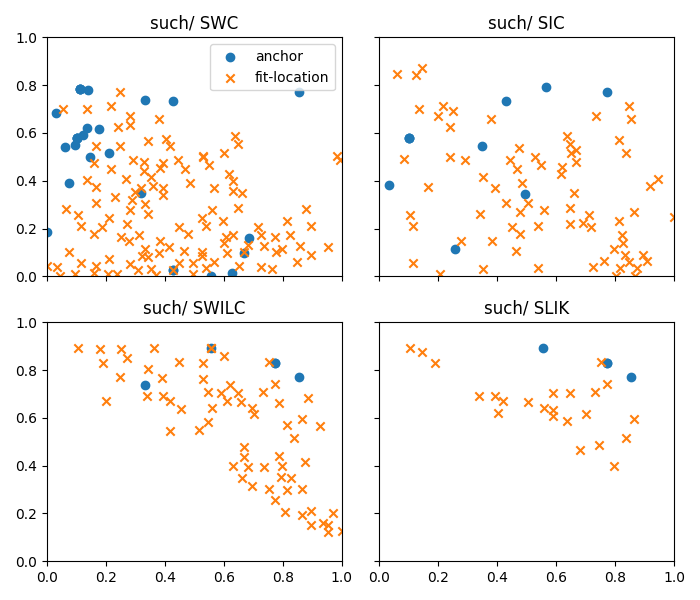}
  \caption{Linguistic Profile locations for four coarsened forms \lf{SWC}\rf, \lf{SIC}\rf, \lf{SWILC}\rf and \lf{SLIK}\rf of the item \{SUCH\}. These coarsened forms correspond to groups of forms in the uncoarsened data, so for example the form \lf{SUCH}\rf itself is coarsened to the group of \lf{SWC}\rf. Anchor profiles are marked with blue circles, while floating profiles locations produced by the fit-technique are shown in orange.}
  \label{fig:SUCH_anchor_locations}
\end{figure}

\subsection{Further background on linguistic profiles}\label{sec:data_background_LP}
The core of the Atlas comprises some 1044 taxonomic inventories, whose contents are the spellings elicited by a fixed-format survey questionnaire; a completed survey questionnaire is a linguistic profile (‘LP’) for an individual scribe. In all, the forms of 623 distinct items were recorded. The items were unevenly surveyed over space. Details of how the list of recorded items developed are recorded in the Atlas. It should perhaps be emphasised that the survey is not concerned with peculiar local vocabulary. The choice of an item was governed by two considerations: (a) it must be represented by at least two co-variant forms of philological interest, and (b) it must be in common use, with a good chance of appearing in whatever text might be available. Most items are hence commonplace vocabulary, like \{THESE\}, \{SHE\}, \{WHICH\}, \{SHALL\}, \{IF\}, \{TOGETHER\}, and \{TWO\}. Grammatical inflexions are almost by definition of common occurrence; among those used in Atlas are the suffixes of the \{present participle\} (so \lf{$\sim$ING}\rf, \lf{$\sim$YnG}\rf, \lf{$\sim$AND}\rf, \lf{$\sim$END}\rf, etc.) and the \{noun plural\} (so \lf{$\sim$ES}\rf, \lf{$\sim$IS}\rf, \lf{$\sim$US}\rf, etc.). The Atlas is in the first instance a survey of written language: much of the variation it records corresponds to variation in the spoken language, but some of it, like that between \lf{$\sim$YnG}\rf and \lf{$\sim$ING}\rf, demonstrably does not. The idea that any given word in English should have one and only one spelling, even if its pronunciation was a constant, was alien to perhaps the majority of scribes; spelling systems had their own generative logic, which could be developed in different ways. Although some scribes were very consistent, it is not unusual to find in the course of a long text that many words, perhaps even a majority, have several forms in either free or conditioned variation. Since the Atlas was intended as a research tool, all variants elicited by the questionnaire, regardless of their presumed significance for the spoken language, are separately recorded; transcribed from manuscript, they are represented in raw state so far as conversion into ASCII characters allows. The contents of the completed survey questionnaires are correspondingly prolific and diverse.

\subsection{Data registration and coarsening}\label{sec:data_background_coarsening}
The number of forms corresponding to a single item can be very large, commonly in the range of 100–200; notorious is \{THROUGH\}, for which the Atlas records over four hundred variants. After analysing this primary data, \citet{haines_simultaneous_2016} adopted a coarsened version, in which the forms of an item under consideration were classified and treated as subsets. For experiments with multi-dimensional scaling as an alternative to the ‘fit’-technique \citep{benskin_numerical_1988}, a segmentation dictionary had been developed in course of making the Atlas. For each item, a template consisting of vowel and consonant segments was devised from which all of its variants could be generated. For example, the template for \{THEM\} is C1+V1+C2+V2+C3. These can be seen as slots into which particles can be dropped. In \{THEM\}, C1 can be null (in the forms \lf{AM}\rf ~and \lf{EM}\rf) or H (so \lf{HAM}\rf, \lf{HEM}\rf, etc.) or TH (so \lf{THEM}\rf, \lf{THAIM}\rf, etc.) or the mediaeval ‘thorn’ (so \lf{yEM}\rf, \lf{yEIM}\rf, etc.) or mediaeval ‘yogh’ (for which \lf{zAM}\rf, \lf{zEOM}\rf, etc.); V1 can be A, E, EE, EO, I, Y, O, OE, U, UE, EI, EY, AI, AIE, AY; C2 is usually M, but often abbreviated (so \lf{HEm}\rf, \lf{THAm}\rf, etc.), rarely MM (for example \lf{HEMM}\rf) or mM (for example \lf{yAmME}\rf), but can be S (seen in \lf{AS}\rf, \lf{ES}\rf, \lf{HIS}\rf, etc.) or N (as \lf{HAN}\rf); V2 is usually null, but may be E (as in \lf{THAIME}\rf); C3 is usually null, but may beN (in \lf{HEMEN}\rf) or {HYMEN}\rf). These do not exhaust the possibilities for the 116 different forms (letters may be written superscript, as for example in \lf{H$\mbox{A}^{\text A}$M}\rf and \lf{$\mbox{Y}^{\text{EM}}$}\rf), but are enough to demonstrate the principle. Via such a template, comparison of LPs in respect of them can be confined to a single segment, and the number of variants reduced to easily managed proportions by identifying different particles within a slot. Thus for C1, a binary contrast of H versus TH and its mediaeval equivalent ‘thorn’ includes most by far of the 116 variants and so removing the distinction between H and TH in C1 of \{THEM\} is a coarsening; the contrast is implicitly phonological, as is that between M (including abbreviated m) and S for C2. Philological judgement, by no means infallible, is called for at every stage, but such restrictions relieve the input data of what is likely to be a great deal of noise. 

Coarsening data is time consuming - so far 71 items have been coarsened. For present purposes, the area investigated was also confined. 

\subsection{LALME model and further figures}
\label{app:experimental-results-LPs}

\subsubsection{Additional background to the data and model} \label{sec:LP-background-expanded}

Data $y_p$ at location $x_p$ is assumed to be generated conditional on the local value $\Phi(x)$ of a multi-dimensional latent spatial field $\Phi: \X \rightarrow [0,1]^{d_\phi}$, $d_\phi = \sum_{i =1}^I F_i$, and a set of other global parameters $\alpha$. Each field component $\phi_{i,f}(x_p)$ represents the probability of an item $i$ to take a particular form $f$, under the constraint that $\sum_{f=1}^{F_i} \phi_{i,f}(x_p) = 1 $ $\forall i \in \I$ and $ \forall p \in \mathcal{P}$.

The proposed observational model for the binary $y_{p,i,f}$'s takes $y_{p,i,f} = I(z_{p,i,f}>0)$, where $z_{p,i,f}$ is the Poisson number of times word/item $i$ in spelling/form $f$ is used in LP $p$, with $z_{p,i,f} \sim Poisson(\mu_i\phi_{i,f}(x_p))$ assumed. Here $\mu_i$ represents the mean number of times item $i$ is used across the corpus, while $\phi_{i,f}(x_p)$ is the probability that LP $p$ uses item $i$ in form $f$, under the constraint $
\sum_{f=1}^{F_i}\phi_{i,f}(x_p)=1$. Selection of form $f$ is just a multinomial thinning of an overall Poisson count. Noting that  
\begin{equation}
p(z_{p,i,f}=0|\mu_i, \phi_{i,f}(x_p)) = e^{-\mu_i \phi_{i,f}(x_p)}
\end{equation}
and adding a parameter $\zeta_i \in [0,1]$ to allow for an observed zero inflation, we obtain
\begin{align}
\label{eq:LP_obs_model}
p(Y_{\mathcal{P}}|\zeta,\mu, \Phi(X_{\mathcal{P}}))
=&\prod_{p\in\mathcal{P}}\prod_{i\in\I}
\prod_{f=1}^{F_i} p(y_{p,i,f}|\zeta_i,\mu_i, \phi_{i,f}(x_p))\\ 
=&\prod_{p\in\mathcal{P}}\prod_{i\in\I}
\prod_{f=1}^{F_i}\bigg[\zeta_i(1-y_{p,i,f})\bigg. \\
&\bigg.+ (1-\zeta_i)(1-e^{-\mu_i\phi_{i,f}(x_p)})^{y_{p,i,f}}(e^{-\mu_i\phi_{i,f}(x_p)})^{1-y_{p,i,f}}\bigg] \nonumber
\end{align}

The priors for item useage rates $\mu_i$ and zero inflation probability $\zeta_i$ are modelled as follows
\begin{align}
    \label{eq:mu-prior}
    p(\mu) &= \prod_{i=1}^I p(\mu_i) = \prod_{i=1}^I \text{Gamma}(a,b)\\
    p(\zeta) &= \prod_{i=1}^I p(\zeta_i) 
    = \prod_{i=1}^I \text{Uni}(0,1)\label{eq:zeta-prior}
\end{align}

The number of spatial fields $\phi_{i,f}(x_p)$ is large (741), and there is also anti-correlation across field values for item forms in each location (indicating a specific dialect prevalence), and positive correlation among item forms across locations (induced by a shared dialect). To reduce the dimensionality of the problem and induce correlation across items and forms at a given location, we parametrise the fields as a (soft-maxed) linear combination of a small number of shared basis fields $\gamma_b:\X\to R,\ b\in\B,\ \B=\{1,\dots,B\}$, with $B$ equal to 10, that only depend on location. To induce correlation across locations, the basis fields are modeled as Gaussian Processes (GPs) with a common exponential quadratic kernel $k(\cdot,\cdot)$ modeling spatial decay of correlation,
\begin{align*}
    \phi_{i,f}(x_p) &= \frac{\text{exp} (\gamma_{i,f}{(x_p)})}{\sum_{f' \in \F_i}\text{exp}(\gamma_{i,f'}{(x_p))}} \\
    \gamma_{i,f}(x_p) &= a_{i,f} + \sum_{b=1}^B\gamma_b(x_p)w_{i,f,b}
    \\
    \gamma_b(x_p) &\sim GP(0,k(\cdot,\cdot)) \quad b \in \B=\{1,\ldots,B\} 
\end{align*}
Finally, the priors for the weights, off-sets and floating locations themselves are represented as
\begin{align*}
    p(X_{\barA}) &= \prod_{p\in \barA}\text{Uni}(x_p \in \X)\\
    p(a) &=\prod_{i=1}^I \prod_{f=2}^{F_i}p(a_{i,f}) = \prod_{i=1}^I \prod_{f=2}^{F_i} \text{Normal}(0,\sigma_a)\\
    p(W) &=\prod_{b=1}^B\prod_{i=1}^I \prod_{f=1}^{F_i}p(w_{i,f,b}) = \prod_{b=1}^B\prod_{i=1}^I \prod_{f=1}^{F_i} \text{Laplace}(0,\sigma_w)
\end{align*}

Indicating all the $\textit{global}$ parameters with $\alpha = (\mu, \zeta, a, W)$, the joint distribution is
\begin{align}
p(Y_{\mathcal{P}},\alpha, \Gamma(X_{\mathcal{P}}), X_{\barA}))
\equiv& p(Y_{\mathcal{P}},\alpha, \Gamma(X_{\mathcal{P}}), X_{\barA})\\
=& p(Y_{A}|\alpha, \Gamma(X_{A}))\\
&\times p(Y_{\barA}|\alpha, \Gamma(X_{\barA})))\\
&\times p(\Gamma(X_A),\Gamma(X_{\barA})|X_{\barA})\\
&\times p(X_{\barA})p(\alpha)
\label{eq:LP-exact-joint-supp}
\end{align}
where for any set of profile labels $\C\subseteq\P$, $\Gamma(X_{\C}) = (\Gamma_b(X_{\C})_{b \in \B} = (\gamma_b(x_p))_{b \in \B}^{p \in \C}$.

For performing inference, we make use of some common approximations used in the GP-literature to circumvent the significant computational burdens coming with computing the joint distribution in Equation \eqref{eq:LP-exact-joint-supp} exactly. This is because for each basis field $b$ the joint distribution of $\Gamma_b(X_{\mathcal{P}})$ is a multivariate Gaussian with covariance matrix of dimension $P\times P$, that will need to be inverted multiple times when targeting the posterior. We therefore target a sparse approximation of Equation \eqref{eq:LP-exact-joint-supp} obtained via augmenting the prior with a set of $U = 121$ inducing points $X_{\U}$, obtained by taking the vertices of a $10\times10$ homogeneous grid of the region of interest \citep{titsias_variational_2009, bonilla_generic_2019, quinonero-candela_unifying_2005}.
Inducing values augment the prior specification and serve as \emph{global parameters} that are particularly convenient for variational approximation \citep{hensman_gaussian_2013, hensman_scalable_2015}.

Given the basis fields at the inducing points, $\Gamma(X_{\U})$, $\Gamma(X_{A})$ are treated as independent of $\Gamma(X_{\barA})$ (clearly an approximation), and thus the prior on basis fields becomes
\begin{align}
p(\Gamma(X_A),\Gamma(X_{\barA}),\Gamma(X_\U)|X_{\barA}) \approx p(\Gamma(X_A)|\Gamma(X_\U))p(\Gamma(X_{\barA})|\Gamma(X_\U),X_{\barA})p(\Gamma(X_\U)),
\end{align}
a product of multivariate Gaussians of lower dimension than $P$. 
Finally, we target a joint distribution integrated over $\Gamma(X_{\mathcal{P}})$ given $\Gamma(X_{\U})$ and $ X_{\barA}$, namely
\begin{align}
p(Y_{\mathcal{P}},\alpha, \Gamma(X_{\U}),X_{\barA})) \simeq \quad & \mathbb{E}_{p(\Gamma(X_{\mathcal{P}}|\Gamma(X_{\U}), X_{\barA})}p(Y_{\mathcal{P}},\alpha, \Gamma(X_{\mathcal{P}}), \Gamma(X_{\U}),X_{\barA})
\nonumber\\ \label{eq:LP-post-expections-Gamma-Abar}
= \quad & \mathbb{E}_{p(\Gamma(X_{A}|\Gamma(X_{\U}))}p(Y_{A}|\alpha, \Gamma(X_{A}), \Gamma(X_{\U}))
\\
\times& \quad \mathbb{E}_{p(\Gamma(X_{\barA}|\Gamma(X_{\U}), X_{\barA})}p(Y_{\barA}|\alpha, \Gamma(X_{\barA}), \Gamma(X_{\U}),X_{\barA}) \nonumber
\\
\times& \quad p(\Gamma(X_\U), X_{\barA}, \alpha) \nonumber\\
= \quad & p(Y_{A}|\Gamma(X_{\U}),\alpha)p(Y_{\barA}|\Gamma(X_{\U}),\alpha,  X_{\barA})p(\Gamma(X_\U),X_{\barA},\alpha) 
\label{eq:LP-integrated-joint}
\end{align}
The integration over $\Gamma(X_\P)$ is common practice in variational work with Gaussian processes and is done to avoid fitting a joint distribution over field values $\Gamma(X_{\barA})$ at random locations $X_{\barA}$.

In practice, we only ever need to evaluate the log-joint rather than the joint distribution and the expectations in \eqref{eq:LP-post-expections-Gamma-Abar} can be numerically unstable. We therefore make an additional approximation to compute the likelihoods for $Y_A$ and $Y_{\barA}$, following \cite{hensman2015scalable}, replacing them with the following lower bounds,
\begin{align}
    \label{eq:LP-approx-joint}
    \text{log } p(Y_{A}| \Gamma(X_\U), \alpha) &\geq E_{p(\Gamma(X_A)|\Gamma(X_\U))}[\text{log }p(Y_{A}|\alpha, \Gamma(X_{A}))]\\
    \text{log } p(Y_{\barA}| \Gamma(X_{\U}),\alpha,X_{\barA}) &\geq  E_{p(\Gamma(X_{\barA})|\Gamma(X_\U),X_{\barA})}[\text{log }p(Y_{\barA}|\alpha, \Gamma(X_{\barA}))],
\end{align}
and these are estimated by averaging over joint samples $\Gamma(X_A) \sim p(\cdot|\Gamma(X_\U))$ and  $\Gamma(X_{\barA}) \sim p(\cdot| \Gamma(X_{\U}), X_{\barA})$. It follows that the joint distribution we target is
\begin{align}
    \label{eq:LP-final-approx-post}
    \log p(Y_{\mathcal{P}},\alpha, \Gamma(X_{\U}),X_{\barA}) 
\simeq &\ E_{p(\Gamma(X_{\barA})|\Gamma(X_\U),X_{\barA})}[\text{log }p(Y_{\barA}|\alpha, \Gamma(X_{\barA}))] \nonumber\\
&\quad + E_{p(\Gamma(X_A)|\Gamma(X_\U))}[\text{log }p(Y_{A}|\alpha, \Gamma(X_{A}))]\nonumber\\ 
  & \quad +  \log(p(\Gamma(X_\U),X_{\barA},\alpha))
\end{align}

Anticipating the results, the model is misspecified. This manifests itself as a ``condensation'' behavior: when the number of items $|\I|$ and floating profiles $|\barA|$ grows we cross a threshold where the posterior for the floating profiles concentrates onto the corners and boundaries of the region and the reconstructed float-locations $X_\barA$ contradict the Atlas values. If $|\I|$ and $|\barA|$ are small, the posterior is well-behaved and the estimated float locations match Atlas values well. Model elaboration would be possible and may answer this issue. However, we can alternatively modify the inference and use SMI to carry out the analysis with all items and all floating profiles.

\begin{table}[hbt!]
\centering
\begin{tabular}{@{}lccc@{}}
\toprule
Parameter     & Dimension       & Full dataset & Small dataset \\ \midrule
$\mu$                   & $I$                     & 71                    & 5                      \\
$\zeta$                 & $I$                     & 71                    & 5                      \\
$a$                     & $\sum f_i$              & 741                   & 51                     \\
$W$                     & $B \sum f_i$            & 7,410                 & 510                    \\
$\Gamma(X_U)$           & $BU$                    & 1,210                 & 1,210                  \\ \midrule
\multicolumn{2}{l}{Subtotal (global params)} & 9,503                 & 1,781                  \\
\\
$X_{\bar{A}}$           & $2M$                    & 494                   & 20                     \\ \midrule
\multicolumn{2}{l}{Subtotal (with floating locations)} & 9,997                 & 1,801                  \\ \\
$\Gamma(X_A)$           & $BN$                    & 1,200                 & 1,200                  \\
$\Gamma(X_{\bar{A}})$   & $BM$                    & 2,470                 & 100                    \\ \midrule
\multicolumn{2}{l}{Total}                     & 13,667                & 3,101                  \\
\end{tabular}
\caption{LALME data: A summary of parameters and their dimensions.}
\label{tab:LP-parameter-dimensions}
\end{table}

\subsubsection{Comparing the VMP to MCMC on a small dataset} \label{sec:LP-MCMC}

In this section we show that the VMP gives good quality posterior approximation in this setting. MCMC on the (Bayes/$\eta=1$) posterior in \eqref{eq:LP-final-approx-post} in Section~\ref{sec:LP-background-expanded} of the Supplementary Material \cite{battaglia2025supplement} is manageable for small data sets only, so although we retain all $|A|=120$ anchor profiles we keep just $|\barA|=10$ floating profiles and 5 randomly chosen items (''vpp'', ``vps13'', ``against'', ``then'', ``which'') coming in $d_\phi=89$ forms. Retained profiles in $\barA$ are selected so that their estimated locations cover $\X$ evenly.

In this section we are demonstrating that the VMP accurately approximates Bayesian inference so we fix prior hyperparameters $\psi$ in all analyses (see Section~\ref{sec:LP-MCMC-small-sup-details} for detail), use MCMC to target \eqref{eq:LP-final-approx-post} in Section~\ref{sec:LP-background-expanded} and fit a VP to \eqref{eq:LP-final-approx-post}. We fit the VMP at all $\eta$-values (so $\tpsi=\{\eta\}$) and then evaluate the VMP at $\eta=1$. We assign both VMP and VP the same hyperparameter tuning budget for the optimiser. The VMP should not be as accurate as the VP
(the amortisation gap) so we wish to see how much we pay for amortising.

We consider the joint and marginal posterior distributions of the 10 floating profiles. Table \ref{tab:WDs} gives Wasserstein Distances between the MCMC baseline the variational approximations for these distributions. Both variational approximations show reasonable agreement to the MCMC output. 
The VP (fit at a single $\eta$) is more accurate than the VMP (a single approximaiton across all $\eta$). The amortisation gap is usually defined in terms of smaller ELBO values for amortised over non-amortised variational posteriors. However, this is the same phenomenon. It is the price we pay for amortising $\eta$.
Figure \ref{fig:MCMC_locations_compare} gives a scatter-plot of MCMC samples, overlaid with the contours of the approximating VMP-posterior (top row) and those of the VP (bottom row).
We initially set up a VMP with the additive conditioner in mentioned at the end of Section~\ref{sec:cond-auto-flow}. Performance is similar to our joint conditioner in \eqref{eq:amortised-conditioner} despite our joint conditioner having $22500$ less variational parameters.
The VMP in ($a$) has comparable performance to a VP trained at $\eta = 1$ ($b$). For this small subset of the data, our estimated $X_\barA$ are in fair agreement with Atlas-estimated locations (red crosses are within contours, except profile 699, where the VMP is at least accurately fitting the true posterior).

\begin{table}[htb]
\centering
\begin{tabular}{lcc}
\toprule
& VMP & VP  \\
\midrule
Joint WD & 0.45 & 0.30 \\
WD & 0.09(.02) & 0.07(.01)  
\\
\bottomrule
\end{tabular}
\caption{Distance between MCMC distribution and VI approximation for the 10 floating profiles. First row: Wasserstein Distance between joint distributions of locations. Second row: mean of Wasserstein Distances (WD) between marginals for individual profile loctions; parentheses give standard deviation  across profiles.}
\label{tab:WDs}
\end{table}

\begin{figure}[H]

\centering
\begin{subfigure}{\linewidth}
\centering
\includegraphics[width=\textwidth]{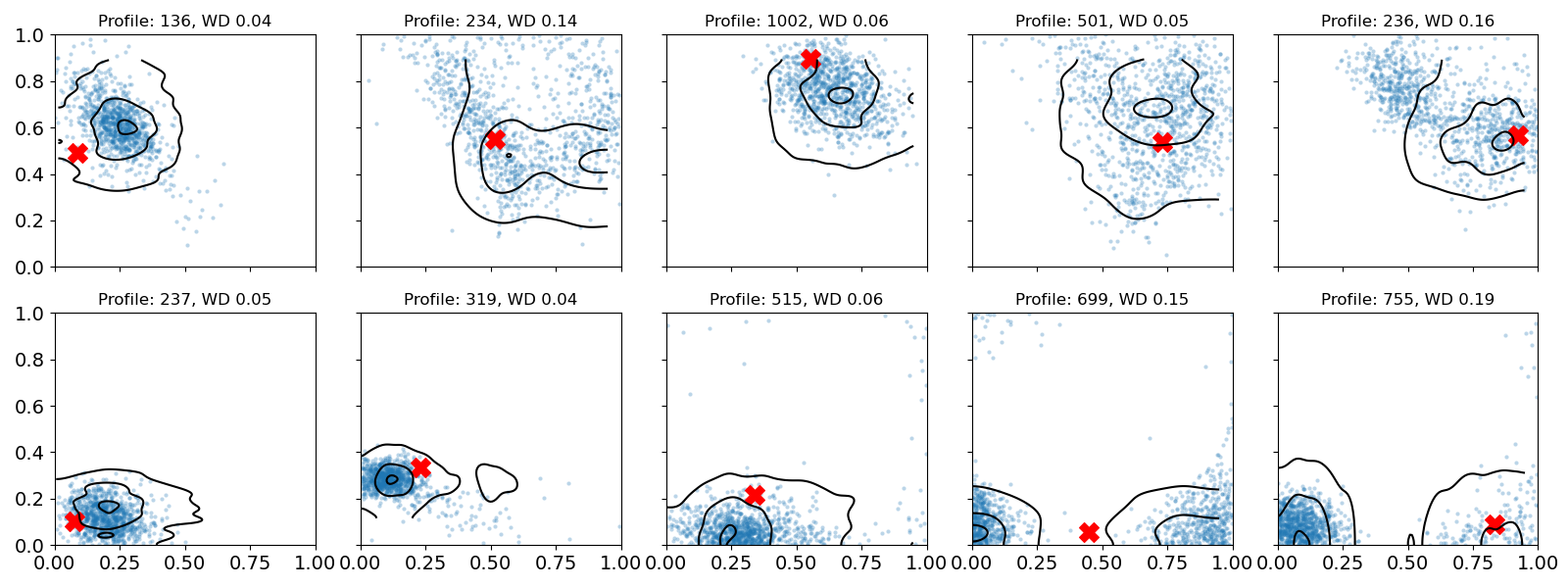}
\caption{VMP at $\eta = 1$}
\end{subfigure}

\begin{subfigure}{\linewidth}
\centering
\includegraphics[width=\textwidth]{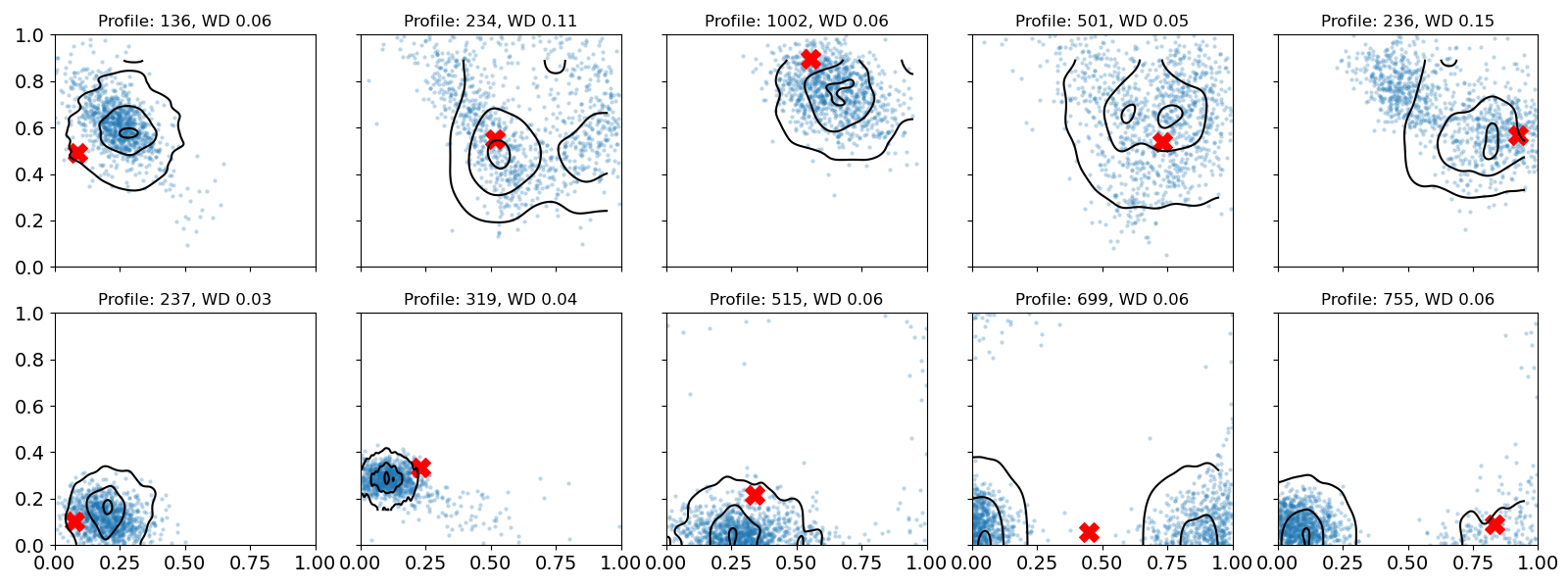}
\caption{VP for $\eta = 1$}
\end{subfigure}

\caption{LALME data in the Bayes setting, small data set of Section~\ref{sec:LP-MCMC}. Scatter of MCMC samples for posterior locations of the 10 floating profiles, overlaid with the level sets of approximation with VMP (top row) and VP (bottom row). The red crosses give the Atlas estimates for the locations of these floating profiles.}
\label{fig:MCMC_locations_compare}
\end{figure}

\subsubsection{Additional information on comparison to MCMC on the small dataset}\label{sec:LP-MCMC-small-sup-details}

The MCMC targets the standard Bayesian posterior as in Equation \ref{eq:LP-approx-joint} after conditioning on $Y_{\mathcal{P}}$. For training, we use a standard No U-Turn Sampler (NUTS) \citep{hoffman2014no} implemented in BlackJAX \citep{cabezas2024blackjax}. Convergence was checked by visual inspection of MCMC traces for a large number of parameters.

For the variational models, we fix the prior hyperparameters to $\sigma_w = 5, \sigma_a = 10, a = 1, b = 0.5, \sigma_k = 0.2, 
 \ell_k= 0.3$. When we fit the VMP to the full data, these values are in the region of hyperparameter values which give the best location prediction in terms of estimated PMSE on held out data (see Figure~\ref{fig:LP-SMI-hps-optim}). We use Hyper-Parameter Optimisation to select the decay factor and peak value for the AdaBelief learning rate schedule \citep{zhuang_adabelief_2020}. The target metric to minimise was the minimum achieved joint Wasserstein Distance to the MCMC distribution over the 10 floating profiles. Search was performed via 40 Bayesian Optimisation (BP) jobs \citep{snoek_practical_2012} implemented with the Weights and Biases software \citep{wandb}. 

Figures \ref{fig:LP-small-mu-compare} and \ref{fig:LP-small-zeta-compare} present additional plots on the performance of the the variational models with respect to MCMC on posterior inference for the $\zeta_i$ and $\mu_i$ parameters. 

\begin{figure}[hbt!]
\centering
\begin{subfigure}{\textwidth}
\centering
\includegraphics[width=0.93\linewidth]{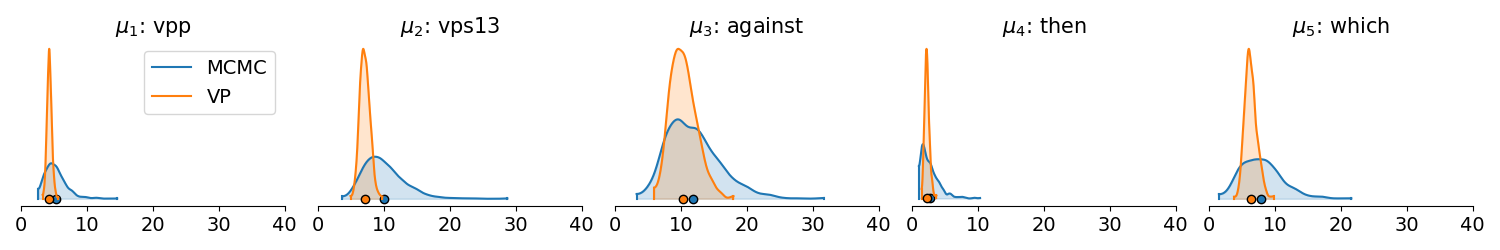}
\caption{VP}
\label{fig:LP-small-mu-compare-VP}
\end{subfigure}
\begin{subfigure}{\textwidth}
\centering
\includegraphics[width=0.93\linewidth]{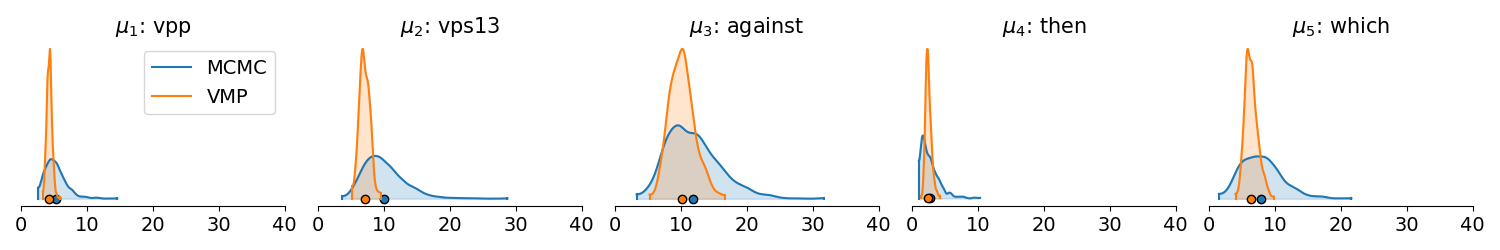}
\caption{VMP}
\label{fig:LP-small-mu-compare-VMP}
\end{subfigure}
\caption{LALME small dataset. Comparison of MCMC to VI samples for $\mu$ on the small dataset across models.}
\label{fig:LP-small-mu-compare}
\end{figure}

\begin{figure}[H]
\centering
\begin{subfigure}{\textwidth}
\centering
\includegraphics[width=0.93\linewidth]{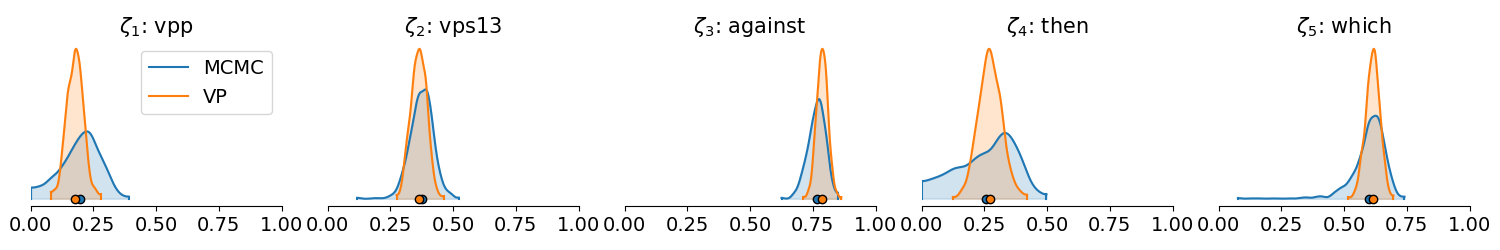}
\caption{VP}
\label{fig:LP-small-zeta-compare-VP}
\end{subfigure}
\begin{subfigure}{\textwidth}
\centering
\includegraphics[width=0.93\linewidth]{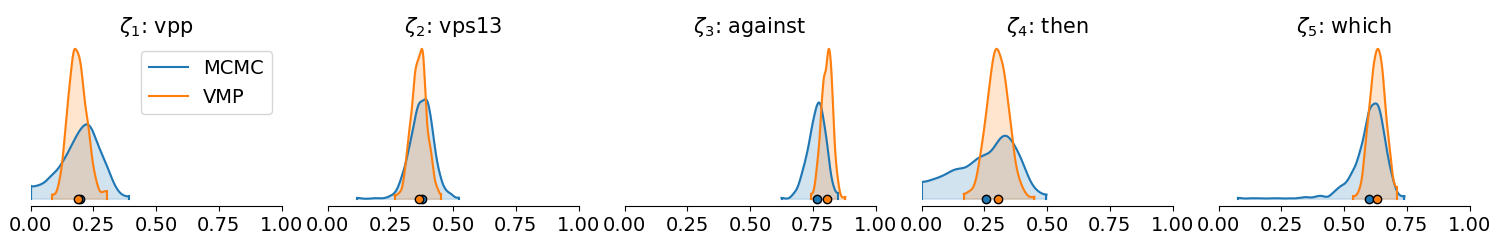}
\caption{VMP}
\label{fig:LP-small-zeta-compare-VMP}
\end{subfigure}
\caption{LALME small dataset. Comparison of MCMC to VI samples for $\zeta$ on the small dataset across VI models.}
\label{fig:LP-small-zeta-compare}
\end{figure}

\newpage
\subsubsection{Additional information on inference on the full dataset}\label{sec:LP-final-float-analysis}

Figure \ref{fig:LP-SMI-floating-compare} shows the VMP posterior approximations for ten randomly chosen floating LPs at optimal prior hyperparameters and varying $\eta$. Once again, we observe that a moderately sized $\eta$ value improves inference by concentrating the distribution around the region approximately assigned by the fitting technique, without introducing excessive dispersion.

\begin{figure}[hbt!]
\centering
\begin{subfigure}{\textwidth}
\centering
\includegraphics[width=0.82\linewidth]{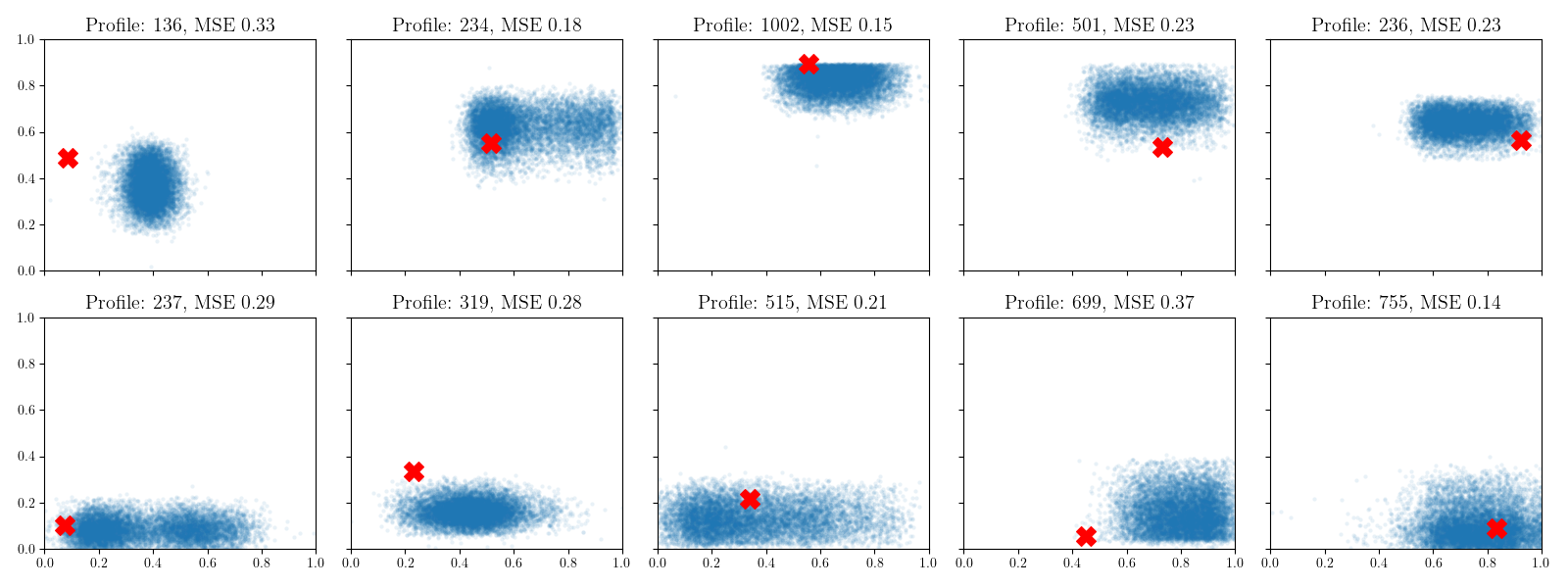}
\caption{$\eta = 0.0$}
\end{subfigure}
\begin{subfigure}{\textwidth}
\centering
\includegraphics[width=0.82\linewidth]{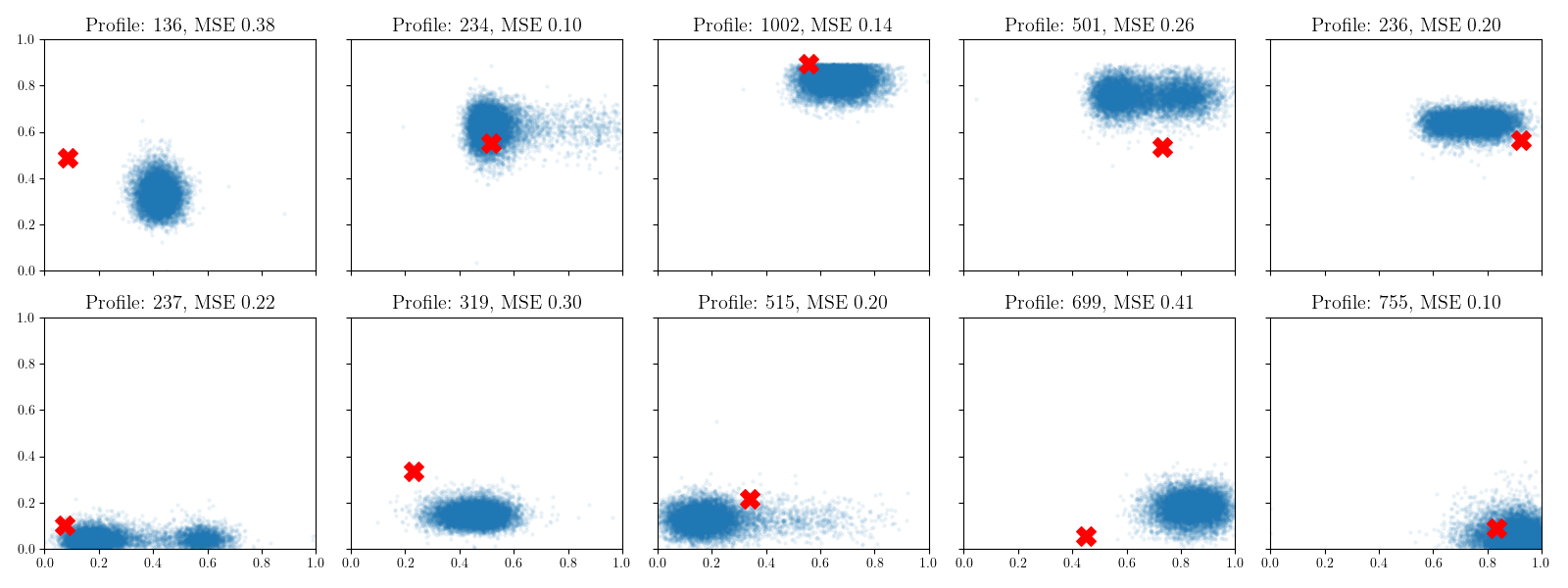}
\caption{$\eta = 0.42$}
\label{fig:LP-SMI-floating-compare-0.4}
\end{subfigure}
\begin{subfigure}{\textwidth}
\centering
\includegraphics[width=0.82\linewidth]{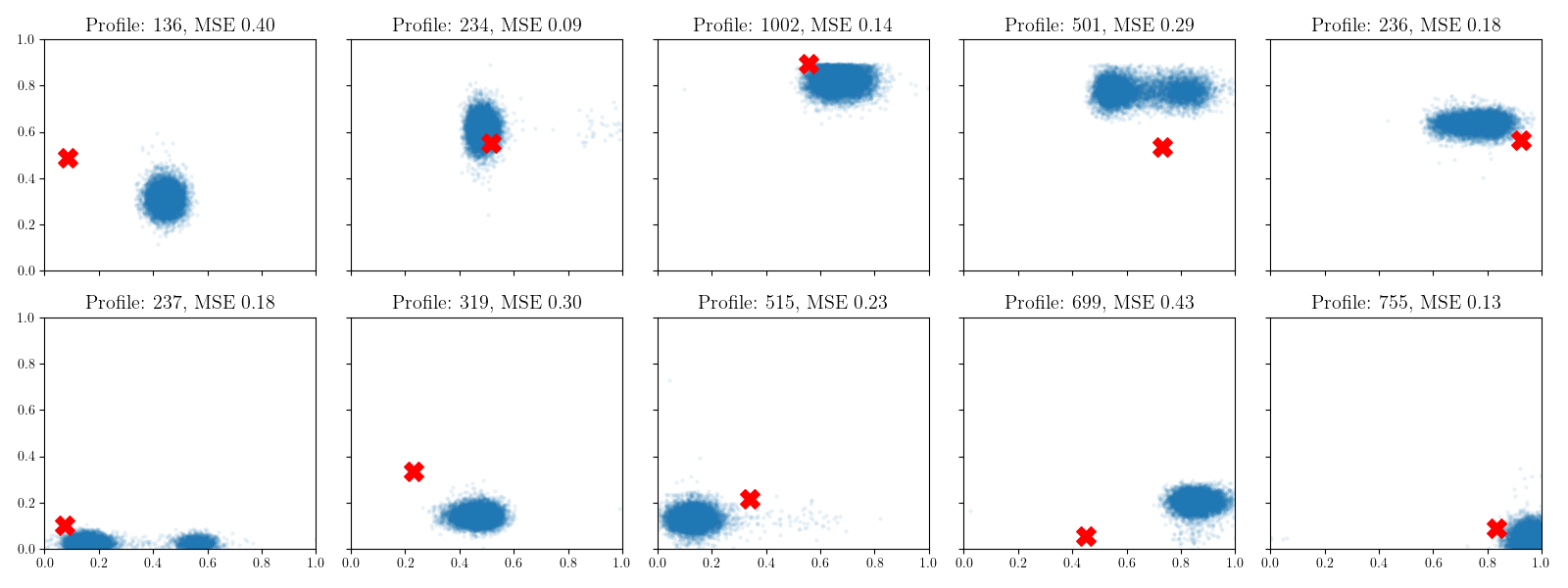}
\caption{$\eta = 1.0$}
\end{subfigure}
\caption{LALME data. VMP posterior approximations 10 floating LPs at optimal $\sigma_a = 5$, $\sigma_w = 10$, $\sigma_k = 0.4$, $\ell_k = 0.2$. Red crosses show fit-technique locations.}
\label{fig:LP-SMI-floating-compare}
\end{figure}

Figure \ref{fig:amortisation-gap} shows how the VP training loss (negative SMI-ELBO) of the fitted VP and VMP vary with $\eta$; let $L_{VP}(\eta,\psi)$ and $L_{VMP}(\eta,\psi)$ denote these losses; we might expect $L_{VP}(\eta;\psi)<L_{VMP}(\eta;\psi)$ (an amortisation gap) as the ``local'' VP fits at fixed $\eta,\psi$ while the VMP is one fit at all $\eta,\psi$. In this study prior hyperparameters $\sigma_a = 5$ and $\sigma_w = 10$ were kept fixed and we amortise the VMP over $\psi=(\sigma_k, \ell_k)$ and $\eta$.
Once the VMP is trained, we fix an $\eta$-value and minimise the PMSE over $\psi$ at that $\eta$ to get $\psi^*_\eta$. This gives a point at $(\eta, L_{VMP}(\eta,\psi^*_\eta))$ on the VMP graph in Figure \ref{fig:amortisation-gap}. The corresponding point $(\eta,L_{VP}(\eta,\psi^*_\eta))$ for the VP is obtained by fitting a VP at $\eta,\psi^*_\eta$ so the VP is fitted six times and the VMP once. We see that the amortisation gap is negligible. In fact, the VMP achieves a slightly smaller loss than the VP for some $\eta$ values, likely given a training advantage due to hyperparameter smoothing in the optimisation routine that compensates for the theoretical gap between the two models. 

\begin{figure}[ht]
    \centering
    \includegraphics[width=0.8\textwidth]{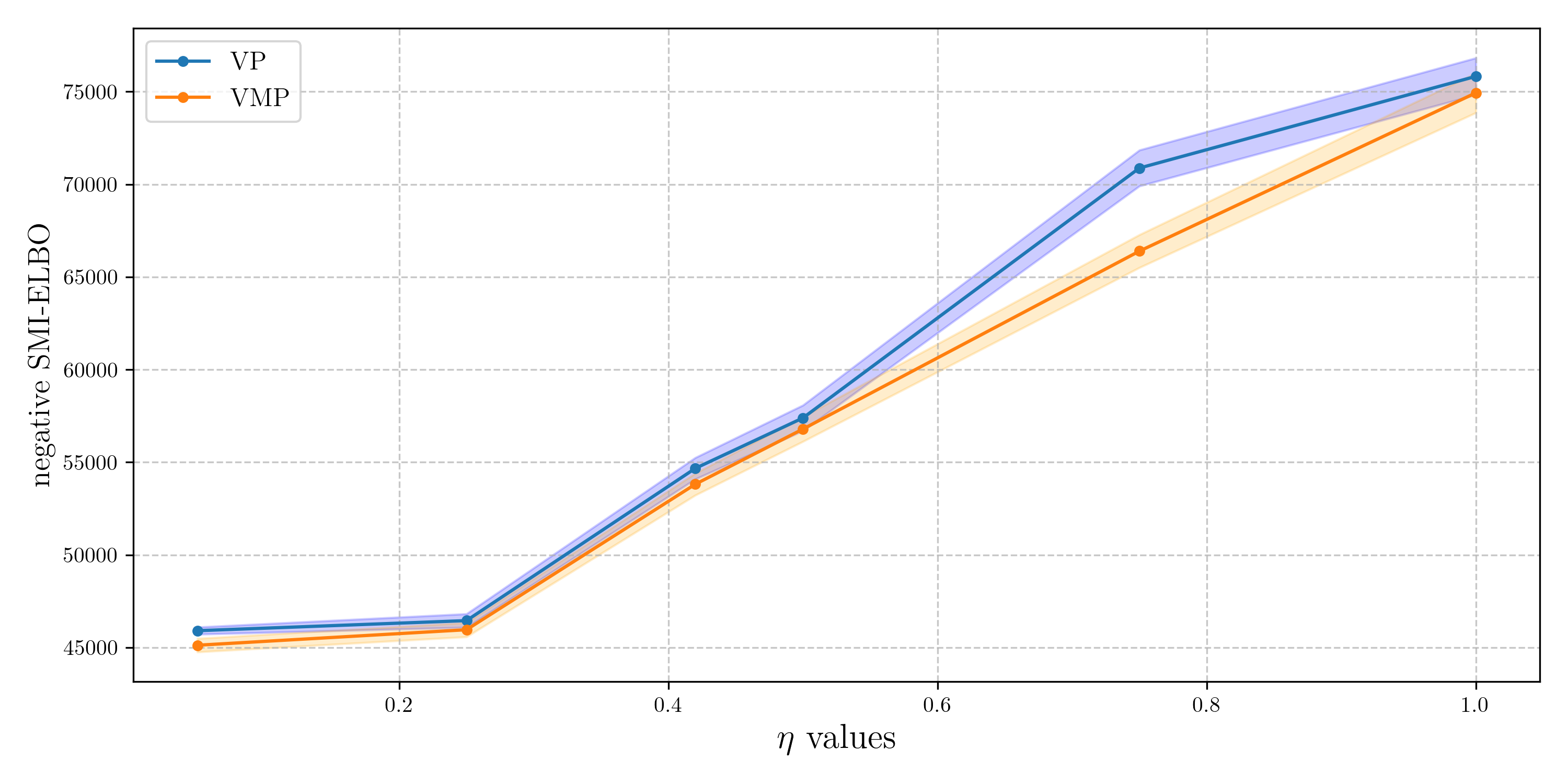}
    \caption{LALME data. Amortisation gap. Mean and 2-$\sigma$ confidence interval for VMP and VP at different $\eta$ values. The VMP line interpolates between the training loss  values obtained by evaluating the negative SMI-ELBO at samples of a single trained VMP at different $\eta$. Negative SMI-ELBO values for the VP are obtained by refitting a different VP at each $\eta$ value. At any given $\eta$, $(\sigma_k, \ell_k)$ set to the optimal values obtained by minimising the PMSE obtained from VMP samples at that $\eta$. $\sigma_a = 5$ and $\sigma_w = 10$ for any $\eta$. }
    \label{fig:amortisation-gap}
\end{figure}

\section{Architecture, hyperparameters and runtime}\label{app:hyperparameters}

The training of the VMP (and of the VP) utilises a variational family leveraging Neural Spline Flows as described in \cite{durkan_neural_2019}, stacking transforms with spline transformers and coupling conditioners. The VMP conditioners take as additional inputs $\rho$-samples for the hyperparameters of interest. We train the variational objective with an Adabelief optimiser \citep{zhuang_adabelief_2020} with a learning rate schedule. Table \ref{tab:hyperpars-flows} presents the hyperparameters used for the algorithms in the different experiments. Table \ref{tab:runtime-iters} reports the runtime and number of iterations for a single run of each algorithm, additionally comparing to MCMC where relevant. The VMP and VP have comparable runtimes per iteration. However, the VMP's key advantage is that it requires only a single run to perform a comprehensive sensitivity analysis over the targeted inference hyperparameters. In contrast, the VP demands multiple runs to accomplish the same task, significantly increasing its overall computational cost.

The experiment infrastructure relies on modifications of the ModularBayes package for SMI \citep{carmona_scalable_2022}, and builds upon Distrax \citep{deepmind2020jax}, a Python package implementing Normalising Flows in JAX \citep{jax2018github}. JAX is an open-source numerical computing library that extends NumPy functionality with automatic differentiation and GPU/TPU support, designed for high-performance machine learning research.  All experiments were run on a Google Cloud Virtual Machine equipped with a v2 TPU with 8 cores, kindly provided by Google  at no cost through the TPU Research Cloud (TRC) program.
\begin{table}[h!]
    \centering
    \footnotesize
    \caption{Hyperparameters for flow training.}
    \begin{tabular}{l c c c c c c c c} 
    \toprule
    Knots & Flow & NN & NN & LR peak  & LR decay & Experiment & Algorithm\\ 
     & Layers & Width &  Depth &value & rate & & \\ 
    \toprule
    10 & 4 & 5 & 3 & 3e-3 & 0.5 & \makecell{Synthetic \\ Synthetic \\ Epidemiology Bayes\\
    Epidemiology SMI} &\makecell{VMP \\ VP \\ VMP \\ VMP} &\\ 
    \specialrule{0.2pt}{0pt}{0pt}
    10 & 8 & 30 & 5 & 0.13 & 6.4e-3 & LP small & VMP \\ 
    \specialrule{0.2pt}{0pt}{0pt}
    10 & 8 & 30 & 5 & 0.01 & 0.19 & LP small & VP \\ 
    \specialrule{0.2pt}{0pt}{0pt} 
    10 & 6 & 30 & 5 & 3e-4 & 0.6 & \makecell{LP all\\LP all } & \makecell{VMP \\VP } &  \\ 
    \bottomrule
    \end{tabular}
    \label{tab:hyperpars-flows}
\end{table}

\begin{table}[h!]
    \centering
    \footnotesize
    \caption{Runtime and iterations for model training.}
    \begin{tabular}{l c c c c } 
    \toprule 
    Experiment & Algorithm & Runtime & Iterations  \\ [0.5ex]
    \toprule 
    Synthetic small & VMP & 10m 26s& 50,000\\ 
    Synthetic small & VP & 10m 48s& 50,000\\
    Synthetic small & MCMC & 1m 3s & 10,000 \\
    Synthetic large & VMP & 16m 18s & 50,000\\ 
    Synthetic large & VP & 16m 10s & 50,000\\
    Synthetic large & MCMC & 29s & 10,000\\
    Epidemiology Bayes  & VMP & 13m 40s & 50,000\\
    Epidemiology Bayes  & MCMC & 2m 12s & 10,000\\
    Epidemiology SMI  & VMP & 14m 3s & 50,000\\
    Epidemiology SMI  & MCMC & 5m 8s & 10,000\\
    LP small  & VMP & 2h 5m 6s & 200,000 \\ 
    LP small  & VP & 2h 14m 40s & 200,000 \\ 
    LP small  & MCMC & 12h 21m 40s & 50,000\\
    LP all  & VMP & 2h 58m 40s & 100,000 \\
    LP all  & VP & 2h 58m 49s & 100,000 \\
    \bottomrule
    \end{tabular}
    \label{tab:runtime-iters}
\end{table}

\putbib                   
\end{bibunit}
\end{supplement}

%






\begin{thebibliography}{103}
\newcommand{\enquote}[1]{``#1''}
\expandafter\ifx\csname natexlab\endcsname\relax\def\natexlab#1{#1}\fi
\expandafter\ifx\csname url\endcsname\relax
  \def\url#1{{\tt #1}}\fi
\expandafter\ifx\csname urlprefix\endcsname\relax\def\urlprefix{URL }\fi
\ifx\endbibitem\undefined \let\endbibitem\relax\fi

\bibitem[{Ambrogioni et~al.(2019)Ambrogioni, G\"{u}\c{c}l\"{u}, Berezutskaya, van~den Borne, G\"{u}\c{c}l\"{u}t\"{u}rk, Hinne, Maris, and van Gerven}]{pmlr-v89-ambrogioni19a}
Ambrogioni, L., G\"{u}\c{c}l\"{u}, U., Berezutskaya, J., van~den Borne, E., G\"{u}\c{c}l\"{u}t\"{u}rk, Y., Hinne, M., Maris, E., and van Gerven, M. (2019).
\newblock \enquote{Forward Amortized Inference for Likelihood-Free Variational Marginalization.}
\newblock In Chaudhuri, K. and Sugiyama, M. (eds.), {\em Proceedings of the Twenty-Second International Conference on Artificial Intelligence and Statistics\/}, volume~89 of {\em Proceedings of Machine Learning Research\/}, 777--786. PMLR.
\newline\urlprefix\url{https://proceedings.mlr.press/v89/ambrogioni19a.html}
\endbibitem

\bibitem[{Ardizzone et~al.(2019{\natexlab{a}})Ardizzone, Kruse, Rother, and Köthe}]{ardizzone2018analyzing}
Ardizzone, L., Kruse, J., Rother, C., and Köthe, U. (2019{\natexlab{a}}).
\newblock \enquote{Analyzing Inverse Problems with Invertible Neural Networks.}
\newblock In {\em International Conference on Learning Representations\/}.
\newline\urlprefix\url{https://openreview.net/forum?id=rJed6j0cKX}
\endbibitem

\bibitem[{Ardizzone et~al.(2019{\natexlab{b}})Ardizzone, L{\"u}th, Kruse, Rother, and K{\"o}the}]{ardizzone2019guided}
Ardizzone, L., L{\"u}th, C., Kruse, J., Rother, C., and K{\"o}the, U. (2019{\natexlab{b}}).
\newblock \enquote{Guided image generation with conditional invertible neural networks.}
\newblock {\em arXiv preprint arXiv:1907.02392\/}.
\endbibitem

\bibitem[{Battaglia et~al.(2025)Battaglia, Carmona, Haines, {Andersen Loake}, Benskin, and Nicholls}]{battaglia2025supplement}
Battaglia, L., Carmona, C.~U., Haines, R.~A., {Andersen Loake}, M., Benskin, M., and Nicholls, G.~K. (2025).
\newblock \enquote{Supplement of Amortising Variational Bayesian Inference over prior hyperparameters with a Normalising Flow.}
\newblock DOI: 10.1214/[provided by typesetter].
\endbibitem

\bibitem[{Bellagente et~al.(2022)Bellagente, Hau{\ss}mann, Luchmann, and Plehn}]{bellagente2022understanding}
Bellagente, M., Hau{\ss}mann, M., Luchmann, M., and Plehn, T. (2022).
\newblock \enquote{Understanding event-generation networks via uncertainties.}
\newblock {\em SciPost Physics\/}, 13(1): 003.
\endbibitem

\bibitem[{Bissiri et~al.(2016)Bissiri, Holmes, and Walker}]{bissiri_general_2016}
Bissiri, P.~G., Holmes, C.~C., and Walker, S.~G. (2016).
\newblock \enquote{A general framework for updating belief distributions.}
\newblock {\em Journal of the Royal Statistical Society: Series B (Statistical Methodology)\/}, 78(5): 1103--1130.
\newline\urlprefix\url{http://doi.wiley.com/10.1111/rssb.12158}
\endbibitem

\bibitem[{Bitzer et~al.(2023)Bitzer, Meister, and Zimmer}]{bitzer2023amortizedinferencegaussianprocess}
Bitzer, M., Meister, M., and Zimmer, C. (2023).
\newblock \enquote{Amortized Inference for Gaussian Process Hyperparameters of Structured Kernels.}
\newline\urlprefix\url{https://arxiv.org/abs/2306.09819}
\endbibitem

\bibitem[{Blei et~al.(2017)Blei, Kucukelbir, and McAuliffe}]{blei2017variational}
Blei, D.~M., Kucukelbir, A., and McAuliffe, J.~D. (2017).
\newblock \enquote{Variational inference: A review for statisticians.}
\newblock {\em Journal of the American statistical Association\/}, 112(518): 859--877.
\endbibitem

\bibitem[{Box(1976)}]{box76}
Box, G. E.~P. (1976).
\newblock \enquote{Science and Statistics.}
\newblock {\em Journal of the American Statistical Association\/}, 71(356): 791--799.
\newline\urlprefix\url{http://www.jstor.org/stable/2286841}
\endbibitem

\bibitem[{Carmona and Nicholls(2020)}]{carmona_semi-modular_2020}
Carmona, C.~U. and Nicholls, G.~K. (2020).
\newblock \enquote{Semi-{Modular} {Inference}: enhanced learning in multi-modular models by tempering the influence of components.}
\newblock In Silvia, C. and Calandra, R. (eds.), {\em Proceedings of the 23rd {International} {Conference} on {Artificial} {Intelligence} and {Statistics}, {AISTATS} 2020\/}, 4226--4235. PMLR.
\newblock ArXiv: 2003.06804.
\newline\urlprefix\url{http://arxiv.org/abs/2003.06804}
\endbibitem

\bibitem[{Carmona and Nicholls(2022)}]{carmona_scalable_2022}
--- (2022).
\newblock \enquote{Scalable {Semi}-{Modular} {Inference} with {Variational} {Meta}-{Posteriors}.}
\newblock ArXiv: 2204.00296.
\newline\urlprefix\url{https://github.com/chriscarmona/modularbayes}
\endbibitem

\bibitem[{Chakraborty et~al.(2022)Chakraborty, Nott, Drovandi, Frazier, and Sisson}]{chakraborty_modularized_2022}
Chakraborty, A., Nott, D.~J., Drovandi, C., Frazier, D.~T., and Sisson, S.~A. (2022).
\newblock \enquote{Modularized {Bayesian} analyses and cutting feedback in likelihood-free inference.}
\newblock ArXiv: 2203.09782.
\newline\urlprefix\url{http://arxiv.org/abs/2203.09782}
\endbibitem

\bibitem[{Chakraborty et~al.(2023)Chakraborty, Nott, Drovandi, Frazier, and Sisson}]{chakraborty2023modularized}
Chakraborty, A., Nott, D.~J., Drovandi, C.~C., Frazier, D.~T., and Sisson, S.~A. (2023).
\newblock \enquote{Modularized Bayesian analyses and cutting feedback in likelihood-free inference.}
\newblock {\em Statistics and Computing\/}, 33(1): 33.
\endbibitem

\bibitem[{Chang et~al.(2024)Chang, Loka, Huang, Remes, Kaski, and Acerbi}]{chang2024amortized}
Chang, P.~E., Loka, N., Huang, D., Remes, U., Kaski, S., and Acerbi, L. (2024).
\newblock \enquote{Amortized probabilistic conditioning for optimization, simulation and inference.}
\newblock {\em arXiv preprint arXiv:2410.15320\/}.
\endbibitem

\bibitem[{Ch{\'e}rief-Abdellatif and Alquier(2020)}]{cherief2020mmd}
Ch{\'e}rief-Abdellatif, B.-E. and Alquier, P. (2020).
\newblock \enquote{MMD-Bayes: Robust Bayesian estimation via maximum mean discrepancy.}
\newblock In {\em Symposium on Advances in Approximate Bayesian Inference\/}, 1--21. PMLR.
\endbibitem

\bibitem[{Cooper et~al.(2024)Cooper, Simpson, Kennedy, Forbes, and Vehtari}]{cooper24}
Cooper, A., Simpson, D., Kennedy, L., Forbes, C., and Vehtari, A. (2024).
\newblock \enquote{{Cross-Validatory Model Selection for Bayesian Autoregressions with Exogenous Regressors}.}
\newblock {\em Bayesian Analysis\/}, 1 -- 25.
\newline\urlprefix\url{https://doi.org/10.1214/23-BA1409}
\endbibitem

\bibitem[{Dayan(2000)}]{dayan2000helmholtz}
Dayan, P. (2000).
\newblock \enquote{Helmholtz machines and wake-sleep learning.}
\newblock {\em Handbook of Brain Theory and Neural Network. MIT Press, Cambridge, MA\/}, 44(0): 1--12.
\endbibitem

\bibitem[{Dellaporta et~al.(2022)Dellaporta, Knoblauch, Damoulas, and Briol}]{dellaporta2022robust}
Dellaporta, C., Knoblauch, J., Damoulas, T., and Briol, F.-X. (2022).
\newblock \enquote{Robust Bayesian inference for simulator-based models via the MMD posterior bootstrap.}
\newblock In {\em International Conference on Artificial Intelligence and Statistics\/}, 943--970. PMLR.
\endbibitem

\bibitem[{Depaoli et~al.(2020)Depaoli, Winter, and Visser}]{depaoli2020importance}
Depaoli, S., Winter, S.~D., and Visser, M. (2020).
\newblock \enquote{The importance of prior sensitivity analysis in Bayesian statistics: demonstrations using an interactive Shiny App.}
\newblock {\em Frontiers in psychology\/}, 11: 608045.
\endbibitem

\bibitem[{Diggle et~al.(2013)Diggle, Moraga, Rowlingson, and Taylor}]{diggle_spatial_2013}
Diggle, P.~J., Moraga, P., Rowlingson, B., and Taylor, B.~M. (2013).
\newblock \enquote{Spatial and {Spatio}-{Temporal} {Log}-{Gaussian} {Cox} {Processes}: {Extending} the {Geostatistical} {Paradigm}.}
\newblock {\em Statistical Science\/}, 28(4).
\newblock ArXiv:1312.6536 [stat].
\newline\urlprefix\url{http://arxiv.org/abs/1312.6536}
\endbibitem

\bibitem[{Diggle and Ribeiro(2007)}]{diggle_model-based_2007}
Diggle, P.~J. and Ribeiro, P.~J. (2007).
\newblock {\em Model-based {Geostatistics}\/}.
\newblock Springer {Series} in {Statistics}. New York, NY: Springer.
\newline\urlprefix\url{http://link.springer.com/10.1007/978-0-387-48536-2}
\endbibitem

\bibitem[{Diggle et~al.(1998)Diggle, Tawn, and Moyeed}]{diggle_model-based_1998}
Diggle, P.~J., Tawn, J.~A., and Moyeed, R.~A. (1998).
\newblock \enquote{Model-based geostatistics.}
\newblock {\em Journal of the Royal Statistical Society: Series C (Applied Statistics)\/}, 47(3): 299--350.
\newline\urlprefix\url{https://onlinelibrary.wiley.com/doi/abs/10.1111/1467-9876.00113}
\endbibitem

\bibitem[{Dinh et~al.(2016)Dinh, Sohl-Dickstein, and Bengio}]{dinh_density_2016}
Dinh, L., Sohl-Dickstein, J., and Bengio, S. (2016).
\newblock \enquote{Density estimation using {Real} {NVP}.}
\newblock In {\em Proceedings of the 5th {International} {Conference} on {Learning} {Representations}, {ICLR} 2017\/}.
\newblock ArXiv: 1605.08803.
\newline\urlprefix\url{http://arxiv.org/abs/1605.08803}
\endbibitem

\bibitem[{Draxler et~al.(2022)Draxler, Schn{\"o}rr, and K{\"o}the}]{draxler2022whitening}
Draxler, F., Schn{\"o}rr, C., and K{\"o}the, U. (2022).
\newblock \enquote{Whitening convergence rate of coupling-based normalizing flows.}
\newblock {\em Advances in Neural Information Processing Systems\/}, 35: 37241--37253.
\endbibitem

\bibitem[{Draxler et~al.(2024{\natexlab{a}})Draxler, Sorrenson, Zimmermann, Rousselot, and K{\"o}the}]{draxler2024free}
Draxler, F., Sorrenson, P., Zimmermann, L., Rousselot, A., and K{\"o}the, U. (2024{\natexlab{a}}).
\newblock \enquote{Free-form flows: Make any architecture a normalizing flow.}
\newblock In {\em International Conference on Artificial Intelligence and Statistics\/}, 2197--2205. PMLR.
\endbibitem

\bibitem[{Draxler et~al.(2024{\natexlab{b}})Draxler, Wahl, Schn{\"o}rr, and K{\"o}the}]{draxler2024universality}
Draxler, F., Wahl, S., Schn{\"o}rr, C., and K{\"o}the, U. (2024{\natexlab{b}}).
\newblock \enquote{On the universality of coupling-based normalizing flows.}
\newblock {\em arXiv preprint arXiv:2402.06578\/}.
\endbibitem

\bibitem[{Durkan et~al.(2019)Durkan, Bekasov, Murray, and Papamakarios}]{durkan_neural_2019}
Durkan, C., Bekasov, A., Murray, I., and Papamakarios, G. (2019).
\newblock \enquote{Neural {Spline} {Flows}.}
\newblock In {\em Proceedings of the 33rd {Conference} on {Neural} {Information} {Processing} {Systems}, {NeurIPS} 2019\/}.
\newblock ArXiv: 1906.04032.
\newline\urlprefix\url{http://arxiv.org/abs/1906.04032}
\endbibitem

\bibitem[{Elsem{\"u}ller et~al.(2024)Elsem{\"u}ller, Olischl{\"a}ger, Schmitt, B{\"u}rkner, Koethe, and Radev}]{elsemuller2024sensitivityaware}
Elsem{\"u}ller, L., Olischl{\"a}ger, H., Schmitt, M., B{\"u}rkner, P.-C., Koethe, U., and Radev, S.~T. (2024).
\newblock \enquote{Sensitivity-Aware Amortized Bayesian Inference.}
\newblock {\em Transactions on Machine Learning Research\/}.
\newline\urlprefix\url{https://openreview.net/forum?id=Kxtpa9rvM0}
\endbibitem

\bibitem[{Fortuin(2022)}]{fortuin2022priors}
Fortuin, V. (2022).
\newblock \enquote{Priors in bayesian deep learning: A review.}
\newblock {\em International Statistical Review\/}, 90(3): 563--591.
\endbibitem

\bibitem[{Frazier et~al.(2024)Frazier, Kelly, Drovandi, and Warne}]{frazier2024statistical}
Frazier, D.~T., Kelly, R., Drovandi, C., and Warne, D.~J. (2024).
\newblock \enquote{The Statistical Accuracy of Neural Posterior and Likelihood Estimation.}
\newblock {\em arXiv preprint arXiv:2411.12068\/}.
\endbibitem

\bibitem[{Frazier and Nott(2023)}]{frazier23posterior}
Frazier, D.~T. and Nott, D.~J. (2023).
\newblock \enquote{Posterior risk of modular and semi-modular Bayesian inference.}
\newline\urlprefix\url{https://arxiv.org/abs/2301.10911}
\endbibitem

\bibitem[{Frazier and Nott(2024)}]{frazier24cutting}
--- (2024).
\newblock \enquote{Cutting feedback and modularized analyses in generalized Bayesian inference.}
\newblock {\em Bayesian Analysis\/}, 1(1): 1--29.
\endbibitem

\bibitem[{Ganguly et~al.(2023)Ganguly, Jain, and Watchareeruetai}]{ganguly2023amortized}
Ganguly, A., Jain, S., and Watchareeruetai, U. (2023).
\newblock \enquote{Amortized variational inference: A systematic review.}
\newblock {\em Journal of Artificial Intelligence Research\/}, 78: 167--215.
\endbibitem

\bibitem[{Gao et~al.(2023)Gao, Deistler, and Macke}]{gao2023generalized}
Gao, R., Deistler, M., and Macke, J.~H. (2023).
\newblock \enquote{Generalized Bayesian Inference for Scientific Simulators via Amortized Cost Estimation.}
\newblock In {\em Thirty-seventh Conference on Neural Information Processing Systems\/}.
\newline\urlprefix\url{https://openreview.net/forum?id=ZARAiV25CW}
\endbibitem

\bibitem[{Gelman and Yao(2020)}]{gelman2020holes}
Gelman, A. and Yao, Y. (2020).
\newblock \enquote{Holes in Bayesian statistics.}
\newblock {\em Journal of Physics G: Nuclear and Particle Physics\/}, 48(1): 014002.
\endbibitem

\bibitem[{Gershman and Goodman(2014)}]{gershman2014amortized}
Gershman, S. and Goodman, N. (2014).
\newblock \enquote{Amortized inference in probabilistic reasoning.}
\newblock In {\em Proceedings of the Annual Meeting of the Cognitive Science Society\/}, volume~36.
\newline\urlprefix\url{https://escholarship.org/uc/item/34j1h7k5}
\endbibitem

\bibitem[{Giordano et~al.(2018)Giordano, Broderick, and Jordan}]{giordano_covariances_2018}
Giordano, R., Broderick, T., and Jordan, M.~I. (2018).
\newblock \enquote{Covariances, robustness, and variational {Bayes}.}
\newblock {\em Journal of Machine Learning Research\/}, 19: 1--49.
\newblock ArXiv: 1709.02536.
\newline\urlprefix\url{http://jmlr.org/papers/v19/17-670.html.}
\endbibitem

\bibitem[{Giordano et~al.(2022)Giordano, Liu, Jordan, and Broderick}]{giordano_evaluating_2022}
Giordano, R., Liu, R., Jordan, M.~I., and Broderick, T. (2022).
\newblock \enquote{Evaluating {Sensitivity} to the {Stick}-{Breaking} {Prior} in {Bayesian} {Nonparametrics}.}
\newblock {\em Bayesian Analysis\/}, -1(-1): 1--67.
\newblock Publisher: International Society for Bayesian Analysis.
\newline\urlprefix\url{https://projecteuclid.org/journals/bayesian-analysis/advance-publication/Evaluating-Sensitivity-to-the-Stick-Breaking-Prior-in-Bayesian-Nonparametrics/10.1214/22-BA1309.full}
\endbibitem

\bibitem[{Giordano et~al.(2019)Giordano, Stephenson, Liu, Jordan, and Broderick}]{giordano2019swiss}
Giordano, R., Stephenson, W., Liu, R., Jordan, M., and Broderick, T. (2019).
\newblock \enquote{A swiss army infinitesimal jackknife.}
\newblock In {\em The 22nd International Conference on Artificial Intelligence and Statistics\/}, 1139--1147. PMLR.
\endbibitem

\bibitem[{Giorgi and Diggle(2015)}]{giorgi_inverse_2015}
Giorgi, E. and Diggle, P.~J. (2015).
\newblock \enquote{On the inverse geostatistical problem of inference on missing locations.}
\newblock {\em Spatial Statistics\/}, 11: 35--44.
\newline\urlprefix\url{https://www.sciencedirect.com/science/article/pii/S2211675314000608}
\endbibitem

\bibitem[{Gl{\"o}ckler et~al.(2022)Gl{\"o}ckler, Deistler, and Macke}]{glockler2022variational}
Gl{\"o}ckler, M., Deistler, M., and Macke, J.~H. (2022).
\newblock \enquote{Variational methods for simulation-based inference.}
\newblock {\em arXiv preprint arXiv:2203.04176\/}.
\endbibitem

\bibitem[{Golinski et~al.(2019)Golinski, Wood, and Rainforth}]{golinski19}
Golinski, A., Wood, F., and Rainforth, T. (2019).
\newblock \enquote{Amortized {M}onte {C}arlo Integration.}
\newblock In Chaudhuri, K. and Salakhutdinov, R. (eds.), {\em Proceedings of the 36th International Conference on Machine Learning\/}, volume~97 of {\em Proceedings of Machine Learning Research\/}, 2309--2318. PMLR.
\newline\urlprefix\url{https://proceedings.mlr.press/v97/golinski19a.html}
\endbibitem

\bibitem[{Grünwald and Ommen(2017)}]{grunwald_inconsistency_2017}
Grünwald, P. and Ommen, T.~v. (2017).
\newblock \enquote{Inconsistency of {Bayesian} {Inference} for {Misspecified} {Linear} {Models}, and a {Proposal} for {Repairing} {It}.}
\newblock {\em Bayesian Analysis\/}, 12(4): 1069--1103.
\newblock Publisher: International Society for Bayesian Analysis.
\newline\urlprefix\url{https://projecteuclid.org/journals/bayesian-analysis/volume-12/issue-4/Inconsistency-of-Bayesian-Inference-for-Misspecified-Linear-Models-and-a/10.1214/17-BA1085.full}
\endbibitem

\bibitem[{Haines(2016)}]{haines_simultan_2016}
Haines, R.~A. (2016).
\newblock \enquote{Simultaneous Reconstruction of Spatial Frequency Fields and Field Sample Locations.}
\newblock PhD thesis, University of Oxford.
\newline\urlprefix\url{http://arxiv.org/abs/1708.08719}
\endbibitem

\bibitem[{Ho et~al.(2020)Ho, Jain, and Abbeel}]{ho2020denoising}
Ho, J., Jain, A., and Abbeel, P. (2020).
\newblock \enquote{Denoising diffusion probabilistic models.}
\newblock {\em Advances in neural information processing systems\/}, 33: 6840--6851.
\endbibitem

\bibitem[{Holmes and Walker(2017)}]{holmes2017assigning}
Holmes, C.~C. and Walker, S.~G. (2017).
\newblock \enquote{Assigning a value to a power likelihood in a general Bayesian model.}
\newblock {\em Biometrika\/}, 104(2): 497--503.
\endbibitem

\bibitem[{Huang et~al.(2018)Huang, Krueger, Lacoste, and Courville}]{huang18}
Huang, C.-W., Krueger, D., Lacoste, A., and Courville, A. (2018).
\newblock \enquote{Neural Autoregressive Flows.}
\newblock In Dy, J. and Krause, A. (eds.), {\em Proceedings of the 35th International Conference on Machine Learning\/}, volume~80 of {\em Proceedings of Machine Learning Research\/}, 2078--2087. PMLR.
\newline\urlprefix\url{https://proceedings.mlr.press/v80/huang18d.html}
\endbibitem

\bibitem[{Ishikawa et~al.(2023)Ishikawa, Teshima, Tojo, Oono, Ikeda, and Sugiyama}]{ishikawa23}
Ishikawa, I., Teshima, T., Tojo, K., Oono, K., Ikeda, M., and Sugiyama, M. (2023).
\newblock \enquote{Universal Approximation Property of Invertible Neural Networks.}
\newblock {\em Journal of Machine Learning Research\/}, 24(287): 1--68.
\newline\urlprefix\url{http://jmlr.org/papers/v24/22-0384.html}
\endbibitem

\bibitem[{Jacob et~al.(2017)Jacob, Murray, Holmes, and Robert}]{jacob_better_2017}
Jacob, P.~E., Murray, L.~M., Holmes, C.~C., and Robert, C.~P. (2017).
\newblock \enquote{Better together? {Statistical} learning in models made of modules.}
\newblock ArXiv: 1708.08719 Pages: 1-31.
\newline\urlprefix\url{http://arxiv.org/abs/1708.08719}
\endbibitem

\bibitem[{Jaini et~al.(2019)Jaini, Selby, and Yu}]{jaini2019sum}
Jaini, P., Selby, K.~A., and Yu, Y. (2019).
\newblock \enquote{Sum-of-squares polynomial flow.}
\newblock In {\em International Conference on Machine Learning\/}, 3009--3018. PMLR.
\endbibitem

\bibitem[{Jerfel et~al.(2021)Jerfel, Wang, Wong-Fannjiang, Heller, Ma, and Jordan}]{jerfel21_FKL_IS}
Jerfel, G., Wang, S., Wong-Fannjiang, C., Heller, K.~A., Ma, Y., and Jordan, M.~I. (2021).
\newblock \enquote{Variational refinement for importance sampling using the forward Kullback-Leibler divergence.}
\newblock In de~Campos, C. and Maathuis, M.~H. (eds.), {\em Proceedings of the Thirty-Seventh Conference on Uncertainty in Artificial Intelligence\/}, volume 161 of {\em Proceedings of Machine Learning Research\/}, 1819--1829. PMLR.
\newline\urlprefix\url{https://proceedings.mlr.press/v161/jerfel21a.html}
\endbibitem

\bibitem[{Jewson and Rossell(2022)}]{jewson2022general}
Jewson, J. and Rossell, D. (2022).
\newblock \enquote{General Bayesian loss function selection and the use of improper models.}
\newblock {\em Journal of the Royal Statistical Society Series B: Statistical Methodology\/}, 84(5): 1640--1665.
\endbibitem

\bibitem[{Jordan et~al.(1999)Jordan, Ghahramani, Jaakkola, and Saul}]{jordan1999introduction}
Jordan, M.~I., Ghahramani, Z., Jaakkola, T.~S., and Saul, L.~K. (1999).
\newblock \enquote{An introduction to variational methods for graphical models.}
\newblock {\em Machine learning\/}, 37: 183--233.
\endbibitem

\bibitem[{Kelly et~al.(2025)Kelly, Warne, Frazier, Nott, Gutmann, and Drovandi}]{kelly2025simulation}
Kelly, R.~P., Warne, D.~J., Frazier, D.~T., Nott, D.~J., Gutmann, M.~U., and Drovandi, C. (2025).
\newblock \enquote{Simulation-based Bayesian inference under model misspecification.}
\newblock {\em arXiv preprint arXiv:2503.12315\/}.
\endbibitem

\bibitem[{Kingma and Dhariwal(2018)}]{kingma2018glow}
Kingma, D.~P. and Dhariwal, P. (2018).
\newblock \enquote{Glow: Generative flow with invertible 1x1 convolutions.}
\newblock {\em Advances in neural information processing systems\/}, 31.
\endbibitem

\bibitem[{Kingma et~al.(2016)Kingma, Salimans, Jozefowicz, Chen, Sutskever, and Welling}]{kingma2016improved}
Kingma, D.~P., Salimans, T., Jozefowicz, R., Chen, X., Sutskever, I., and Welling, M. (2016).
\newblock \enquote{Improved variational inference with inverse autoregressive flow.}
\newblock {\em Advances in neural information processing systems\/}, 29.
\endbibitem

\bibitem[{Kingma and Welling(2014)}]{kingma2014auto}
Kingma, D.~P. and Welling, M. (2014).
\newblock \enquote{Auto-encoding variational Bayes.}
\newblock {\em Internation Conference on Learning Representations\/}.
\endbibitem

\bibitem[{Kruse et~al.(2021)Kruse, Detommaso, Scheichl, and K{\"o}the}]{kruse21}
Kruse, J., Detommaso, G., Scheichl, R., and K{\"o}the, U. (2021).
\newblock \enquote{HINT: Hierarchical Invertible Neural Transport for Density Estimation and Bayesian Inference.}
\newblock In {\em Proceedings of the 35th AAAI Conference on Artificial Intelligence (AAAI-21)\/}, volume~35 of {\em Proceedings of the 35th AAAI Conference on Artificial Intelligence (AAAI-21)\/}, 8191--8199. Association for the Advancement of Artificial Intelligence (AAAI).
\endbibitem

\bibitem[{Lee et~al.(2025)Lee, Liu, and Nicholls}]{lee2025bayesianinferencelearningrate}
Lee, J.~E., Liu, S., and Nicholls, G.~K. (2025).
\newblock \enquote{Bayesian inference for the learning rate in Generalised Bayesian inference.}
\newline\urlprefix\url{https://arxiv.org/abs/2506.12532}
\endbibitem

\bibitem[{Lipman et~al.(2023)Lipman, Chen, Ben-Hamu, Nickel, and Le}]{lipman2023flow}
Lipman, Y., Chen, R. T.~Q., Ben-Hamu, H., Nickel, M., and Le, M. (2023).
\newblock \enquote{Flow Matching for Generative Modeling.}
\newblock In {\em The Eleventh International Conference on Learning Representations\/}.
\newline\urlprefix\url{https://openreview.net/forum?id=PqvMRDCJT9t}
\endbibitem

\bibitem[{Liu et~al.(2020)Liu, Sun, Ramadge, and Adams}]{liu2020task}
Liu, S., Sun, X., Ramadge, P.~J., and Adams, R.~P. (2020).
\newblock \enquote{Task-agnostic amortized inference of gaussian process hyperparameters.}
\newblock {\em Advances in Neural Information Processing Systems\/}, 33: 21440--21452.
\endbibitem

\bibitem[{Liu and Goudie(2022)}]{liu_general_2022}
Liu, Y. and Goudie, R. J.~B. (2022).
\newblock \enquote{A {General} {Framework} for {Cutting} {Feedback} within {Modularized} {Bayesian} {Inference}.}
\newblock ArXiv:2211.03274 [math, stat].
\newline\urlprefix\url{http://arxiv.org/abs/2211.03274}
\endbibitem

\bibitem[{Lunn et~al.(2009)Lunn, Best, Spiegelhalter, Graham, and Neuenschwander}]{lunn2009combining}
Lunn, D., Best, N., Spiegelhalter, D., Graham, G., and Neuenschwander, B. (2009).
\newblock \enquote{Combining MCMC with ‘sequential’PKPD modelling.}
\newblock {\em Journal of Pharmacokinetics and Pharmacodynamics\/}, 36: 19--38.
\endbibitem

\bibitem[{Lyddon et~al.(2019)Lyddon, Holmes, and Walker}]{lyddon2019general}
Lyddon, S.~P., Holmes, C., and Walker, S. (2019).
\newblock \enquote{General Bayesian updating and the loss-likelihood bootstrap.}
\newblock {\em Biometrika\/}, 106(2): 465--478.
\endbibitem

\bibitem[{Margossian and Blei(2023)}]{margossian2023amortized}
Margossian, C.~C. and Blei, D.~M. (2023).
\newblock \enquote{Amortized Variational Inference: When and Why?}
\endbibitem

\bibitem[{Matsubara et~al.(2022)Matsubara, Knoblauch, Briol, and Oates}]{matsubara2022robust}
Matsubara, T., Knoblauch, J., Briol, F.-X., and Oates, C.~J. (2022).
\newblock \enquote{Robust generalised Bayesian inference for intractable likelihoods.}
\newblock {\em Journal of the Royal Statistical Society Series B: Statistical Methodology\/}, 84(3): 997--1022.
\endbibitem

\bibitem[{Maucort-Boulch et~al.(2008)Maucort-Boulch, Franceschi, Plummer, and Group}]{maucort2008international}
Maucort-Boulch, D., Franceschi, S., Plummer, M., and Group, I. H. P. S.~S. (2008).
\newblock \enquote{International correlation between human papillomavirus prevalence and cervical cancer incidence.}
\newblock {\em Cancer Epidemiology Biomarkers \& Prevention\/}, 17(3): 717--720.
\endbibitem

\bibitem[{McIntosh et~al.(1986)McIntosh, Samuels, Benskin, Laing, and Williamson}]{mcintosh1986linguistic}
McIntosh, A., Samuels, M.~L., Benskin, M., Laing, M., and Williamson, K. (1986).
\newblock {\em A linguistic atlas of late mediaeval English\/}.
\newblock Aberdeen University Press.
\endbibitem

\bibitem[{McLatchie et~al.(2024)McLatchie, Fong, Frazier, and Knoblauch}]{mclatchie2024predictive}
McLatchie, Y., Fong, E., Frazier, D.~T., and Knoblauch, J. (2024).
\newblock \enquote{Predictive performance of power posteriors.}
\newblock {\em arXiv preprint arXiv:2408.08806\/}.
\endbibitem

\bibitem[{Meng(1994)}]{meng94}
Meng, X.-L. (1994).
\newblock \enquote{{Multiple-Imputation Inferences with Uncongenial Sources of Input}.}
\newblock {\em Statistical Science\/}, 9(4): 538 -- 558.
\newline\urlprefix\url{https://doi.org/10.1214/ss/1177010269}
\endbibitem

\bibitem[{Mikkola et~al.(2024)Mikkola, Acerbi, and Klami}]{mikkola2024preferential}
Mikkola, P., Acerbi, L., and Klami, A. (2024).
\newblock \enquote{Preferential Normalizing Flows.}
\newblock {\em arXiv preprint arXiv:2410.08710\/}.
\endbibitem

\bibitem[{Mittal et~al.(2025)Mittal, Bracher, Lajoie, Jaini, and Brubaker}]{mittal2025forwardVreverseKL}
Mittal, S., Bracher, N.~L., Lajoie, G., Jaini, P., and Brubaker, M. (2025).
\newblock \enquote{Amortized In-Context Bayesian Posterior Estimation.}
\newline\urlprefix\url{https://arxiv.org/abs/2502.06601}
\endbibitem

\bibitem[{Nicholls et~al.(2022)Nicholls, Lee, Wu, and Carmona}]{nicholls_valid_2022}
Nicholls, G.~K., Lee, J.~E., Wu, C.-H., and Carmona, C.~U. (2022).
\newblock \enquote{Valid belief updates for prequentially additive loss functions arising in {Semi}-{Modular} {Inference}.}
\newblock ArXiv:2201.09706 [stat].
\newline\urlprefix\url{http://arxiv.org/abs/2201.09706}
\endbibitem

\bibitem[{Nicholson et~al.(2022)Nicholson, Blangiardo, Briers, Diggle, Fjelde, Ge, Goudie, Jersakova, King, Lehmann, Mallon, Padellini, Teh, Holmes, and Richardson}]{Nicholson22}
Nicholson, G., Blangiardo, M., Briers, M., Diggle, P.~J., Fjelde, T.~E., Ge, H., Goudie, R. J.~B., Jersakova, R., King, R.~E., Lehmann, B. C.~L., Mallon, A.-M., Padellini, T., Teh, Y.~W., Holmes, C., and Richardson, S. (2022).
\newblock \enquote{{Interoperability of Statistical Models in Pandemic Preparedness: Principles and Reality}.}
\newblock {\em Statistical Science\/}, 37(2): 183 -- 206.
\newline\urlprefix\url{https://doi.org/10.1214/22-STS854}
\endbibitem

\bibitem[{Pacchiardi et~al.(2024)Pacchiardi, Khoo, and Dutta}]{pacchiardi2024generalized}
Pacchiardi, L., Khoo, S., and Dutta, R. (2024).
\newblock \enquote{Generalized Bayesian likelihood-free inference.}
\newblock {\em Electronic Journal of Statistics\/}, 18(2): 3628--3686.
\endbibitem

\bibitem[{Papamakarios et~al.(2021)Papamakarios, Nalisnick, Rezende, Mohamed, and Lakshminarayanan}]{papamakarios21}
Papamakarios, G., Nalisnick, E., Rezende, D.~J., Mohamed, S., and Lakshminarayanan, B. (2021).
\newblock \enquote{Normalizing Flows for Probabilistic Modeling and Inference.}
\newblock {\em J. Mach. Learn. Res.\/}, 22(1).
\endbibitem

\bibitem[{Papamakarios et~al.(2017)Papamakarios, Pavlakou, and Murray}]{papamakarios2017masked}
Papamakarios, G., Pavlakou, T., and Murray, I. (2017).
\newblock \enquote{Masked autoregressive flow for density estimation.}
\newblock {\em Advances in neural information processing systems\/}, 30.
\endbibitem

\bibitem[{Plummer(2015)}]{plummer_cuts_2015}
Plummer, M. (2015).
\newblock \enquote{Cuts in {Bayesian} graphical models.}
\newblock {\em Statistics and Computing\/}, 25(1): 37--43.
\newline\urlprefix\url{http://link.springer.com/10.1007/s11222-014-9503-z}
\endbibitem

\bibitem[{Pompe and Jacob(2021)}]{pompe2021asymptotics}
Pompe, E. and Jacob, P.~E. (2021).
\newblock \enquote{Asymptotics of cut distributions and robust modular inference using Posterior Bootstrap.}
\newblock {\em arXiv preprint arXiv:2110.11149\/}.
\endbibitem

\bibitem[{Radev et~al.(2021)Radev, Graw, Chen, Mutters, Eichel, B{\"a}rnighausen, and K{\"o}the}]{radev2021outbreakflow}
Radev, S.~T., Graw, F., Chen, S., Mutters, N.~T., Eichel, V.~M., B{\"a}rnighausen, T., and K{\"o}the, U. (2021).
\newblock \enquote{OutbreakFlow: Model-based Bayesian inference of disease outbreak dynamics with invertible neural networks and its application to the COVID-19 pandemics in Germany.}
\newblock {\em PLoS computational biology\/}, 17(10): e1009472.
\endbibitem

\bibitem[{Radev et~al.(2020)Radev, Mertens, Voss, Ardizzone, and K{\"o}the}]{radev2020bayesflow}
Radev, S.~T., Mertens, U.~K., Voss, A., Ardizzone, L., and K{\"o}the, U. (2020).
\newblock \enquote{BayesFlow: Learning complex stochastic models with invertible neural networks.}
\newblock {\em IEEE transactions on neural networks and learning systems\/}, 33(4): 1452--1466.
\endbibitem

\bibitem[{Radev et~al.(2023)Radev, Schmitt, Pratz, Picchini, K{\"o}the, and B{\"u}rkner}]{radev2023jana}
Radev, S.~T., Schmitt, M., Pratz, V., Picchini, U., K{\"o}the, U., and B{\"u}rkner, P.-C. (2023).
\newblock \enquote{JANA: Jointly amortized neural approximation of complex Bayesian models.}
\newblock In {\em Uncertainty in Artificial Intelligence\/}, 1695--1706. PMLR.
\endbibitem

\bibitem[{Rezende and Mohamed(2015)}]{rezende2015variational}
Rezende, D. and Mohamed, S. (2015).
\newblock \enquote{Variational inference with normalizing flows.}
\newblock In {\em International conference on machine learning\/}, 1530--1538. PMLR.
\endbibitem

\bibitem[{Robbins(1956)}]{robbins56}
Robbins, H. (1956).
\newblock \enquote{An Empirical Bayes Approach to Statistics.}
\newblock In {\em Proceedings of the Third Berkeley Symposium on Mathematical Statistics and Probability, 1954–1955, Vol. I\/}, 157--163. Berkeley/Los Angeles: Univ. of California Press.
\endbibitem

\bibitem[{Ruggeri et~al.(2005)Ruggeri, Insua, and Mart{\'\i}n}]{ruggeri2005robust}
Ruggeri, F., Insua, D.~R., and Mart{\'\i}n, J. (2005).
\newblock \enquote{Robust bayesian analysis.}
\newblock {\em Handbook of statistics\/}, 25: 623--667.
\endbibitem

\bibitem[{Savitsky et~al.(2011)Savitsky, Vannucci, and Sha}]{savitsky_variable_2011}
Savitsky, T., Vannucci, M., and Sha, N. (2011).
\newblock \enquote{Variable {Selection} for {Nonparametric} {Gaussian} {Process} {Priors}: {Models} and {Computational} {Strategies}.}
\newblock {\em Statistical Science\/}, 26(1).
\newline\urlprefix\url{https://projecteuclid.org/journals/statistical-science/volume-26/issue-1/Variable-Selection-for-Nonparametric-Gaussian-Process-Priors--Models-and/10.1214/11-STS354.full}
\endbibitem

\bibitem[{Schmitt et~al.(2024)Schmitt, Pratz, K{\"o}the, B{\"u}rkner, and Radev}]{schmitt2024consistency}
Schmitt, M., Pratz, V., K{\"o}the, U., B{\"u}rkner, P.-C., and Radev, S. (2024).
\newblock \enquote{Consistency models for scalable and fast simulation-based inference.}
\newblock {\em Advances in Neural Information Processing Systems\/}, 37: 126908--126945.
\endbibitem

\bibitem[{Sharrock et~al.(2022)Sharrock, Simons, Liu, and Beaumont}]{sharrock2022sequential}
Sharrock, L., Simons, J., Liu, S., and Beaumont, M. (2022).
\newblock \enquote{Sequential neural score estimation: Likelihood-free inference with conditional score based diffusion models.}
\newblock {\em arXiv preprint arXiv:2210.04872\/}.
\endbibitem

\bibitem[{Siahkoohi et~al.(2023)Siahkoohi, Rizzuti, Orozco, and Herrmann}]{siahkoohi2023reliable}
Siahkoohi, A., Rizzuti, G., Orozco, R., and Herrmann, F.~J. (2023).
\newblock \enquote{Reliable amortized variational inference with physics-based latent distribution correction.}
\newblock {\em Geophysics\/}, 88(3): R297--R322.
\endbibitem

\bibitem[{Song et~al.(2021)Song, Sohl-Dickstein, Kingma, Kumar, Ermon, and Poole}]{song2021scorebased}
Song, Y., Sohl-Dickstein, J., Kingma, D.~P., Kumar, A., Ermon, S., and Poole, B. (2021).
\newblock \enquote{Score-Based Generative Modeling through Stochastic Differential Equations.}
\newblock In {\em International Conference on Learning Representations\/}.
\newline\urlprefix\url{https://openreview.net/forum?id=PxTIG12RRHS}
\endbibitem

\bibitem[{Syring and Martin(2019)}]{syring2019calibrating}
Syring, N. and Martin, R. (2019).
\newblock \enquote{Calibrating general posterior credible regions.}
\newblock {\em Biometrika\/}, 106(2): 479--486.
\endbibitem

\bibitem[{Tabak and Turner(2013)}]{tabak2013family}
Tabak, E.~G. and Turner, C.~V. (2013).
\newblock \enquote{A family of nonparametric density estimation algorithms.}
\newblock {\em Communications on Pure and Applied Mathematics\/}, 66(2): 145--164.
\endbibitem

\bibitem[{Tabak and Vanden-Eijnden(2010)}]{tabak2010density}
Tabak, E.~G. and Vanden-Eijnden, E. (2010).
\newblock \enquote{Density estimation by dual ascent of the log-likelihood.}
\newblock {\em Communications in Mathematical Sciences\/}, 8(1): 217--233.
\endbibitem

\bibitem[{Vehtari et~al.(2017)Vehtari, Gelman, and Gabry}]{vehtari_practical_2017}
Vehtari, A., Gelman, A., and Gabry, J. (2017).
\newblock \enquote{Practical {Bayesian} model evaluation using leave-one-out cross-validation and {WAIC}.}
\newblock {\em Statistics and Computing\/}, 27(5): 1413--1432.
\newblock ArXiv: 1507.04544 Publisher: Springer US ISBN: 1507.04544.
\newline\urlprefix\url{http://link.springer.com/10.1007/s11222-016-9696-4}
\endbibitem

\bibitem[{von Krause et~al.(2022)von Krause, Radev, and Voss}]{von2022mental}
von Krause, M., Radev, S.~T., and Voss, A. (2022).
\newblock \enquote{Mental speed is high until age 60 as revealed by analysis of over a million participants.}
\newblock {\em Nature human behaviour\/}, 6(5): 700--708.
\endbibitem

\bibitem[{Wainwright et~al.(2008)Wainwright, Jordan et~al.}]{wainwright2008graphical}
Wainwright, M.~J., Jordan, M.~I., et~al. (2008).
\newblock \enquote{Graphical models, exponential families, and variational inference.}
\newblock {\em Foundations and Trends{\textregistered} in Machine Learning\/}, 1(1--2): 1--305.
\endbibitem

\bibitem[{Watanabe(2010)}]{watanabe2010asymptotic}
Watanabe, S. (2010).
\newblock \enquote{Asymptotic equivalence of Bayes cross validation and widely applicable information criterion in singular learning theory.}
\newblock {\em Journal of machine learning research\/}, 11(12).
\endbibitem

\bibitem[{Wildberger et~al.(2023)Wildberger, Dax, Buchholz, Green, Macke, and Sch{\"o}lkopf}]{wildberger2023flow}
Wildberger, J., Dax, M., Buchholz, S., Green, S., Macke, J.~H., and Sch{\"o}lkopf, B. (2023).
\newblock \enquote{Flow matching for scalable simulation-based inference.}
\newblock {\em Advances in Neural Information Processing Systems\/}, 36: 16837--16864.
\newline\urlprefix\url{https://openreview.net/forum?id=D2cS6SoYlP}
\endbibitem

\bibitem[{Winter et~al.(2023)Winter, Melikechi, and Dunson}]{winter2023sequential}
Winter, S., Melikechi, O., and Dunson, D.~B. (2023).
\newblock \enquote{Sequential Gibbs Posteriors with Applications to Principal Component Analysis.}
\newblock {\em arXiv preprint arXiv:2310.12882\/}.
\endbibitem

\bibitem[{Wu et~al.(2020)Wu, Choi, Goodman, and Ermon}]{wu2020meta}
Wu, M., Choi, K., Goodman, N., and Ermon, S. (2020).
\newblock \enquote{Meta-Amortized Variational Inference and Learning.}
\newblock {\em Proceedings of the AAAI Conference on Artificial Intelligence\/}, 34(04): 6404--6412.
\newline\urlprefix\url{https://ojs.aaai.org/index.php/AAAI/article/view/6111}
\endbibitem

\bibitem[{Wu and Martin(2023)}]{wu23}
Wu, P.-S. and Martin, R. (2023).
\newblock \enquote{{A Comparison of Learning Rate Selection Methods in Generalized Bayesian Inference}.}
\newblock {\em Bayesian Analysis\/}, 18(1): 105 -- 132.
\newline\urlprefix\url{https://doi.org/10.1214/21-BA1302}
\endbibitem

\bibitem[{Yu et~al.(2023)Yu, Nott, and Smith}]{yu-nott-cutvi23}
Yu, X., Nott, D.~J., and Smith, M.~S. (2023).
\newblock \enquote{{Variational Inference for Cutting Feedback in Misspecified Models}.}
\newblock {\em Statistical Science\/}, 38(3): 490 -- 509.
\newline\urlprefix\url{https://doi.org/10.1214/23-STS886}
\endbibitem

\bibitem[{Zammit-Mangion et~al.(2024)Zammit-Mangion, Sainsbury-Dale, and Huser}]{zammit2024neural}
Zammit-Mangion, A., Sainsbury-Dale, M., and Huser, R. (2024).
\newblock \enquote{Neural methods for amortized inference.}
\newblock {\em Annual Review of Statistics and Its Application\/}, 12.
\endbibitem

\end{thebibliography}


\begin{thebibliography}{37}
\newcommand{\enquote}[1]{``#1''}
\expandafter\ifx\csname natexlab\endcsname\relax\def\natexlab#1{#1}\fi
\expandafter\ifx\csname url\endcsname\relax
  \def\url#1{{\tt #1}}\fi
\expandafter\ifx\csname urlprefix\endcsname\relax\def\urlprefix{URL }\fi
\ifx\endbibitem\undefined \let\endbibitem\relax\fi

\bibitem[{Battaglia et~al.(2025)Battaglia, Carmona, Haines, {Andersen Loake}, Benskin, and Nicholls}]{battaglia2025supplement}
Battaglia, L., Carmona, C.~U., Haines, R.~A., {Andersen Loake}, M., Benskin, M., and Nicholls, G.~K. (2025).
\newblock \enquote{Supplement of Amortising Variational Bayesian Inference over prior hyperparameters with a Normalising Flow.}
\newblock DOI: 10.1214/[provided by typesetter].
\endbibitem

\bibitem[{Benskin(1982)}]{benskin1982letter}
Benskin, M. (1982).
\newblock \enquote{The letter {\textless}þ{\textgreater} and {\textless}y{\textgreater} in later middle {English}, and some related matters.}
\newblock {\em Journal of the Society of Archivists\/}, 7(1): 13--30.
\newblock Publisher: Routledge.
\newline\urlprefix\url{https://doi.org/10.1080/00379818209514199}
\endbibitem

\bibitem[{Benskin(1988)}]{benskin_numerical_1988}
--- (1988).
\newblock \enquote{The numerical classification of languages, and dialect maps for the past.}
\newblock In Reenen, P.~v. and Reenen-Stein, K.~v. (eds.), {\em Distributions spatiales et temporelles, constellations des manuscrits/{Spatial} and {Temporal} {Distributions}, {Manuscript} {Constellations}: {Etudes} de variation linguistiques offertes à {Anthonij} {Dees} à l'occasion de son 60ème anniversaire / {Studies} in language variation offered to {Anthonij} {Dees} on the occasion of his 60th birthday\/}. John Benjamins Publishing Company.
\newline\urlprefix\url{https://benjamins.com/catalog/z.37.07ben}
\endbibitem

\bibitem[{Benskin(2004{\natexlab{a}})}]{benskin2004chancery}
--- (2004{\natexlab{a}}).
\newblock \enquote{Chancery standard.}
\newblock {\em Amsterdam studies in the theory and history of linguistic science, Series 4\/}, 252: 1--40.
\endbibitem

\bibitem[{Benskin(2004{\natexlab{b}})}]{benskin_chancery_2004}
--- (2004{\natexlab{b}}).
\newblock \enquote{Chancery {Standard}.}
\newblock In Kay, C., Hough, C., and Wotherspoon, I. (eds.), {\em New {Perspectives} on {English} {Historical} {Linguistics} {Volume} {II}: {Lexis} and {Transmission}\/}, Current {Issues} in {Linguistic} {Theory}, 21--26. Amsterdam/Philadelphia: John Benjamins Publishing Company.
\newline\urlprefix\url{https://benjamins.com/catalog/cilt.252.03ben}
\endbibitem

\bibitem[{Benskin and Laing(1981{\natexlab{a}})}]{benskin1981translations}
Benskin, M. and Laing, M. (1981{\natexlab{a}}).
\newblock \enquote{Translations and mischsprachen in Middle English manuscripts.}
\newblock In {\em So meny people longages and tonges: Philological essays in Scots and mediaeval English presented to Angus McIntosh\/}, 55--106. University of Edinburgh.
\endbibitem

\bibitem[{Benskin and Laing(1981{\natexlab{b}})}]{benskin_translations_1981}
--- (1981{\natexlab{b}}).
\newblock \enquote{Translations and {Mischsprachen} in {Middle} {English} {Manuscripts}.}
\newblock In {\em So meny people longages and tonges: philological essays in {Scots} and mediaeval {English} presented to {Angus} {McIntosh}\/}, 55--106. University of Edinburgh.
\newblock ISBN: 0950693820.
\newline\urlprefix\url{https://www.research.ed.ac.uk/en/publications/translations-and-mischsprachen-in-middle-english-manuscripts}
\endbibitem

\bibitem[{Benskin et~al.(2013)Benskin, Laing, Karaiskos, and Williamson}]{benskin_electronic_2013}
Benskin, M., Laing, M., Karaiskos, V., and Williamson, K. (2013).
\newblock \enquote{An {Electronic} {Version} of {A} {Linguistic} {Atlas} of {Late} {Mediaeval}.}
\newblock Url: https://www.lel.ed.ac.uk/ihd/elalme/elalme.html.
\newline\urlprefix\url{https://www.lel.ed.ac.uk/ihd/elalme/elalme.html}
\endbibitem

\bibitem[{Biewald(2020)}]{wandb}
Biewald, L. (2020).
\newblock \enquote{Experiment Tracking with Weights and Biases.}
\newblock Software available from wandb.com.
\newline\urlprefix\url{https://www.wandb.com/}
\endbibitem

\bibitem[{Bonilla et~al.(2019)Bonilla, Krauth, and Dezfouli}]{bonilla_generic_2019}
Bonilla, E.~V., Krauth, K., and Dezfouli, A. (2019).
\newblock \enquote{Generic inference in latent {Gaussian} process models.}
\newblock {\em Journal of Machine Learning Research\/}, 20.
\newblock ArXiv: 1609.00577.
\newline\urlprefix\url{http://arxiv.org/abs/1609.00577}
\endbibitem

\bibitem[{Bradbury et~al.(2018)Bradbury, Frostig, Hawkins, Johnson, Leary, Maclaurin, Necula, Paszke, Vander{P}las, Wanderman-{M}ilne, and Zhang}]{jax2018github}
Bradbury, J., Frostig, R., Hawkins, P., Johnson, M.~J., Leary, C., Maclaurin, D., Necula, G., Paszke, A., Vander{P}las, J., Wanderman-{M}ilne, S., and Zhang, Q. (2018).
\newblock \enquote{{JAX}: composable transformations of {P}ython+{N}um{P}y programs.}
\newline\urlprefix\url{http://github.com/jax-ml/jax}
\endbibitem

\bibitem[{Cabezas et~al.(2024)Cabezas, Corenflos, Lao, and Louf}]{cabezas2024blackjax}
Cabezas, A., Corenflos, A., Lao, J., and Louf, R. (2024).
\newblock \enquote{BlackJAX: Composable {B}ayesian inference in {JAX}.}
\endbibitem

\bibitem[{Carmona and Nicholls(2020)}]{carmona_semi-modular_2020}
Carmona, C.~U. and Nicholls, G.~K. (2020).
\newblock \enquote{Semi-{Modular} {Inference}: enhanced learning in multi-modular models by tempering the influence of components.}
\newblock In Silvia, C. and Calandra, R. (eds.), {\em Proceedings of the 23rd {International} {Conference} on {Artificial} {Intelligence} and {Statistics}, {AISTATS} 2020\/}, 4226--4235. PMLR.
\newblock ArXiv: 2003.06804.
\newline\urlprefix\url{http://arxiv.org/abs/2003.06804}
\endbibitem

\bibitem[{Carmona and Nicholls(2022)}]{carmona_scalable_2022}
--- (2022).
\newblock \enquote{Scalable {Semi}-{Modular} {Inference} with {Variational} {Meta}-{Posteriors}.}
\newblock ArXiv: 2204.00296.
\newline\urlprefix\url{https://github.com/chriscarmona/modularbayes}
\endbibitem

\bibitem[{DeepMind et~al.(2020)DeepMind, Babuschkin, Baumli, Bell, Bhupatiraju, Bruce, Buchlovsky, Budden, Cai, Clark, Danihelka, Dedieu, Fantacci, Godwin, Jones, Hemsley, Hennigan, Hessel, Hou, Kapturowski, Keck, Kemaev, King, Kunesch, Martens, Merzic, Mikulik, Norman, Papamakarios, Quan, Ring, Ruiz, Sanchez, Sartran, Schneider, Sezener, Spencer, Srinivasan, Stanojevi\'{c}, Stokowiec, Wang, Zhou, and Viola}]{deepmind2020jax}
DeepMind, Babuschkin, I., Baumli, K., Bell, A., Bhupatiraju, S., Bruce, J., Buchlovsky, P., Budden, D., Cai, T., Clark, A., Danihelka, I., Dedieu, A., Fantacci, C., Godwin, J., Jones, C., Hemsley, R., Hennigan, T., Hessel, M., Hou, S., Kapturowski, S., Keck, T., Kemaev, I., King, M., Kunesch, M., Martens, L., Merzic, H., Mikulik, V., Norman, T., Papamakarios, G., Quan, J., Ring, R., Ruiz, F., Sanchez, A., Sartran, L., Schneider, R., Sezener, E., Spencer, S., Srinivasan, S., Stanojevi\'{c}, M., Stokowiec, W., Wang, L., Zhou, G., and Viola, F. (2020).
\newblock \enquote{The {D}eep{M}ind {JAX} {E}cosystem.}
\newline\urlprefix\url{http://github.com/deepmind}
\endbibitem

\bibitem[{Durkan et~al.(2019)Durkan, Bekasov, Murray, and Papamakarios}]{durkan_neural_2019}
Durkan, C., Bekasov, A., Murray, I., and Papamakarios, G. (2019).
\newblock \enquote{Neural {Spline} {Flows}.}
\newblock In {\em Proceedings of the 33rd {Conference} on {Neural} {Information} {Processing} {Systems}, {NeurIPS} 2019\/}.
\newblock ArXiv: 1906.04032.
\newline\urlprefix\url{http://arxiv.org/abs/1906.04032}
\endbibitem

\bibitem[{Haines(2016)}]{haines_simultaneous_2016}
Haines, R.~A. (2016).
\newblock \enquote{Simultaneous {Reconstruction} of {Spatial} {Frequency} {Fields} and {Field} {Sample} {Locations}.}
\newblock Ph.D. thesis, University of Oxford.
\endbibitem

\bibitem[{Hensman et~al.(2013)Hensman, Fusi, and Lawrence}]{hensman_gaussian_2013}
Hensman, J., Fusi, N., and Lawrence, N.~D. (2013).
\newblock \enquote{Gaussian {Processes} for {Big} {Data}.}
\newblock In {\em Proceedings of the 29th {Conference}, of {Uncertainty} in {Artificial} {Intelligence}, {UAI} 2013\/}, 282--290.
\newblock ArXiv: 1309.6835.
\newline\urlprefix\url{http://arxiv.org/abs/1309.6835}
\endbibitem

\bibitem[{Hensman et~al.(2015{\natexlab{a}})Hensman, Matthews, and Ghahramani}]{hensman_scalable_2015}
Hensman, J., Matthews, A., and Ghahramani, Z. (2015{\natexlab{a}}).
\newblock \enquote{Scalable {Variational} {Gaussian} {Process} {Classification}.}
\newblock In {\em Proceedings of the 18th {International} {Conference} on {Artificial} {Intelligence} and {Statistics}, {AISTATS} 2015\/}.
\newblock ArXiv: 1411.2005.
\newline\urlprefix\url{http://arxiv.org/abs/1411.2005}
\endbibitem

\bibitem[{Hensman et~al.(2015{\natexlab{b}})Hensman, Matthews, and Ghahramani}]{hensman2015scalable}
--- (2015{\natexlab{b}}).
\newblock \enquote{Scalable variational Gaussian process classification.}
\newblock In {\em Artificial Intelligence and Statistics\/}, 351--360. PMLR.
\endbibitem

\bibitem[{Hoffman et~al.(2014)Hoffman, Gelman et~al.}]{hoffman2014no}
Hoffman, M.~D., Gelman, A., et~al. (2014).
\newblock \enquote{The No-U-Turn sampler: adaptively setting path lengths in Hamiltonian Monte Carlo.}
\newblock {\em J. Mach. Learn. Res.\/}, 15(1): 1593--1623.
\endbibitem

\bibitem[{Jacob et~al.(2020)Jacob, O’Leary, and Atchadé}]{jacob20}
Jacob, P.~E., O’Leary, J., and Atchadé, Y.~F. (2020).
\newblock \enquote{{Unbiased Markov Chain Monte Carlo Methods with Couplings}.}
\newblock {\em Journal of the Royal Statistical Society Series B: Statistical Methodology\/}, 82(3): 543--600.
\newline\urlprefix\url{https://doi.org/10.1111/rssb.12336}
\endbibitem

\bibitem[{Liu and Goudie(2022)}]{liu_stochastic_2022}
Liu, Y. and Goudie, R. J.~B. (2022).
\newblock \enquote{Stochastic approximation cut algorithm for inference in modularized {Bayesian} models.}
\newblock {\em Statistics and Computing\/}, 32(1): 7.
\newblock ArXiv: 2006.01584.
\newline\urlprefix\url{http://arxiv.org/abs/2006.01584}
\endbibitem

\bibitem[{McIntosh(1963)}]{mcintosh_new_1963}
McIntosh, A. (1963).
\newblock \enquote{A new approach to {Middle} {English} dialectology.}
\newblock {\em English Studies\/}, 44(1-6): 1--11.
\newblock Publisher: Routledge.
\newline\urlprefix\url{https://doi.org/10.1080/00138386308597152}
\endbibitem

\bibitem[{McIntosh(1974)}]{McIntosh1974}
--- (1974).
\newblock \enquote{Towards an inventory of Middle English scribes.}
\newblock {\em Neuphilologische Mitteilungen\/}, 75: 602--624.
\newblock Reprinted in \textit{Middle English Dialectology: Essays on some Principles and Problems}, ed. M. Laing (Aberdeen: Aberdeen University Press, 1989), pp. 46--63.
\endbibitem

\bibitem[{McIntosh(1975)}]{McIntosh1975}
--- (1975).
\newblock \enquote{Scribal profiles from Middle English texts.}
\newblock {\em Neuphilologische Mitteilungen\/}, 76: 218--235.
\newblock Reprinted in \textit{Middle English Dialectology: Essays on some Principles and Problems}, ed. M. Laing (Aberdeen: Aberdeen University Press, 1989), pp. 32--45.
\endbibitem

\bibitem[{McIntosh et~al.(1986)McIntosh, Samuels, and Benskin}]{mcintosh_linguistic_1986}
McIntosh, A., Samuels, M.~L., and Benskin, M. (1986).
\newblock {\em A {Linguistic} {Atlas} of {Late} {Mediaeval} {English}\/}.
\newblock Aberdeen: Aberdeen University Press.
\endbibitem

\bibitem[{Orton and Dieth(1962)}]{orton_survey_1962}
Orton, H. and Dieth, E. (1962).
\newblock {\em Survey of {English} {Dialects}\/}.
\newblock University of Leeds.
\newblock Google-Books-ID: FqrJzgEACAAJ.
\endbibitem

\bibitem[{Plummer(2015)}]{plummer_cuts_2015}
Plummer, M. (2015).
\newblock \enquote{Cuts in {Bayesian} graphical models.}
\newblock {\em Statistics and Computing\/}, 25(1): 37--43.
\newline\urlprefix\url{http://link.springer.com/10.1007/s11222-014-9503-z}
\endbibitem

\bibitem[{Quiñonero-Candela and Rasmussen(2005)}]{quinonero-candela_unifying_2005}
Quiñonero-Candela, J. and Rasmussen, C.~E. (2005).
\newblock \enquote{A unifying view of sparse approximate {Gaussian} process regression.}
\newblock {\em Journal of Machine Learning Research\/}, 6: 1939--1959.
\endbibitem

\bibitem[{Rezende and Mohamed(2015)}]{rezende2015variational}
Rezende, D. and Mohamed, S. (2015).
\newblock \enquote{Variational inference with normalizing flows.}
\newblock In {\em International conference on machine learning\/}, 1530--1538. PMLR.
\endbibitem

\bibitem[{Samuels(1963)}]{Samuels1963}
Samuels, M.~L. (1963).
\newblock \enquote{Some applications of Middle English dialectology.}
\newblock {\em English Studies\/}, 44: 81--94.
\newblock Reprinted in \textit{Middle English Dialectology: Essays on some Principles and Problems}, ed. M. Laing (Aberdeen: Aberdeen University Press, 1989), pp. 64--80.
\endbibitem

\bibitem[{Samuels(1981)}]{Samuels1981}
--- (1981).
\newblock \enquote{Spelling and dialect in the late- and post-Middle English periods.}
\newblock In Benskin, M. and Samuels, M.~L. (eds.), {\em So meny people longages and tonges: Philological essays in Scots and Middle English presented to Angus McIntosh\/}, 43--54. Edinburgh: the editors.
\endbibitem

\bibitem[{Snoek et~al.(2012)Snoek, Larochelle, and Adams}]{snoek_practical_2012}
Snoek, J., Larochelle, H., and Adams, R.~P. (2012).
\newblock \enquote{Practical {Bayesian} {Optimization} of {Machine} {Learning} {Algorithms}.}
\newblock In {\em Advances in {Neural} {Information} {Processing} {Systems}\/}, volume~25. Curran Associates, Inc.
\newline\urlprefix\url{https://papers.nips.cc/paper/2012/hash/05311655a15b75fab86956663e1819cd-Abstract.html}
\endbibitem

\bibitem[{Titsias(2009)}]{titsias_variational_2009}
Titsias, M. (2009).
\newblock \enquote{Variational {Learning} of {Inducing} {Variables} in {Sparse} {Gaussian} {Processes}.}
\newblock In van Dyk, D. and Welling, M. (eds.), {\em Proceedings of the 12th {International} {Conference} on {Artificial} {Intelligence} and {Statistics}, {AISTATS} 2009\/}, volume~5, 567--574. Hilton Clearwater Beach Resort, Clearwater Beach, Florida USA: PMLR.
\newblock Series Title: Proceedings of Machine Learning Research.
\newline\urlprefix\url{http://proceedings.mlr.press/v5/titsias09a.html}
\endbibitem

\bibitem[{Williamson(2008)}]{Williamson2008}
Williamson, K. (2008).
\newblock \enquote{A Linguistic Atlas of Older Scots, 1380--1500.}
\newblock \url{http://www.lel.ed.ac.uk/ihd/laos1/laos1.html}.
\newblock Online resource.
\endbibitem

\bibitem[{Zhuang et~al.(2020)Zhuang, Tang, Ding, Tatikonda, Dvornek, Papademetris, and Duncan}]{zhuang_adabelief_2020}
Zhuang, J., Tang, T., Ding, Y., Tatikonda, S., Dvornek, N., Papademetris, X., and Duncan, J.~S. (2020).
\newblock \enquote{{AdaBelief} {Optimizer}: {Adapting} {Stepsizes} by the {Belief} in {Observed} {Gradients}.}
\newblock ArXiv:2010.07468 [cs, stat].
\newline\urlprefix\url{http://arxiv.org/abs/2010.07468}
\endbibitem

\end{thebibliography}
\end{document}